%% file: ms.tex
\documentclass[12pt,preprint]{aastex}

\def\ltsima{$\; \buildrel < \over \sim \;$}
\def\simlt{\lower.5ex\hbox{\ltsima}}
\def\gtsima{$\; \buildrel > \over \sim \;$}
\def\simgt{\lower.5ex\hbox{\gtsima}}

\shorttitle{Integrated Light GC Abundances}
\shortauthors{McWilliam \& Bernstein}

\begin{document}

\newcommand{\znh}{[{\rm Zn/H}]}
\newcommand{\msol}{M_\odot}
\newcommand{\etal}{et al.\ }
\newcommand{\delv}{\Delta v}
\newcommand{\kms}{km~s$^{-1}$ }
\newcommand{\cm}[1]{\, {\rm cm^{#1}}}
\newcommand{\N}[1]{{N({\rm #1})}}
\newcommand{\e}[1]{{\epsilon({\rm #1})}}
\newcommand{\f}[1]{{f_{\rm #1}}}
\newcommand{\rAA}{{\AA \enskip}}
\newcommand{\sci}[1]{{\rm \; \times \; 10^{#1}}}
\newcommand{\ltk}{\left [ \,}
\newcommand{\ltp}{\left ( \,}
\newcommand{\ltb}{\left \{ \,}
\newcommand{\rtk}{\, \right  ] }
\newcommand{\rtp}{\, \right  ) }
\newcommand{\rtb}{\, \right \} }
\newcommand{\ohf}{{1 \over 2}}
\newcommand{\nohf}{{-1 \over 2}}
\newcommand{\rhf}{{3 \over 2}}
\newcommand{\smm}{\sum\limits}
\newcommand{\perd}{\;\;\; .}
\newcommand{\cmma}{\;\;\; ,}
\newcommand{\intl}{\int\limits}
\newcommand{\mkms}{{\rm \; km\;s^{-1}}}
\newcommand{\ew}{W_\lambda}


\title{Globular Cluster Abundances from High-Resolution Integrated
  Light Spectra, I: 47 Tuc}

\author{Andrew McWilliam}
\affil{The Observatories of the Carnegie Institute of Washington, \\
813 Santa Barbara St., Pasadena, CA 91101--1292}
\email{andy@ociw.edu}

\and 

\author{Rebecca A. Bernstein}
\affil{Department of Astronomy and Astrophysics, 1156 High Street,\\
UCO/Lick Observatory, University of California, Santa Cruz, CA 95064}
\email{rab@ucolick.org}

\begin{abstract}

  We describe the detailed chemical abundance analysis of a
  high-resolution (R$\sim$35,000), integrated-light (IL), spectrum of
  the core of the Galactic globular cluster 47~Tuc, obtained using the
  du~Pont echelle at Las Campanas.  This is the first paper describing
  the abundance analysis of a set of Milky Way clusters of which we
  have obtained integrated light spectra; we are analyzing these to
  develop and demonstrate an abundance analysis strategy that can be
  applied to spatial unresolved, extra-galactic clusters.

  We have computed abundances for 
  Na, Mg, Al, Si, Ca, Sc, Ti, V, Cr, Mn, Fe, Co, Ni, Cu, Y, Zr, Ba, La, Nd
  and Eu.  For an analysis with the known color-magnitude
  diagram ({\em cmd}) for 47~Tuc we obtain a mean [Fe/H] value 
  of $-$0.75 $\pm$0.026$\pm$0.045 dex (random and systematic error), 
  in good agreement with the mean of 5 recent 
  high resolution abundance studies, at $-$0.70 dex.  Typical random errors
  on our mean [X/Fe] ratios are 0.07--0.10 dex, similar
  to studies of individual stars in 47~Tuc.
  Only Na and Al appear
  anomalous, compared to abundances from individual stars: they are
  enhanced in the IL spectrum, perhaps due to proton burning
  in the most luminous cluster stars.
  
  Our IL abundance analysis with an unknown {\em cmd} employed
  theoretical Teramo isochrones; however, the observed frequency of AGB bump 
  stars in 47~Tuc is approximately three times higher than theoretical predictions,
  so we adopt zero-point abundance corrections to account for this deficiency.
  The spectra, alone, provide mild constraints on the cluster age, ruling-out
  ages younger than $\sim$2 Gyr.  However, when we include theoretical IL B$-$V colors
  and metallicity derived from the
  Fe~I lines, the age is constrained to 10--15 Gyr and we obtain
  [Fe/H]=$-$0.70$\pm$0.021$\pm$0.052 dex.


  We also find that trends of IL Fe line abundances with equivalent width
  and excitation potential can constrain the horizontal
  branch morphology of an unresolved cluster, as well as age.  Lastly,
  our spectrum synthesis of 5.4 million TiO lines indicates that the
  7300--7600\AA\ TiO window should be useful for estimating the effect of M giants
  on the IL abundances, and important for clusters more metal-rich than 47~Tuc.

\end{abstract}


\section{Introduction}

Detailed high-resolution chemical abundance analysis of individual
stars in Galactic globular clusters (GCs) has been pursued for 
30 years (e.g. Cohen 1978; Pilachowski, Canterna \& Wallerstein 1980).
Such abundance studies are a critical tool for probing the chemical
evolution of the Galaxy as well as stellar evolution up the giant
branch (e.g. Sneden et al. 1991; Briley, Smith \& Lambert 1994) and
complemented earlier work on nearby field stars (e.g. Wallerstein
1962, Luck \& Bond 1985).  Unfortunately, similar studies have never
been possible in other galaxies beyond the Local Group dwarfs,
as even the brightest red giant branch stars are too faint to be seen
in distant galaxies.  Indeed, only recently have abundances of
individual GC red giant stars in the closest members of the Local
Group become available (e.g. Johnson, Ivans \& Stetson 2006; Letarte
et al. 2006).

For extra-galactic GCs, integrated-light metallicities have been
estimated for systems ranging from M31 to the Virgo cluster of galaxies,
using broad-band photometric colors (e.g. Forte, Strom \&
Strom 1981; Geisler et al. 1996) and low-resolution spectra
(e.g. Racine, Oke \& Searle 1978; Brodie \& Huchra 1990).  The low
resolution spectra typically employ the ``Lick'' index system 
(e.g. Faber 1973), which is based on correlations between strong absorption
features at low resolution and detailed abundances obtained for individual
Galactic GC stars, at high spectral resolution.
All of these indices contain numerous lines from several elements, although many
are dominated by particular species that can be empirically calibrated to
give very approximate composition information (e.g. the Mg$_2$ index can be
calibrated for approximate [Mg/Fe] or [$\alpha$/Fe]).  An example of
this is the study of NGC~5128 by Peng, Ford \& Freeman (2004); in this
case even quite crude composition information provides a powerful investigative
tool.  Results from such studies include the discovery of bimodal GC
metallicity/color distribution in extra-galactic systems (e.g. Elson \& Santiago 1996;
Whitmore et al. 1995), reminiscent of the bimodal GCs in the Galaxy (e.g. Zinn 1985).  

Because of their relative homogeneity, ages ($>>$ 1 Gyr), and high luminosity,
GCs can be used to probe the chemical evolution history of galaxies.  The most 
luminous GCs presumably only trace the major star forming events, including mergers.
As indicated above, even basic metallicity provides interesting information 
for comparison with the Milky Way galaxy.  However, detailed chemical composition
of GCs could potentially provide a plethora of information on galaxy evolution,
because the chemical elements are produced by a variety of stars, with
varying sensitivity to stellar mass and metallicity.

\begin{figure}
\plotone{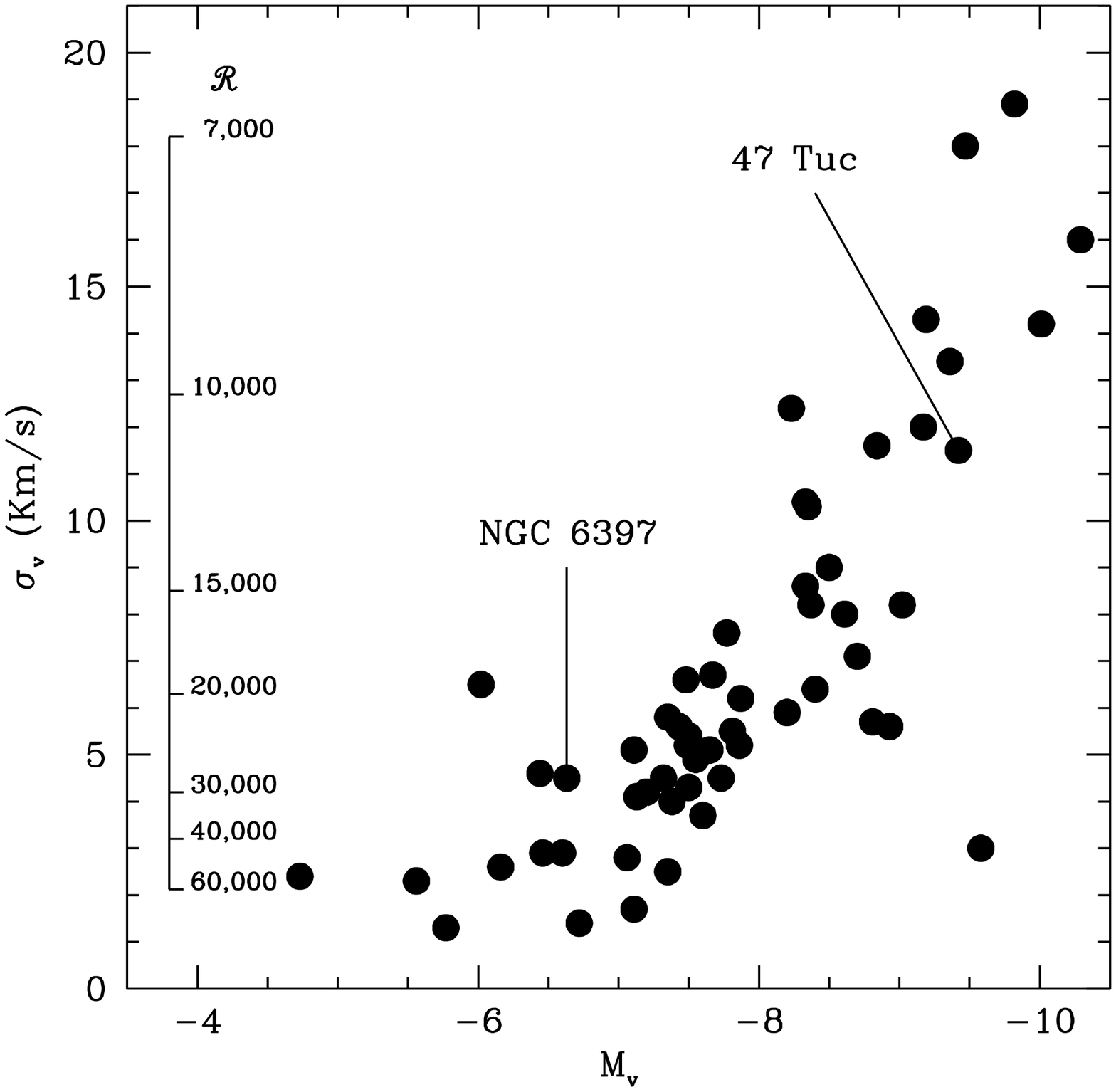}
\caption{The relation between velocity dispersion and globular cluster
  absolute visual magnitude for Galactic globular clusters, based on
  Pryor \& Meylan (1993); 47~Tuc (the globular cluster analyzed in
  this paper) and NGC~6397 (a metal-poor cluster of low luminosity)
  are identified.}
\label{fig-sigmv}
\end{figure}

We are developing a method for measuring detailed chemical abundances
of GCs using high resolution spectra of their integrated light
(McWilliam \& Bernstein 2002; Bernstein \& McWilliam 2005).  Due to
low the velocity dispersions of GCs ($\sigma$$\sim$1--20 Km/s), the
intrinsic line widths of the integrated light spectra are small enough
that individual lines are well resolved and blending is not much more
problematic than for individual red giant stars.  In
Figure~\ref{fig-sigmv}, we demonstrate this point with a plot of
velocity dispersion as a function of M$_{\rm v}$ for Galactic GCs. Also
shown in that figure is the line width parameter, $R$=$\lambda$/FWHM,
that corresponds to the plotted velocity dispersions.  It is clear
from this plot that spectral resolutions of R$\geq$10,000 are
necessary to resolve the spectral lines in even the brightest GCs;
R$\geq$30,000 is necessary for the average Galactic GCs.  With
spectrograph resolving power somewhat larger than these values of $R$,
line profiles can be fully characterized.  These narrow line widths
are in stark contrast to those of giant galaxies (elliptical and
spiral), which have velocity dispersions in the range 100-300 Km/s
(e.g. Faber \& Jackson 1976) and line width parameters, $R$, of order
1,000--3,000.  The Lick system, which has been used to measure the
ages and metallicities of stellar systems from GCs to giant elliptical
galaxies, is based on spectra with resolving power
$\lambda/\Delta\lambda\sim$600 (e.g. Faber et al. 1985; Worthey et
al. 1994; Trager 2004).  While the sampling of this index system is
justified for faint giant galaxies, a low resolution system such as
this does not utilize the available information for bright GCs.  As we
show in this paper, it is possible to use high dispersion
integrated-light spectra of GCs to reveal a wealth of abundance
information from the weak lines of numerous elements that are lost in
low resolution, low S/N, spectra. Indeed, high resolution spectra,
employing diagnostics similar to those used in the spectral analysis
of single stars, can be used to break the age-metallicity degeneracy
that is troublesome at low spectral resolution.

In this paper we demonstrate the possibilities and some of the
practical challenges for using IL spectra of GCs to measure detailed
chemical composition. This analysis holds significant potential for
galaxy evolution constraints.  Due to the high luminosity of GCs,
high-resolution spectra of sufficient quality can be obtained for GCs
at extra-galactic distances and used to probe the chemical evolution
of distant galaxies.  The luminosities of the brightest GCs are
comparable to young supergiant stars, for which abundances have been
measured in Local Group galaxies (e.g. Venn et al. 2001; Kaufer et al.
2004) using spectra from large telescopes.  However, unlike short-lived,
supergiant stars that reveal only recent gas compositions, 
GCs ages cover the full range of galactic history.

In particular, this paper focuses on the analysis of 
one cluster (NGC 104) from a sample of ``training set''
clusters that we have observed and analyzed with the goal of 
developing and demonstrating a technique for abundance analysis of a 
single age, chemically homogeneous, stellar population. 
Our ``training set''  is comprised of a sample
of Galactic and LMC GCs.  Of these, the Galactic GCs are 
well studied, with photometric and detailed
abundance studies of individual stars in the literature.  Because they
are well studied, distance and reddening parameters are readily
available for use in our analysis, as are ``fiducial'' abundances
obtained from individual stars which we can use to check our final
abundance results.  We begin with the luminous GC 47~Tuc
([Fe/H]$\sim$$-$0.7 e.g. Carretta et al. 2004) for several reasons.
47~Tuc is very bright, making IL spectra easy to obtain, and it has
very low reddening (near E(B$-$V)=0.03 magnitudes), which will be
useful for the analysis.  Furthermore, the relatively high metallicity
results in easy visibility of lines from many elements.

\section{Observations and Reductions}

We acquired high resolution, integrated light spectra of our ``training
set'' GCs using the echelle spectrograph on the Las Campanas 2.5m
du Pont telescope during lunar dark time in July 2000.  We limit the
discussion here to 47~Tuc; the full training set
will be discussed in a later paper (Bernstein et al 2008). The
observations were facilitated by a modification to the telescope
guider program, kindly performed by S. Shectman, which enabled 
a uniform scan of the echelle slit across a $32\times32$ arcsecond square
region of sky, from $16''$ South-West to $16''$ North-East of the
field center.  Since the cluster regions were scanned once per
exposure, clear skies were necessary to ensure an unbiased weighting
of the cluster light.  The entire 1$\times$4 arcsecond echelle slit was
filled with cluster light during these scans of the cluster core, and
significant sky flux outside the telluric emission lines was only
detected near twilight. Nevertheless sky exposures were taken
separately to allow subtraction of the sky signal from the science
exposures.  We obtained three exposures of roughly 1 hour each on the
47~Tuc core and five 20 minute exposures on the sky.

We performed the basic data reduction steps using the echelle package
in IRAF, including the routines for overscan, bias subtraction, and
flat-field division.  The sky spectra were scaled and subtracted from
the individual integrated-light exposures using simple arithmetic
routines.  

The mean line widths of the Th-Ar comparison spectra, at 2.6 pixels,
correspond to a spectral resolving power ($R$=$\lambda$/FWHM) of 34,760,
but this value varies over the CCD by $\sigma$$\sim$4\%, due to focus 
variations in the camera optics.

While our du~Pont echelle integrated-light spectra cover the wavelength
interval 3700--7800\AA , the useful range is limited at the red end, because 
the orders are too close in the cross dispersion direction for good extraction, 
and due to blending with telluric absorption lines.  The utility of the blue
side is limited by the reduced flux and increased line blending.  Thus, the
useful wavelength coverage of our 47~Tuc spectrum is from $\sim$5000 to 7570\AA .

\begin{figure}
\plotone{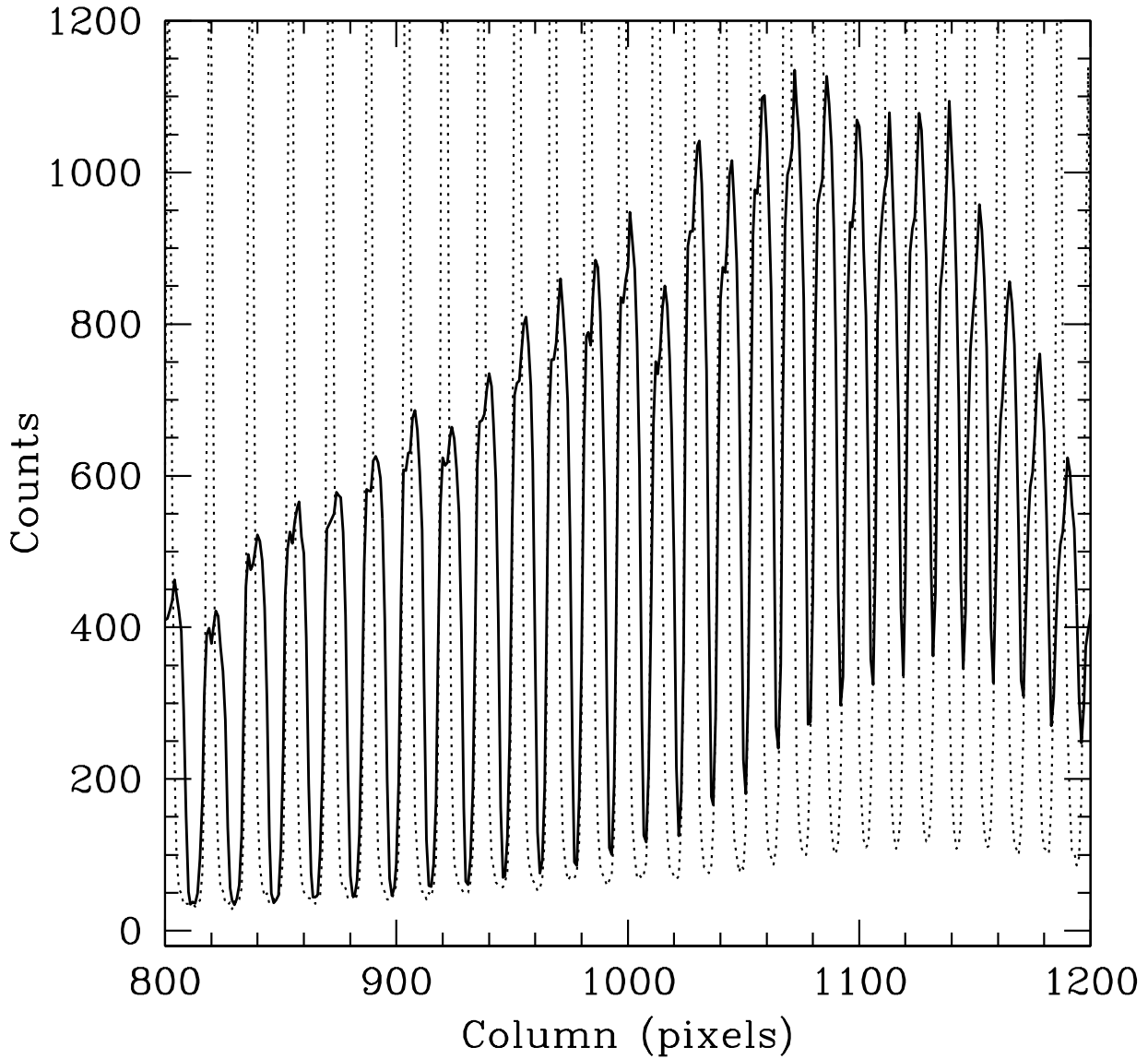}
\caption{A cut along the cross-dispersion direction of a 47~Tuc echelle 
integrated-light exposure (1$\times$4 arc sec slit; heavy solid line)
compared with a scaled echelle spectrum of a single red giant star 
through a short slit (0.75$\times$0.75 arc sec; dotted line).  
The short slit flux was scaled to the total flux in the 47~Tuc spectrum 
in the order near column 1055.  Note the similarity of the inter-order
light of the two spectra below column 910.  The sharp minima of the
47~Tuc spectra at larger column positions (redder wavelengths) suggests
that the inter-order light there suffers contamination from the
order wings.  Where the GC IL scattered-light cannot be measured directly
we employ the ratio of total flux to scattered-light
in the short slit spectrum to determine the scattered-light
for the 47~Tuc spectrum.}
\label{fig-scatt}
\end{figure}

In spectra from the du Pont echelle, the wings of adjacent orders in
the red region of the spectrum ($\lambda$$\geq$6000\AA) overlap
slightly, making it difficult to identify the local scattered-light
background levels; this problem is particularly acute when the source
fills the slit, as in our observations, thus widening the order profile
relative to the spectrum of a point source.  It was therefore necessary to
estimate the scattered-light background using an empirical model.  To
do so, we used the two-dimensional image of a spectrum of a bright red
giant star, taken with the smallest possible slit (0.75$\times$0.75
arcseconds square).  The bright giant star flux distribution is
similar to that of the integrated-light spectrum of an old cluster
like 47~Tuc.  
This short slit gave well separated spectral orders, with
much less inter-order light in the red region (where the orders are
closest) than for the cluster integrated-light spectra.  The
relatively flat bottomed troughs of the short slit inter-order
light suggests scattered-light rather than overlapping wings of
adjacent orders.  In Figure~\ref{fig-scatt} we show a comparison of the GC
spectral orders and the orders of the short slit star spectrum;
essentially, we use the ratio of total flux in an order to the
inter-order flux for the short slit spectrum to define the scattered
light in the GC spectrum.  It is notable that the scaled short slit
inter-order flux in Figure~\ref{fig-scatt} is consistent with the GC
inter-order light for column positions less than 910 pixels.
We found that the scattered light is roughly $\sim 6$\% of the total
stellar flux, in a given order, and is a function of wavelength redder
than $5000$\AA.  To
measure the total flux in each order of the science spectra, we
extracted the orders using the IRAF echelle routine {\em apall} with
wide apertures and without background subtraction.  The scattered light flux
in the model was scaled according to the total flux and the size of the
extraction aperture and then subtracted from the object spectrum.

We noticed that the scattered light followed approximately the continuous
flux distribution of the object spectrum, in the wavelength and
cross-dispersion directions.  
To gain a greater understanding of the scattered-light, we constructed a
simple analytical model for the scattering and performed numerical 
experiments to try to mimic the shape of the scattered-light along the
dispersion and cross-dispersion directions on the CCD. This analytical
model included only two parameters: the amount of light scattered and a
scattering scale length, based on a Gaussian scattered-light distribution.
Our results indicated that the scattered-light scale length was
approximately 200 pixels.  At any given pixel, the scattered light therefore
came from a $\sim$10--20 \AA\ region in the dispersion direction and roughly
8 or 9 echelle orders in the cross-dispersion direction.  We used only the
empirical scattered-light model derived from the bright star spectrum through
the (0.75$\times$0.75 arc sec) short slit, not the analytical model,
in the reduction of our science exposures.

Although the empirical scattered-light model appeared to work quite
well for the red region of the spectrum, it did not always compare
well with the measured scattered light in the blue, where the orders
are well spaced and scattered light can be measured directly.
We believe that this is because the scattered-light in
the blue portion of the CCD was overwhelmed by scattered light from
the bright red part of the spectrum; essentially, a small fraction of
light scattered from the far red end of the spectrum could 
overwhelm the local scattered light between the blue spectral orders.  In 
this case, a color difference between 47~Tuc and the star used for the
empirical scattered-light model could make the empirical scattering
model unreliable in the blue spectral region.  Fortunately, the bluer
orders of the IL spectrum are well separated, so that it is possible
to measure the scattered light directly from the GC IL spectrum at these
wavelengths.  Therefore, we chose to use the scattered-light model
only for the red orders where direct measurement of the scattered
light was not possible.

After extraction and wavelength calibration, multiple exposures were
combined using IRAF {\em combine} routines, with the {\em crreject}
algorithm to eliminate pixels affected by cosmic ray strikes.  For the
purpose of measuring equivalent widths, we performed a simple
normalization which removes the shape of the echelle blaze function
from each order. The approximate shape of the blaze function was
obtained by tracing the continuum flux of a bright giant in NGC~6397.

\begin{figure}
\plotone{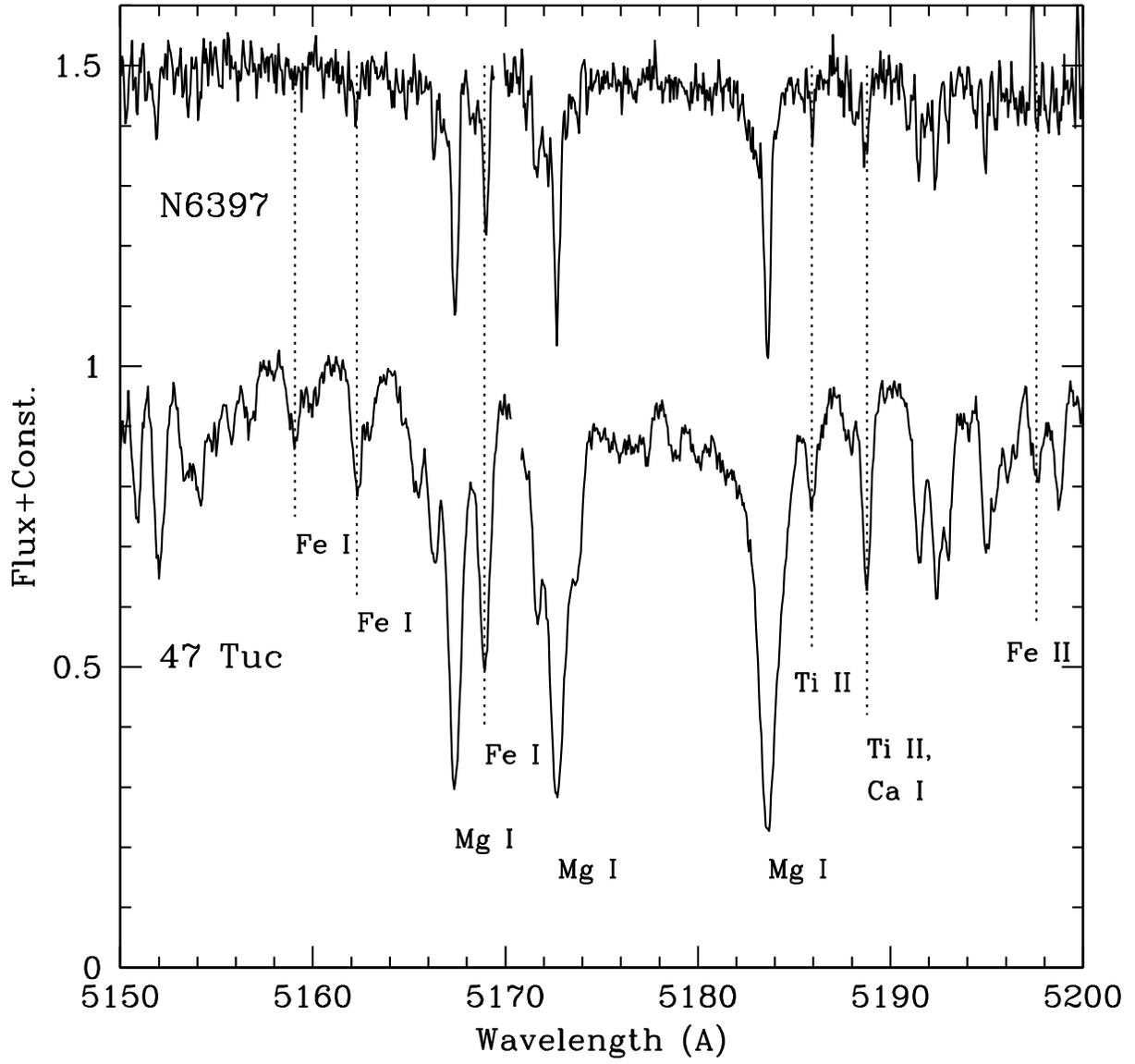}
\caption{The integrated-light spectrum of the Mgb line region in NGC
  6397 and 47 Tuc.  }
\label{fig-mgb}
\end{figure}

\begin{figure}
\plotone{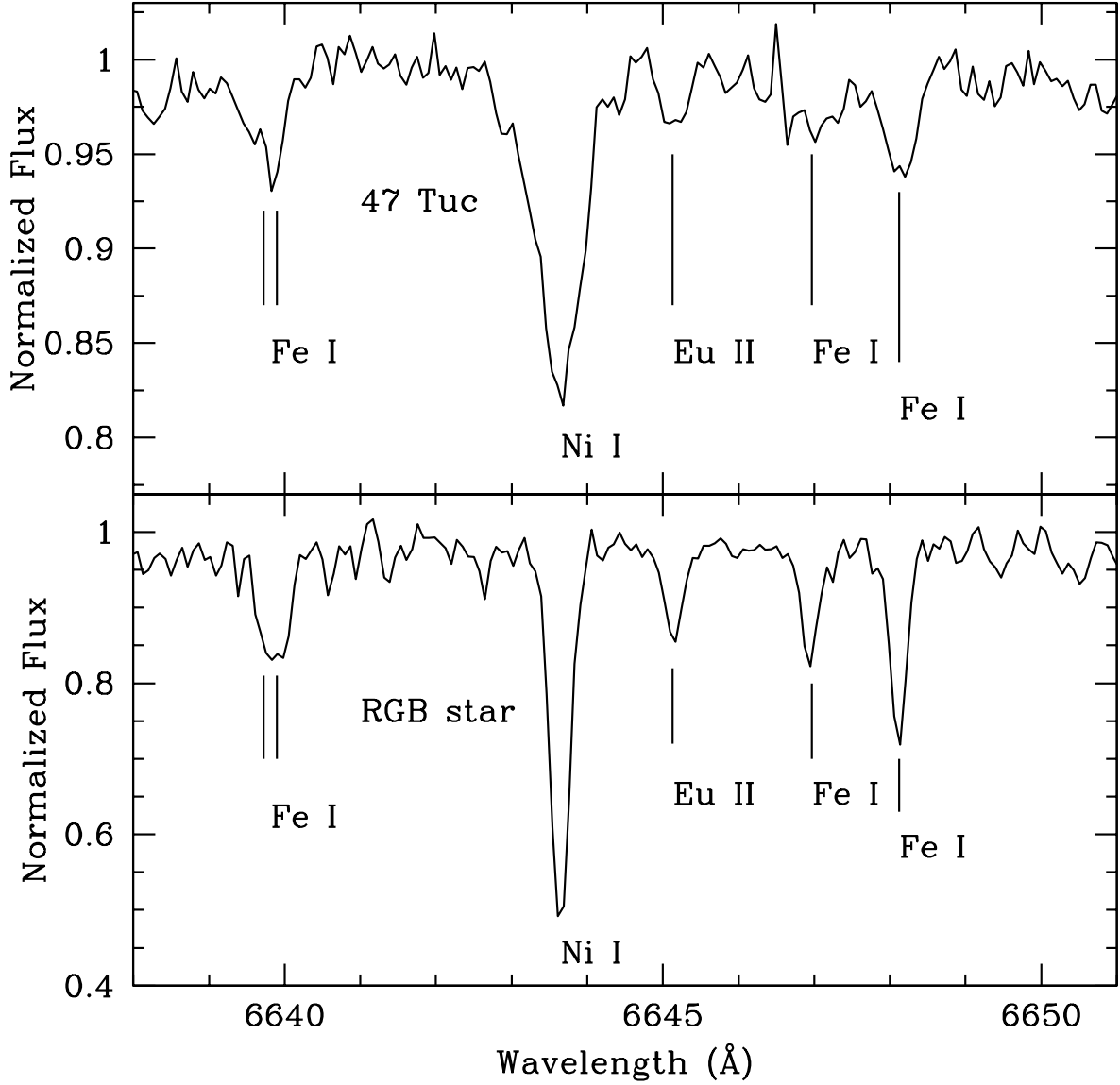}
\caption{The integrated-light spectrum of the Eu~II line region in 47
  Tuc (above), compared to the same region in an RGB star (below).  }
\label{fig-eu}
\end{figure}

The S/N of the final spectrum is listed at three wavelengths in
Table~\ref{table1}.  Figures~\ref{fig-mgb} and \ref{fig-eu} show
examples of the integrated-light spectra for GCs NGC~6397 and 47~Tuc.
The average line widths of weak lines in the 5000--6500\AA\ region of
the 47~Tuc spectrum was measured at $\lambda$/FWHM=10,552, corresponding
to an intrinsic $\lambda$/FWHM=11,065 once the spectrograph is taken into
account. This line width indicates
a velocity dispersion of 11.5 $\pm$0.2 Km/s, exactly equal to the result of
Prior \& Meylan (1993).  From Figures~\ref{fig-mgb} and \ref{fig-eu} 
it is clear that abundance information for numerous elemental species is
present in the high resolution integrated-light spectra, even for the relatively
broad lines of 47~Tuc ($M_{\rm v}=-9.4$).  Lines in GC integrated-light 
spectra are weaker and wider than for individual red giants, as a
result of both the velocity broadening and the contribution of weaker lines 
from stars warmer and less luminous than the bright cluster giants.  For
this reason, greater S/N is required to obtain reliable abundances
from integrated-light spectra than for individual red giant stars.

\section{Analysis}

Our goal in this paper is to explore strategies for the detailed
abundance analysis of high-resolution IL spectra of GCs.  To simplify
the issues, we have used two methods that isolate different
aspects of the problem to some degree.  In the first method, we employ
extant photometry of 47~Tuc and characterize the stellar populations
by regions on the color-magnitude diagram ({\em cmd}).  We begin by
splitting the observed {\em cmd} into 27 regions (``boxes''), each with a small
range in color and magnitude (see Figure \ref{fig-47tuccmd}).  Our
analysis then involves synthesizing a theoretical equivalent width
(EW) for each spectral line for each of these boxes; we then combine
these EWs, weighted by the continuum flux at each line and the total
flux in each box to obtain a theoretical, flux weighted, IL EW.
Obviously, it is not possible to employ this particular technique for
very distant GCs that are spatially unresolved, because the {\em cmds}
are not available; however, the method could be used to study GCs in nearby
galaxies (e.g. M31) using space-based photometry.  

The objective for this part of the investigation is to first, provide
an independent abundance analysis of the integrated-light spectrum
of 47~Tuc. The second, and more important, goal is to check our
abundance analysis algorithms with minimal uncertainty in the 
adopted isochrone (or {\em cmd}). 
Thus, with this method we simply use the available information
regarding the stellar population of the cluster to determine whether
an IL spectrum synthesis method will, ultimately, be able to measure
detailed chemical abundances from high resolution GC spectra.  If IL
spectra combined with the known color-magnitude diagrams of the
clusters do not provide reliable abundance results, then it will
certainly not be possible in the situation where a theoretical {\em
  cmd} must be assumed.

Our second method for analyzing IL spectra employs theoretical
isochrones in place of the resolved cluster photometry.  The accuracy
of the derived abundances will clearly depend on how well the
predicted isochrones match the luminosity function and temperatures of
the stars in the clusters, and whether critical isochrone parameters
can be constrained using the IL spectra alone.  Since we do not know
{\em a priori} what isochrone to use when analyzing unresolved GCs,
we also develop diagnostics that enable us to
constrain the isochrone parameters for the abundance analysis.

\subsection{Measurement of the Spectral Lines}

The first step in our analysis is uniform, repeatable measurement of
the absorption line equivalent widths (EWs) in the 47~Tuc integrated-light
spectrum.  We measured the EWs by fitting the lines with
Gaussian profiles using the semi-automated program GETJOB (see
McWilliam et al. 1995a).  We select our list of lines from those used
by McWilliam \& Rich (1994), McWilliam et al.  (1995a,b), and
Smecker-Hane \& McWilliam (2002).  There are fewer useful lines in
globular cluster integrated-light spectra than for stars because the
GC lines are broader and weaker than in the spectrum of individual red
giant stars, and thus more difficult to measure.  A complete list of
our lines and EWs is given in
Table~\ref{table2}.  These line measurements are used for both of the
analyses outlined below.

\subsection{Integrated-Light Abundances with the Resolved CMD}
\label{sec:cmd_analysis}

As a preliminary step in developing a way to analyze unresolved
globular clusters, we first consider the abundance analysis of the
integrated light spectra using the available photometry for this
resolved Milky Way cluster.  The analysis involves two general stages:
characterizing the stellar population in the cluster and synthesizing
the strength of the spectral lines.  

To begin the first stage of the analysis using the observed {\em cmd},
we divided the observed {\em cmd} into small boxes containing stars
with similar photometric properties and we used standard relations to
derive the flux-weighted ``average'' atmosphere parameters of the stars in each
box.  The boxes run from the main-sequence to the tip of the red giant
branch, then along the horizontal branch/red clump region and the
AGB. Two additional boxes are included for the blue straggler
population.  In this analysis, the mean stellar atmosphere parameters
for stars in each box are estimated from the flux-weighted photometry.
We then computed the absolute visual luminosities using reddening
corrections and the distance modulus according to the equation
\begin{equation}
            M_{\rm v} = V - (m-M) .                                   
\end{equation}
The flux-weighted temperatures appropriate for each {\em cmd} box was
computed using standard color-temperature relations.  With initial
estimates for [Fe/H], T$_{\rm eff}$ and $\log{g}$, the bolometric
corrections, BC, were interpolated from the Kurucz grid (2002
unpublished)\footnote{available from
  http://kurucz.harvard.edu/grids.html}.  Bolometric magnitudes and
luminosities were computed using the expressions
\begin{equation}
            M_{\rm bol} = M_{\rm v} + BC                            
\end{equation}
and 
\begin{equation}
    \log L/L{_\odot} = -0.4(M_{\rm bol} - M_{\rm bol\odot}),    
\end{equation}
where M$_{\rm bol \odot}$ = 4.74.  Gravities were computed using
the expression
\begin{equation}
\log{g}  = \log{g}_{\odot} + \log M/M_{\odot} - 
        log L/L_{\odot} + 4log T_{\rm eff}/T_{\rm eff\odot} ,     
\end{equation}
assuming a solar effective temperature of T$_{\rm eff \odot}$=5777K,
solar gravity of $\log{g}_{\odot}$=4.4378, and a mass for the cluster
stars of 0.8M$_{\odot}$.  We re-evaluated the specific gravity and
bolometric corrections iteratively until convergence in $\log{g}$ was
obtained to within 0.05 dex.  Because 47~Tuc is old, and because our
photometry included only a small portion of the main-sequence, we note
that a negligible error in the final result would occur by assuming
that the masses of all the stars in our {\em cmd} were equal to the
turnoff mass.  Microturbulent velocities were assumed to fit a linear
regression through 1.00 Km/s for the sun at $\log{g}=4.44$, to 1.60
Km/s for Arcturus, at $\log{g}=1.60$ (Fulbright, McWilliam \& Rich
2006):
\begin{equation}
\xi = \xi_{\odot} + \frac{(logg_{\odot} - logg)}
      {(logg_{\odot}-logg_{\alpha Boo})} 
      (\xi_{\alpha Boo} - \xi_{\odot}) .          
\end{equation}
Finally, stellar radii were computed for each {\em cmd} box according
to the expression
\begin{equation}
R/R_{\odot} =  (T_{\rm eff}\odot/T_{\rm eff})^2  \sqrt[]{L/L_{\odot}}   
\end{equation}
These radii are needed in order to compute the total flux from each {\em cmd} 
box.

\begin{figure}
\plotone{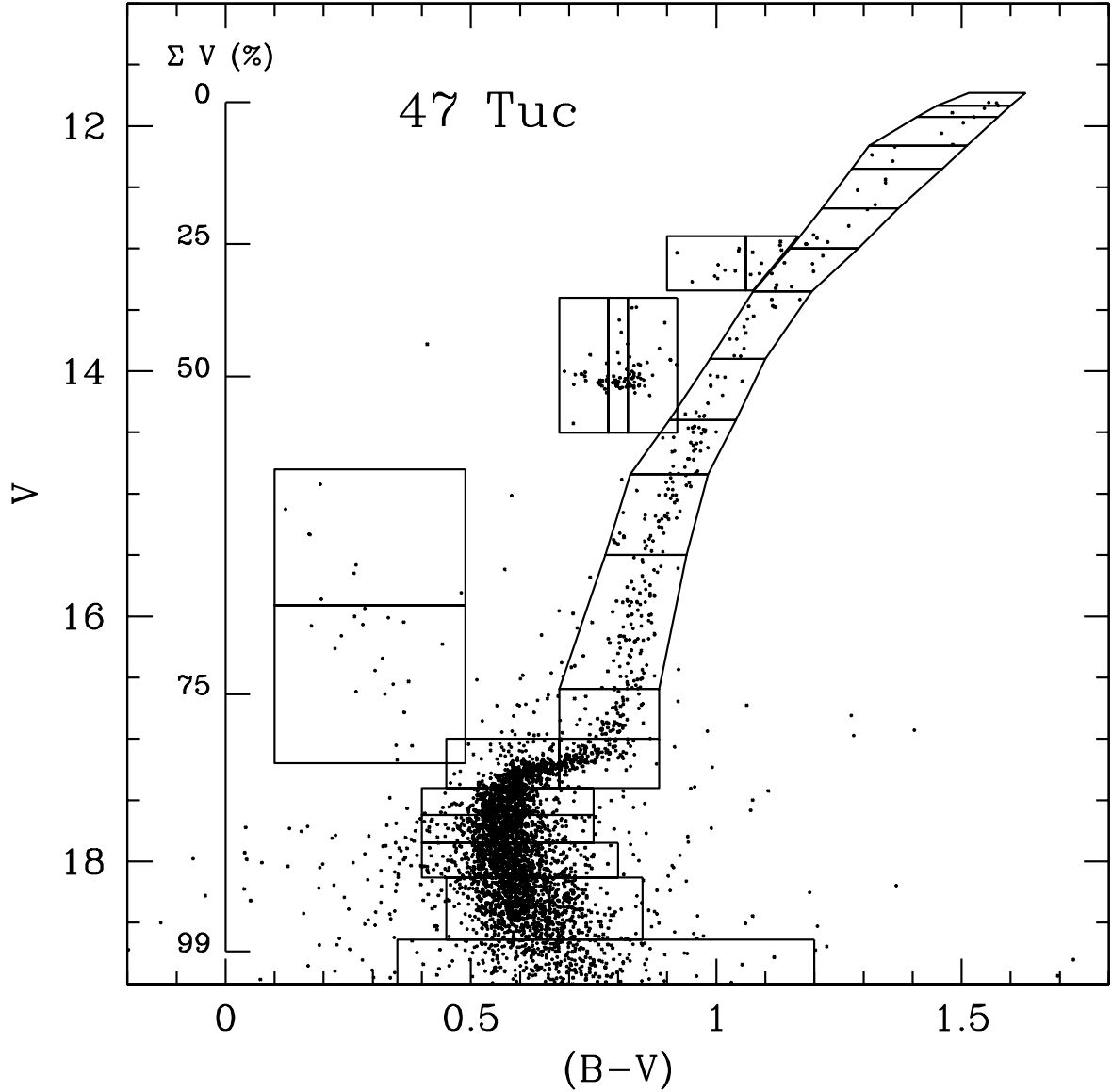}
\caption{CMD of the scanned region of 47 Tucanae (inner 32x32 arc
  sec), from HST images of Guhathakurta et al. (1992), Howell et
  al. (2000); kindly provided by R.~Schiavon.  Boxes indicate how the
  CMD was split into 27 sub-regions for the abundance analysis.  }
\label{fig-47tuccmd}
\end{figure}

In Figure~\ref{fig-47tuccmd} we show the 47~Tuc V, B$-$V {\em cmd} (from
Guhathakurta et al. 1992; Howell et al. 2000) for
the central, 32$\times$32 arc seconds, scanned region.  This region contains
4192 stars, which we have divided into 27 boxes; an additional 229 stars
lay beyond the {\it cmd} boxes, but contribute insignificant flux to
the total.  As outlined above, we computed the flux-weighted V and
B$-$V values for each {\em cmd} box, and used those values to
calculate mean effective temperatures and gravities with the Alonso et
al. (1999) color-temperature relations. Ancillary assumptions included
[Fe/H]=$-$0.7 to compute the bolometric corrections and a turnoff mass
of 0.8 M$_{\odot}$ for the gravity. 

Because the B$-$V color--temperature relation and bolometric correction
are sensitive to the stellar metallicity, it is necessary to adopt an
initial [Fe/H] for the calculation of T$_{\rm eff}$ and BC.  As a consequence 
the abundance analysis must be iterated until the adopted [Fe/H] value is
consistent with the abundances derived from the Fe lines.  
Fortunately, the iterations converge extremely quickly: for 47~Tuc
if we derive temperatures and BC values starting with
an assumed metallicity of [Fe/H]=0.0 the first iteration obtains [Fe/H]=$-$0.65
from the lines, and the second iteration gives [Fe/H] within 0.02 dex of the
self-consistent value.  However, for other
colors, such as V$-$I and V$-$K, the sensitivity of the color--T$_{\rm eff}$
relations and BC to metallicity are extremely small, so iteration is not
required.

Table~\ref{table3}
provides the adopted stellar atmosphere parameters for the 47~Tuc BV
{\em cmd} boxes.  It is interesting to note that  50\%
of the V-band flux comes from giant stars at the red clump luminosity
and brighter in this cluster (see Figure~\ref{fig-47tuccmd}).

The second stage in the abundance analysis then involves spectrum
synthesis to compute theoretical EWs of each line for each {\em cmd}
box using a model atmosphere.  The final EWs of lines in the
integrated light of a cluster can then be computed by averaging the
synthesized EWs together, weighted by the flux in each {\em cmd} box.
In the line synthesis, the only adjustable parameter is the input
abundances.  All other parameters are fixed by the average stellar
atmosphere parameters of the stars in each {\em cmd} box and the properties 
of the spectral line being calculated.  The spectrum synthesis was
performed using the program MOOG (Sneden 1973) modified to be called
as a subroutine and to provide the continuum flux, $F_c$, for each
line, within a larger program, ILABUNDS.  Our code employs the
alpha-enhanced Kurucz models (Castelli \& Kurucz 2004)\footnote{The models
  are available from Kurucz's website at 
  http://kurucz.harvard.edu/grids.html} with the latest opacity
distribution function (AODFNEW), linearly interpolated to arbitrary
T$_{\rm eff}$, logg and [Fe/H].\footnote{The use of solar
  alpha-element ratios does not significantly alter the Fe~I
  abundances derived, but lines sensitive to electron density, like
  [O~I], Fe~II, are affected by the choice of alpha enhancement.  If
  analysis of the $\alpha$-elements indicates solar abundances are
  appropriate for a given unresolved cluster, models with
  self-consistent $\alpha$-abundances would be used.}  The
flux-weighted, average line equivalent width for the integrated-light
of the cluster was computed according to the following expression:
\begin{equation}
\overline{EW} = \frac{ \sum _{i=1}^n EW_i w_i } {\sum _{i=1}^n w_i },  
\end{equation}
where $EW_i$ is the equivalent width of the line for a given box ($i$)
and $w_i$ are the weights for each {\em cmd} box. We compute $w_i$
values using the radii, $R/R_{\odot}$, number of stars in the {\em
  cmd} box, $N_*$, and the emergent continuum fluxes (computed using
MOOG) for the box, $F_c$, according to
\begin{equation}
w = R^2 N_* F_c  .                                    
\end{equation}

Abundances were determined by iteratively adjusting the assumed
abundance in the line synthesis until the synthetic flux-weighted EW
matched the observed IL EW for each line. We continued to iterate with
small adjustments to the input abundance until the observed and
predicted IL EWs agreed to one percent.

The use of an observed {\em cmd} limits the utility of this technique
for detailed abundance analysis to globular clusters within the Local
Group of galaxies, where ground or space-based telescopes are able to
at least partially resolve individual cluster stars. While future 30
meter class telescopes, equipped with adaptive optics systems, may be
able to resolve GC stars to even greater distances, ultimately our
objective is to measure detailed composition from unresolved globular
clusters. Analysis using observed photometry is a step toward that
goal and confirms that the basic strategy of computing light weighted
EWs is sound.

\subsubsection{Iron Abundances}
\label{sec:Iron_Abundances}

\begin{figure}
\plotone{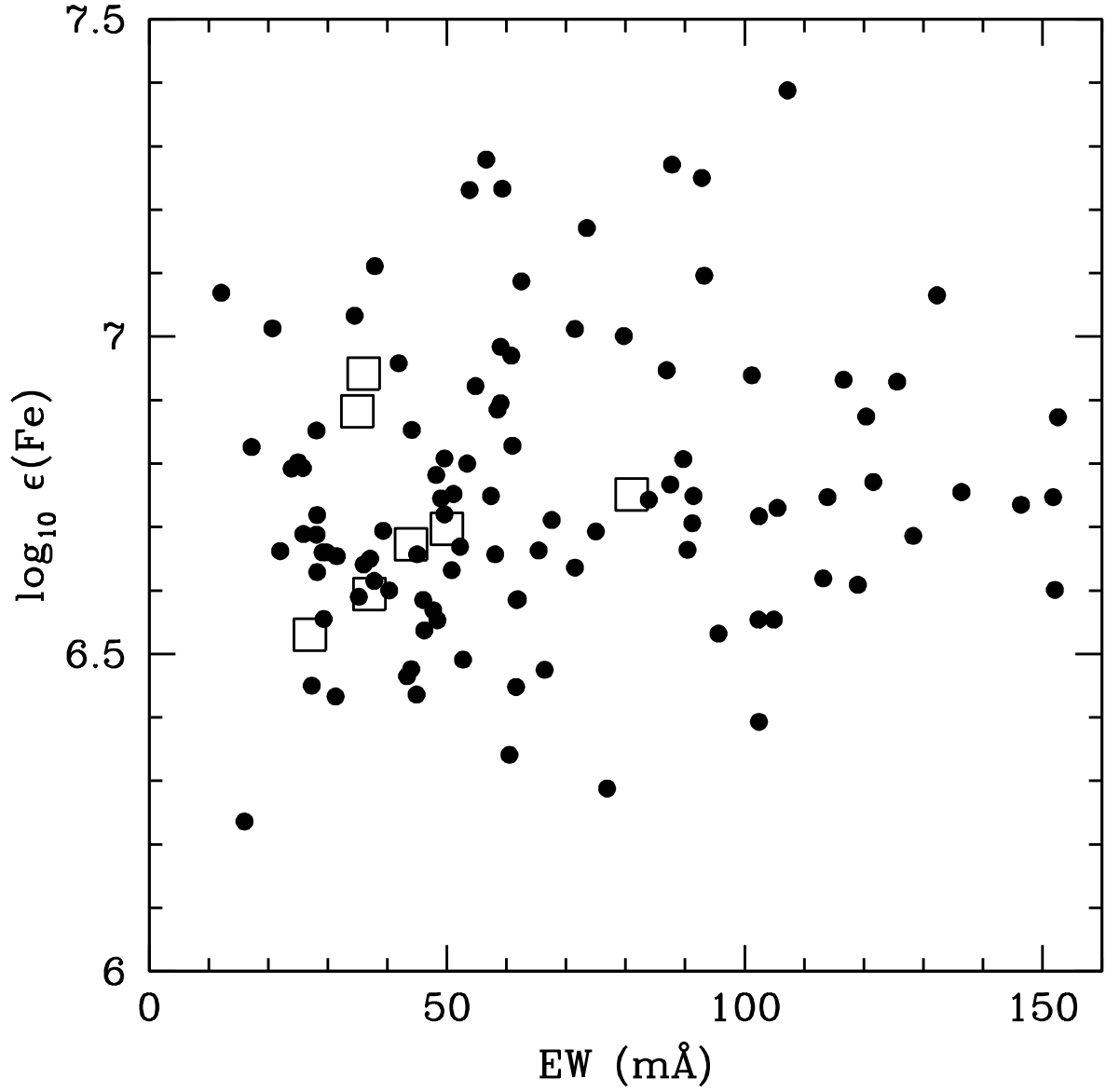}
\caption{Iron abundance versus line equivalent width, in m\AA .
  Filled circles represent Fe~I lines, open squares are Fe~II lines.
  The absence of a trend of abundance with EW validates our choice of
  the microturbulent velocity law.}
\label{fig-47tuc.ewab.bv}
\end{figure}

\begin{figure}
\plotone{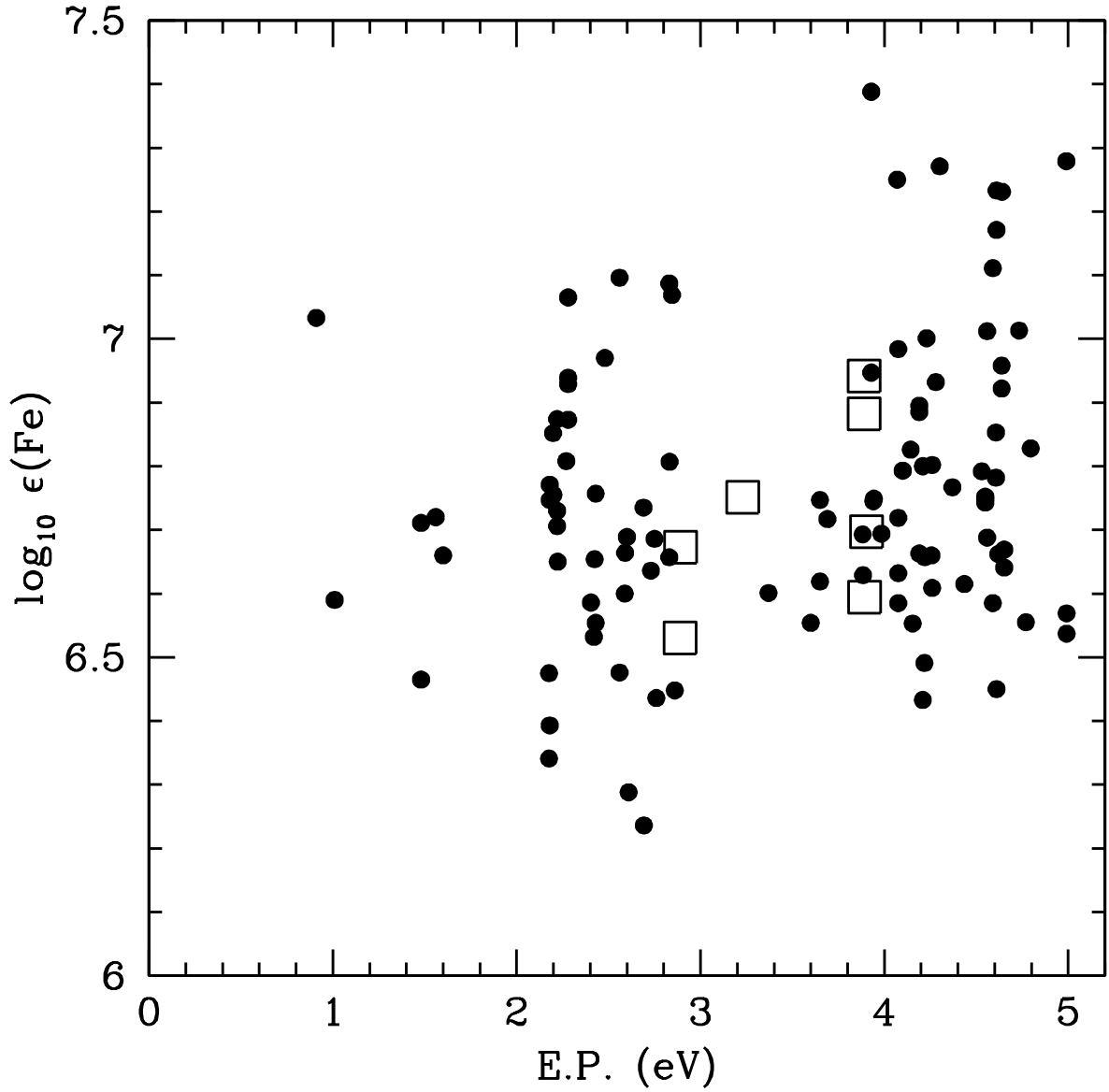}
\caption{Iron abundance versus line excitation potential, in eV.
  Filled circles represent Fe~I lines, open squares are Fe~II lines.
  The abundance is approximately independent of excitation potential,
  which suggests that the atmosphere temperatures are approximately
  correct.  The Fe~I and Fe~II abundances show good overlap, although
  the average $\epsilon$(Fe II) is higher by 0.09 dex.}
\label{fig-47tuc.epab.bv}
\end{figure}

\begin{figure}
\plotone{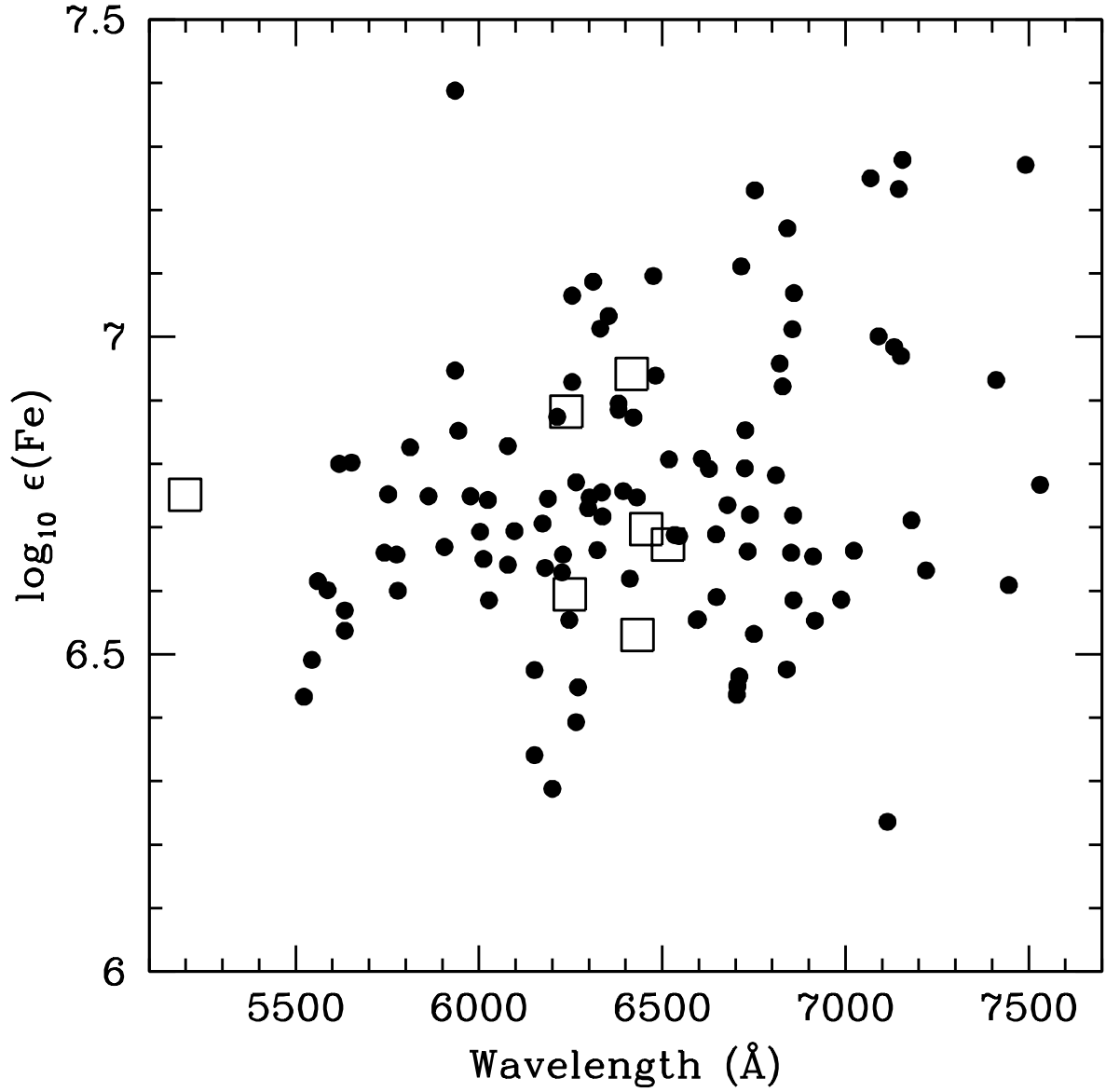}
\caption{Iron abundance versus line wavelength, in m\AA .  Filled
  circles represent Fe~I lines, open squares are Fe~II lines. }
\label{fig-47tuc.wab.bv}
\end{figure}

We begin our abundance study with the analysis for iron, because its
numerous lines provide useful diagnostics of stellar atmosphere 
parameters.  In our IL abundance analysis we employ the atmospheric
parameters from the observed {\em cmd} listed in Table~\ref{table3}, 
the EWs and atomic parameters given in Table~\ref{table2}, and the prescription
outlined above in \S\ref{sec:cmd_analysis}.  The atomic line parameters
listed in Table~\ref{table2} were taken from McWilliam \& Rich (1994)
and McWilliam et al. (1995).  In
Figures~\ref{fig-47tuc.ewab.bv}--\ref{fig-47tuc.wab.bv} we show
diagnostic plots, using the abundances derived from the iron lines,
similar to the diagnostics used for standard abundance analysis of
single stars.  Figure~\ref{fig-47tuc.ewab.bv} shows that
$\epsilon$(Fe) is independent of EW, which indicates that the assumed
microturbulent velocity law that we have employed is sufficiently
accurate; a positive slope would have suggested microturbulent
velocities that are too low.  The plot of $\epsilon$(Fe) versus
excitation potential in Figure~\ref{fig-47tuc.epab.bv} is sensitive to
the temperatures of the {\em cmd} boxes; the near-zero slope indicates
that the adopted temperatures for the cmd boxes was approximately
correct.  Figure~\ref{fig-47tuc.wab.bv} shows that our iron abundances
are independent of wavelength.  This provides a general check on our
abundance analysis, from consistency of the EW measurements to the mix
of stellar types and the continuous opacity subroutines in our
spectrum synthesis code.  The increased upper envelope and scatter of
iron abundances for lines redder than 7000\AA\ led us to suspect that
the scattered-light subtraction was flawed in this region, but it
might instead be due to the effects of blends or poor $gf$ values.  In
general the trend of abundance with wavelength should provide a probe
of the existence of sub-populations of varying temperature, such as
hot stars on the blue horizontal giant branch.

From 102 measurements of 96 Fe I lines, we obtain a mean Fe~I
abundance of 6.77 $\pm$0.03 dex, with rms scatter about the mean of
0.26 dex.  From 7 Fe~II lines, we obtain a mean Fe~II abundance of
6.73 $\pm$0.06 dex, with rms scatter about the mean of 0.16 dex,

The recent estimate of the solar iron abundance by Asplund, Grevesse
\& Sauval (2005) indicates a value of 7.45$\pm$0.05 dex, based on a 3D
hydrodynamical solar model.  Lodders (2003) found a solar iron
abundance of 7.54 dex.  While we believe that the Asplund et
al. (2005) solar iron abundance value is probably the best current
estimate, it is more reasonable for us to obtain a consistent and
differential [Fe/H] value for 47~Tuc relative to the sun, by using the
same lines, $gf$ values, grid of Kurucz 1D atmospheres and abundance
analysis program for the sun and 47~Tuc.  In this way, our Fe lines
indicate a solar iron abundance of $\epsilon$(Fe)=7.52 for Fe~I lines and
$\epsilon$(Fe)=7.45 for Fe~II lines.  For [X/Fe] abundance ratios we
adopt a solar iron abundance of $\epsilon$(Fe)=7.50, which is the
average of many studies in recent years.  

Our differential iron abundance values from Fe~I and
Fe~II lines in 47~Tuc are $-$0.75$\pm$0.026 and $-$0.72$\pm$0.056 dex respectively,
independent of adopted $gf$ values, where the uncertainties represent the
1$\sigma$ random error on the mean.  We consider the 1$\sigma$ systematic uncertainties 
on our derived [Fe/H] values from the model atmospheres (0.03 and 0.03 dex for 
Fe~I and II respectively), $gf$ scale zero-point (0.03 and 0.04 dex), and the 
effect of $\alpha$-enhancement (0.015 and 0.04 dex).  We refer the reader to 
Koch \& McWilliam (2008) for details of these systematic uncertainties.  Thus, 
we estimate the total 1$\sigma$ random + systematic uncertainty on the [Fe/H] 
values of 0.052 and 0.085 dex for Fe~I and Fe~II respectively.  Note that these
uncertainties do not include potential uncertainty due to non-LTE; we prefer to
simply provide the condition that our abundances are calculated using the LTE
assumption.  The 0.03 dex difference between our mean Fe~II and Fe~I abundances is 
less than half of the formal uncertainty on the mean Fe~II abundance.  

Our [Fe~I/H] value of $-$0.75 dex is consistent with other values in the
literature., Brown \& Wallerstein (1992) obtained [Fe/H]=$-$0.81, based
on a differential analysis of echelle spectra; Carretta \& Gratton
(1997) found [Fe/H]=$-$0.70; the Kraft \& Ivans (2003) reanalysis of
various literature EWs gave [Fe/H]=$-$0.63; echelle analysis of
turnoff stars by Carretta et al.  (2004) yields [Fe/H]=$-$0.67.  In
addition to these high resolution results, we note that there 
are several calcium triplet results based on the calibration by 
Kraft \& Ivans (2003), who found [Fe/H]=$-$0.79 for their
calibration against Kurucz model
atmosphere results.  Wylie et al. (2006) obtained [Fe~I/H]=$-$0.60 and
[Fe~II/H]=$-$0.64 dex.  Finally, Koch and McWilliam (2008)
find [Fe/H]=$-$0.76~$\pm$0.01~$\pm$0.04 dex (random and systematic error respectively)
for 47~Tuc, based on a
robust differential analysis of 8 individual red giant stars relative
to Arcturus.

We omit here a discussion of the individual systematic
differences between the literature studies listed above that might
yield a consensus best estimate of the [Fe/H].  We prefer the following
simple conclusion: from the comparison of reported results,  our
integrated-light abundance analysis for the core of 47 Tuc provides an
iron abundance close (within 0.10 dex) to that obtained by detailed
analysis of individual cluster stars.

\begin{figure}
\plotone{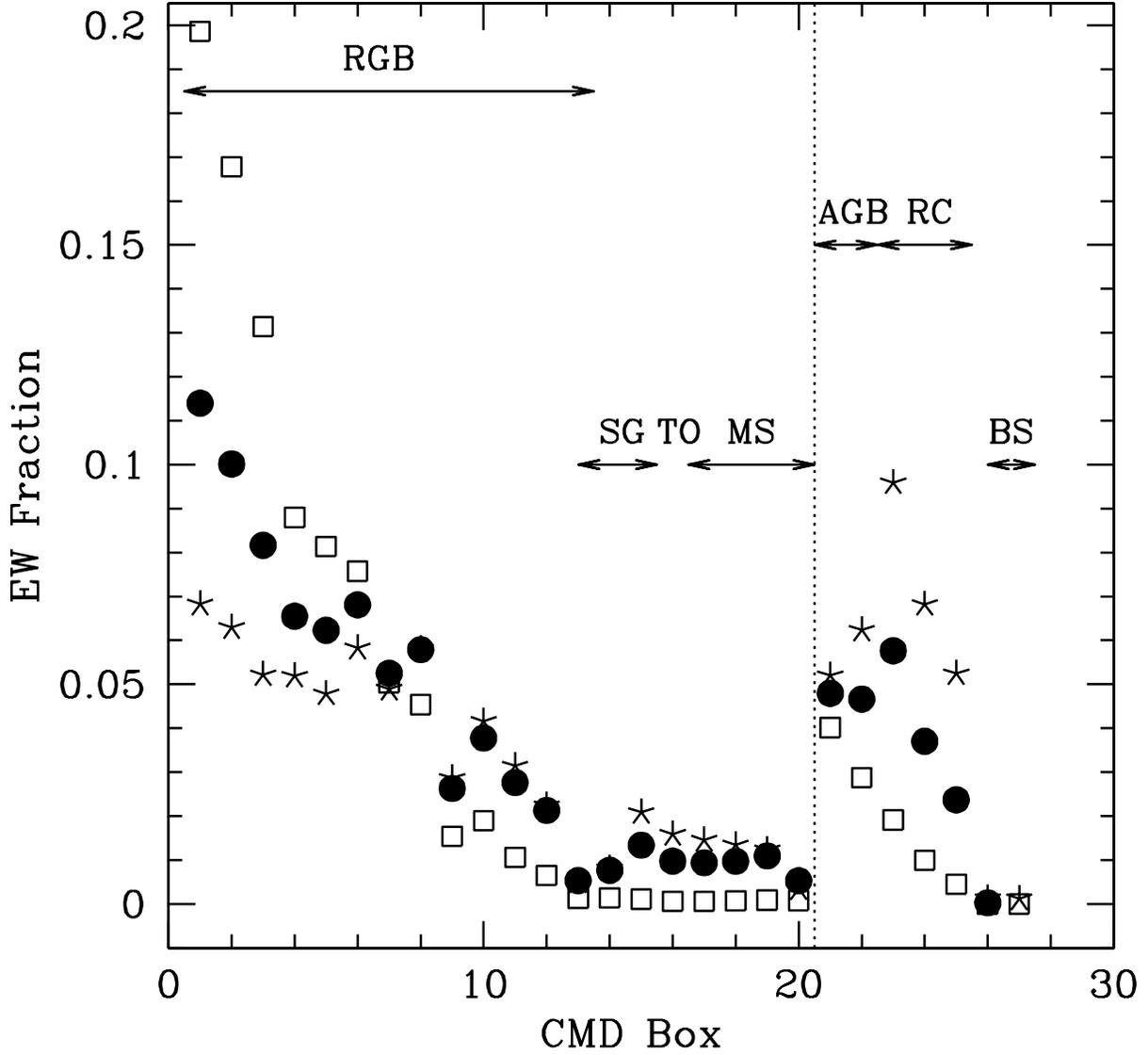}
\caption{The fractional contribution to the total equivalent width
  from each {\em cmd} box.  Temperature increases for the {\em cmd}
  boxes from left to right, in the two segments of the figure.  The
  figure shows the growth of one Fe~I line of high excitation at
  6597\AA (4.8 eV, filled circles), an Fe~I line of low excitation
  (1.0 eV, open squares) at 6648\AA, and an Fe~II line, at 6432\AA\
  (five pointed stars).  All three lines are weak, being close to
  30m\AA\ in 47 Tuc.  The Fe~I lines are predominately formed by stars
  on the Red Giant Branch, with some contribution from the AGB and red
  clump (RC); however, very little contribution to the Fe~I line EWs
  occurs in the subgiant branch (SG), turnoff (TO) and main sequence
  (MS).  The most important contribution to the Fe~II line strength
  comes from the red clump and AGB.  The coolest {\em cmd} boxes do
  not contribute as much to the total Fe~II EW as they do for the Fe~I
  lines due to the low ionization in these cool stars.  The blue
  stragglers make no significant contribution to the EW of our iron
  lines.  }\label{fig-47tuc.grow}
\end{figure}

Figure~\ref{fig-47tuc.grow} shows the fractional contribution to the
total equivalent width of each {\em cmd} box, for three Fe lines:
one high excitation potential Fe~I line, one low excitation potential Fe~I
line, and an Fe~II line.  All three lines are weak, being close to
30m\AA\ in the 47~Tuc IL spectrum.  It is clear from this figure that the
Fe~I lines are predominately formed by stars on the Red Giant Branch, with
some contribution from the AGB and red clump (RC); however, very little
contribution to the Fe~I line EWs occurs in the subgiant branch (SG),
turnoff (TO) and main sequence (MS).  The most important contribution
to the Fe~II line strength comes from the red clump and AGB.  The
coolest {\em cmd} boxes do not contribute as much to the total Fe~II
EW, unlike the Fe~I lines, presumably because of the low
ionization fraction in these cool stars.  The differences between Fe~I
and Fe~II formation, seen in Figure~\ref{fig-47tuc.grow}, indicates
that the Fe~II/Fe~I differences may provide constraints on the giant
branch luminosity function.  Figure~\ref{fig-47tuc.grow} also shows
that blue stragglers make no significant contribution to the EW of our
iron lines.

Although not shown in Figure~\ref{fig-47tuc.grow}, our calculations
also indicate that very strong Fe~I lines have significant formation
across all {\em cmd} boxes, rather than being strongly skewed to the
top of the RGB. This is partly due to the fact that the lines in the
RGB stars are strongly saturated.  Thus, for strong and weak lines to
produce the same abundance it is necessary to correctly account for
the stars at the lower end of the luminosity function.  This may form
the basis of a probe of the lower luminosity population, but it will
be sensitive to the adopted microturbulent velocity law, damping
constants, and the coolest parts of the model atmospheres.

\subsubsection{M Giants, B$-$V colors and the Tip of the Giant Branch}

One difficulty with an abundance technique that relies upon measured
EWs, as outlined above, is that the line and continuum could suffer
from line blanketing.  Line and continuum blending is particularly
acute for the M stars, which are characterized by heavy blanketing
from the TiO molecule over the entire optical region. The continuum
and line regions in these stars can be heavily depressed by TiO
absorption that would alter their contribution to the IL EWs of a GC.

While much of the red giant branch contains stars of G and K spectral
type, with identifiable continuum regions, the very tip of the giant
branch may include a number of M giants which are difficult to
identify from photometry, as we discuss further below.  The TiO
density in the atmospheres depends, approximately, on the square
metallicity; thus, the exact fraction of M giants that lie on the
giant branch is a function of the metal content, with more M giants in
the more metal-rich clusters. The actual number of M stars
contributing to the IL flux also depends on the luminosity of the
cluster region, because lower luminosity clusters may not contain
enough stars to completely populate the top of their giant branches.
Because of the potentially significant uncertainty in the identification
of the continuum level in the IL spectrum, due to the M star population,
we have taken some care to quantify the
effect of these stars in our IL analysis. In the following
section, we evaluate how many M giants were present in our 47~Tuc
spectrum, develop a crude method to include the M giants in our IL EW
abundance analysis, and determine the effect that these giants have on
the derived abundances using the EW technique.

The tip of the 47~Tuc giant branch is well known to contain numerous M
giant stars.  In order to assess the consequences of the presence of M
giants in our IL spectrum, we first need to correctly identify the M
giant population in the 47~Tuc core, included in our spectrum.
Unfortunately, the TiO blanketing in M giants reduces both the V and
B-band fluxes, such that the M giant B$-$V colors are similar to
hotter K giants.  This blanketing lowers the V magnitude of the
coolest M giants more than the early M giants.  Thus, while our B$-$V
{\em cmd} can be used to estimate the run of stellar parameters for
giants earlier (hotter) than M0, the extant M giants are confused for
K giants, so we cannot use the HST B$-$V photometry to find the M
giants in the 47~Tuc core.

The degeneracy of the K and M giants does not occur for the (V$-$I)
color-magnitude diagram, presumably because the V and I band blanketing are 
less saturated than the B band.  The Kaluzny et al. (1998) (V$-$I)
versus I {\em cmd}, for an outer region of 47~Tuc, clearly shows the M
giant population at the tip of the giant branch.  Based on the M-star
to Clump-star number ratio in the Kaluzny (1997) V$-$I {\em cmd} and
the frequency of clump stars in the HST B$-$V {\em cmd} we expect 2.0
M giants in the 32$\times$32 arc second scanned core of 47~Tuc.  
Direct evidence of M giants in the 47~Tuc core is also seen in the
list of long period variables (LPVs) from Lebzelter \& Wood
(2005). Their coordinates show that there {\em are} two LPVs (LW11 and
LW12) that lie within the 32$\times$32 arc second core
included in our spectrum.  Thus, for our {\em cmd} abundance analysis of 
our IL spectrum, these two M giants are the only ones that we need to 
consider.  

The V$-$K colors of these two M giants are given by Lebzelter \& Wood
(2005) as 5.82 and 4.55 respectively.  These very red colors are due,
mainly, to the effect of TiO blanketing on the V-band.  If
[Fe/H]=$-$0.7 dex is adopted for 47~Tuc, the metallicity-dependent
theoretical color-temperature relations of Houdashelt et al. (2000)
and Ku$\check{c}$inskas et al. (2006) indicates that these two M stars
have effective temperatures of 3350K and 3600K respectively.  However,
because 47~Tuc has an enhanced, halo-like, Ti/Fe ratio, the TiO bands
must be stronger than solar composition M stars of the same
temperature. Since the theoretical color-temperature relations of
Houdashelt et al. (2000) and Ku$\check{c}$inskas et al. (2006) were
computed based on assumed solar composition, the actual temperatures
of the 47~Tuc M giants must be slightly higher than their relations
indicate, but probably not much higher.  For the coolest solar
metallicity M giants, inspection of the individual synthetic M giant
star spectra of Houdashelt et al. (2000) indicates that the flux is so
strongly blanketed in the 6000--7000\AA\ region that individual atomic
lines, useful for EW abundance analysis, are almost completely
obliterated.  Also, the blanketing reduces the optical flux
contributed by these cool solar-metallicity M stars to insignificance
compared to slightly warmer K giants.  The TiO band-strength and
blending decreases gradually with increasing temperature.  Because of
this gradual change in TiO blanketing it is difficult to give an exact
temperature limit, below which the M stars would contribute negligible
V-band flux to a GC IL optical spectrum.  However, it is more important to
account for the spectral flux coming from the earlier (warmer,
e.g. M0-M2) M giants than the coolest M giants.  
The temperature below which the M star phenomenon occurs depends on
TiO formation, which in turn depends on metallicity; at the metallicity
of 47~Tuc, the M0 stars begin below roughly 3600K, but perhaps as high
as 3700K.  For comparison, M0 giants occur starting below $\sim$3900K at
solar metallicity.

\begin{figure}
\plotone{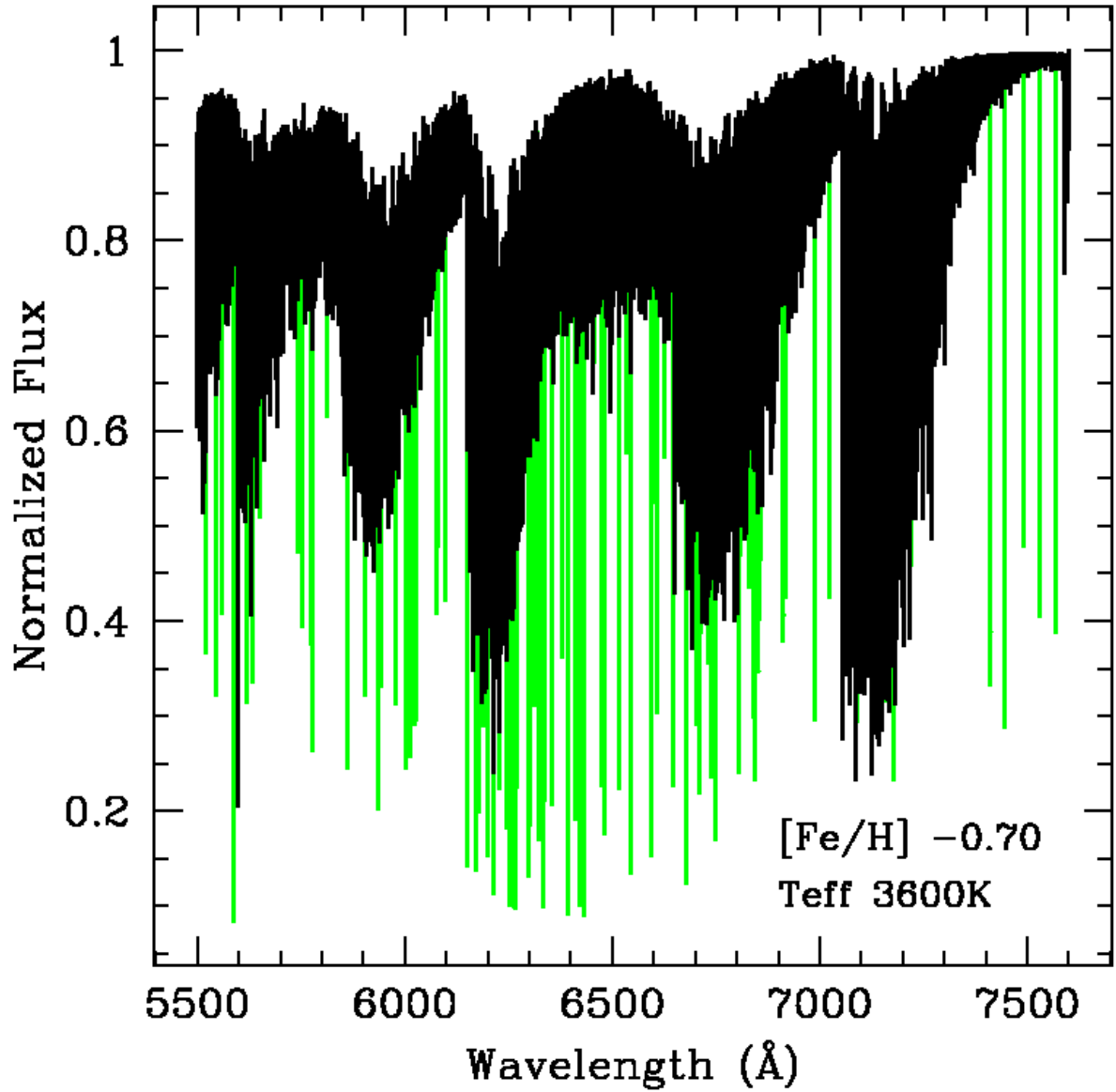}
\caption{ A synthetic spectrum from 5500-7600\AA\ for the M giant with
  T$_{\rm eff}$=3600K, in the core of 47~Tuc, including 5.4 million
  TiO lines, and Fe lines measured for the EW analysis of the IL
  spectrum.  The green color indicates the synthesis performed with
  TiO plus Fe lines, and the black color indicates synthesis of TiO
  only.  The TiO band heads of the alpha system are clearly visible.
}
\label{fig-tio3600}
\end{figure}

\begin{figure}
\plotone{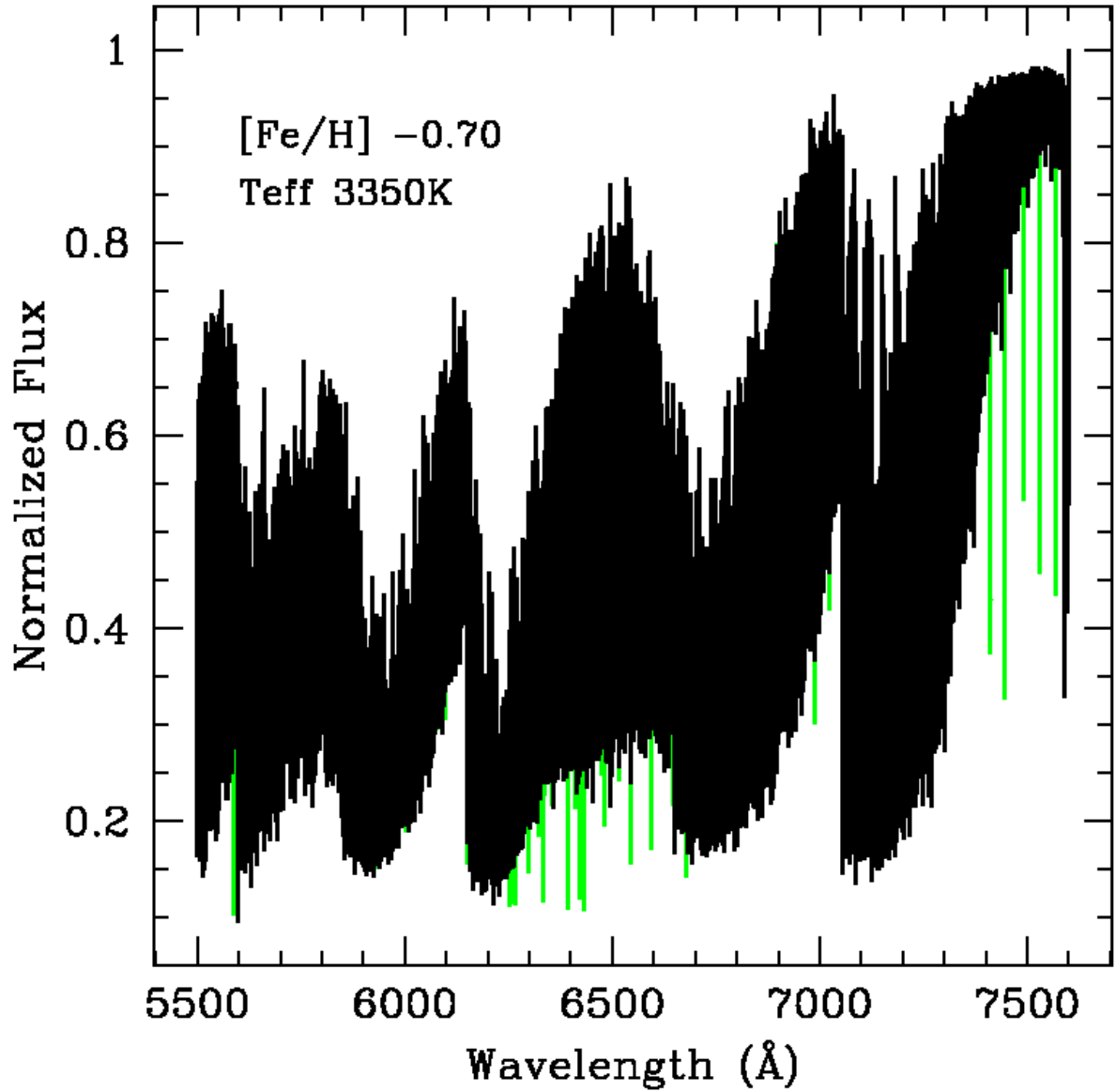}
\caption{ A synthetic spectrum from 5500-7600\AA\ for the M giant with
  T$_{\rm eff}$=3350K, in the core of 47~Tuc, including 5.4 million
  TiO lines, and Fe lines measured for the EW analysis of the IL
  spectrum.  Colors are the same as for Figure~\ref{fig-tio3600}.
  Notice the heavy TiO blanketing below $\sim$7300\AA\ , that
  depresses the continuum and obliterates most atomic lines; the M
  giant window in the interval 7300--7600\AA\ is relatively unaffected
  by TiO blanketing .  }
\label{fig-tio3350}
\end{figure}

To understand the effect on the cluster core IL spectrum caused by
blanketing and blending in the M star spectra, we present
Figures~\ref{fig-tio3600} and \ref{fig-tio3350}, which show synthetic
spectra in the region 5500--7600\AA\ for the two M giants known to be
in the core of 47~Tuc.  The atomic lines measured in our IL EW
spectrum analysis are also included in these figures.  The synthetic
spectra were computed using the 2004 version of the synthesis program
MOOG (modified to take large line lists) and 5.4 million TiO lines
from the 2006 version\footnote{see
  ftp://saphir.dstu.univ-montp2.fr/GRAAL/plez/TiOdata/} of the list by
Plez (1998).  We have not included other molecules
or atomic lines in our calculations other than the list of Fe lines that were measured in
the IL spectrum.  We employed Kurucz alpha-enhanced model atmospheres
with [Fe/H]=$-$0.70 dex and [O/Fe] and [Ti/Fe] = $+$0.3 dex.  The
model atmospheres were derived from the alpha-enhanced Kurucz grid;
the cooler M giant model was extrapolated using a cubic spline
technique, whereas the warmer M giant model was obtained by linear
interpolation.  Figures~\ref{fig-tio3600} and \ref{fig-tio3350} show
syntheses computed both with and without our Fe line list. The
comparison demonstrates the effect of the TiO blends and pseudo
continuum on the detectability of these atomic lines.  Note that the
TiO blanketing is significantly less than would be expected of
solar-metallicity M stars of the same temperature (see Houdashelt et
al. 2000) due to reduced TiO formation at the low metallicity of
47~Tuc.

It is clear from these figures that the TiO blanketing in the coolest
of the 47~Tuc core M giants has severely attenuated the apparent
continuum flux, and obliterated many of the atomic lines below 7300\AA
.  Thus, pseudo equivalent widths for many of the atomic lines may be
very small, suggesting that the coolest M star in the core of 47~Tuc
probably makes very little contribution to the integrated-light EW of
lines below 7300\AA .  The warmer of the two M giants in the core of
47~Tuc, with intermediate strength TiO bands, has greater optical flux
and potentially a more serious effect on IL line EWs.  Because TiO
formation and the M giant fraction increases with metallicity, the EW
abundance method used here will be less reliable for GCs more
metal-rich than 47~Tuc.

The effect of the M giants on the integrated light EW of Fe~I lines is
difficult to assess without detailed computations due to two
competing effects: First, while the neutral metal line strength increase for
cooler stars, so does the TiO line blanketing, which effectively
reduces the apparent continuum flux level.  As indicated above, the
continuum line blanketing is dominant for the coolest M giants.
Second, the M giants constitute $\sim$8.0\% of the total I band light, but
only $\sim$2.5\% of the total V band light for the cluster.  Thus, the
inclusion of the M giants would seem to be important for lines in the
I band $\sim$8000--9000\AA , but much less important in the V band,
near 5500\AA .

To test the effect of the M stars on our abundance results, we
made two changes to the analysis described so far based on
the HST photometry.  First, we reduced the number of K giants by two
and, second, we included the two M giant LPVs indicated by Lebzelter
\& Wood (2005).  In order to include the M stars in the EW IL
abundance analysis, we generated pseudo continuum fluxes and pseudo
EWs for the two known 47~Tuc core M giants over the 5500--7600\AA\
region, based on the spectrum synthesis calculations that we generated
for Figures~\ref{fig-tio3600} and \ref{fig-tio3350}.  As mentioned
earlier, the syntheses included the Fe lines for which we had EW
measurements, plus the 5.4 million TiO lines.  The abundance input for
the synthesis of each M giant was scaled to [Fe/H]=$-$0.70 dex, with
[Ti/Fe] and [O/Fe]=$+$0.3 dex.  The pseudo continuum flux levels were
determined for the synthetic spectrum by integrating the theoretical
spectrum synthesis flux over the wavelength bounds of the actual
continuum windows used for measuring the EWs from the IL spectrum.
For each iron line we averaged the synthetic fluxes in the adjacent
continuum windows and then we calculated a pseudo EW by integrating
the synthetic flux over 0.4\AA , centered on the line.  In most cases
the pseudo EWs were smaller than if there had been no TiO, due to the
pseudo continuum TiO blanketing, but in a few cases TiO features
increased the apparent line EW.  For yet other
lines the pseudo EWs were negative, due to heavy blanketing in the
continuum regions, but not in the lines.  The ILABUNDS program was
then enhanced to permit the use of pre-calculated pseudo EWs for the M
giants to be included in the calculation of the total mean
integrated-light EW.

To test the reliability of the calculations, we substituted pre-computed
pseudo-EWs for a {\em cmd} box containing three K giants; the method
reproduced the mean [Fe/H] within 0.006 dex of the mean determined
without the approximation.  In the nominal method, omission of the
three K giants would have altered the derived mean iron abundance by 0.06
dex; thus, the approximation was accurate to about 10\%.

When we compare the IL abundance results from the {\em cmd} with the
two M-giants to abundances determined for a {\em cmd} in which the two
M giants were excluded, the [Fe/H] from neutral and ionized species
increased by 0.02 and 0.01 dex respectively.  Furthermore, there was
no obvious effect on the Fe lines redder than 7000\AA , so the small
discrepancy between lines redder and bluer than this wavelength does
not appear to have been due to the M giants.  Most likely the problem
resulted from difficulties associated with extraction of the data for
orders that are very close together on the CCD.  With corrections of
only 0.01 to 0.02 dex we can conclude that for 47~Tuc we are fortunate
that the M giants in the {\em cmd}, with their heavy TiO blanketing,
make a negligible difference to the derived abundances and the IL
spectrum.

We note also that wavelength interval 7300--7600\AA\ has vastly lower
TiO opacity than surrounding regions, as can be readily seen in
Figures~\ref{fig-tio3600} and \ref{fig-tio3350}.  This window has been
used for abundance analysis of individual M giants (e.g. Smith \&
Lambert 1985).  Because the M giant continuum flux is unimpeded by TiO
absorption in this window, a comparison of EW abundance results from
lines in this region with line abundances from bluer wavelengths
(where the continuum is heavily blanketed in M giants) will be
sensitive to the M giant fraction in the cluster.  The M giant window
not only provides a potential probe for unresolved GC M giants, but the
EW abundance method could be employed more reliably in this wavelength
interval for metal-rich GCs, without the need to account for TiO
blanketing effects.  This is not a good option in our du~Pont data,
because there is some evidence that the orders might suffer from
background subtraction problems at these wavelengths.

Because globular clusters more metal-rich than 47~Tuc have a larger
fraction of M giants, they would require bigger corrections to the EW
analysis result than the 0.01--0.02 dex found for 47~Tuc.  For
clusters with significant M giant populations, and very large
abundance corrections to the EW technique (using pseudo EWs), it will
be necessary to abandon the EW analysis technique entirely, and
instead use spectrum synthesis profile matching for every line and
continuum region not in the M giant window.  This would significantly
increase the effort required and the uncertainty of the results.
Further computational studies are necessary to determine the
metallicity beyond which it will be important to employ this spectrum
synthesis profile matching abundance method.

In summary, we have explored the effect of the M giant population on
the IL spectrum and abundance analysis for 47~Tuc using an approximate
method of pseudo-EWs for these stars, including synthesis of millions
of TiO lines.  We find that the derived overall GC IL abundances
change by only 0.02 to 0.01 dex for Fe~I and Fe~II lines respectively.
Thus, the M giants could safely be ignored in an abundance analysis of
47~Tuc, but for higher metallicity GCs, with a larger fraction of M
stars, the effect will be greater, possibly enough to require use of a
laborious, and less reliable, synthesis profile matching abundance
technique.  For GC IL abundance studies at high metallicities, the M
giant window at 7300--7600\AA\ would constitute a valuable check
on the abundance results from lines outside this window.

\subsubsection{Abundances of Elements Other than Iron}

Abundances for elements other than Fe were computed using the EWs and
atomic parameters listed in Table~\ref{table2}.  Due to the
velocity broadening in 47~Tuc, various lines that are normally clean in red
giant stars were not usable.  A notable example is the strongest line
of a Mg~I triplet at 6318.708\AA , which is blended with a Ca~I line 0.1
\AA\ to the blue.  The [O~I] line near 6300\AA\ is another unfortunate
case, as it is blended with the very strong telluric emission line. 
In GCs with large systemic velocities, however, the oxygen feature may be usable.

Elements with odd numbers of protons or neutrons, or elements with
significant fraction of odd-numbered isotopes, suffer from hyperfine
splitting (hfs) of the energy levels.  For lines with large hfs
splittings, the line is split into many non-overlapping, or partially
overlapping, sub-components that reduce or eliminate saturation of the
feature.  Desaturation by hfs can significantly lower the computed
abundance compared to a single line.  However, for the case of unsaturated,
weak, lines (e.g. $\leq$20--30m\AA ), the hfs treatment gives the
same result as a single line approximation.

We have computed hfs abundances for lines of Ba, Co, Cu, Eu, La, Mn, Nd,
V and Zr, by synthesis of each line including all hfs components.
Previous experience with lines of Al, Na, Sc and Y indicated that the
hfs effect is too small to make a difference, so we neglected to perform 
hfs abundance calculations for those species.  The A and B hfs constants
for each line were taken from references indicated by McWilliam \&
Rich (1994) and McWilliam et al. (1995), and the wavelengths and
strengths of the hfs components were computed using standard formulae.
For the lines considered here we find typical hfs abundance
corrections for Cu, Co, Mn, La, V and Nd of $-$0.8, $-$0.3, $-$0.3, $-$0.15, 
$-$0.1 and $-$0.08 dex respectively; abundances derived with hfs are always 
lower than abundances derived assuming a single line.  We note that for the 
La~II line at 6774\AA\ hfs constants were available only for one level in the 
transition; however, we are still able to use the line because its small
equivalent width ensured that the single line treatment was reliable.

In Table~\ref{table4} we list the average
integrated-light abundances for all measured elements in 47~Tuc.  For
abundance ratios relative to Fe we employed the solar abundance
distribution of Asplund et al. (2005), except for Fe, for which we
adopted a solar abundance of 7.50 dex.
Table~\ref{table4} also contains comparisons with other
analyses for 47~Tuc: Brown \& Wallerstein (1992), Carretta et
al. (2004), Alves-Brito et al. (2005), Wylie et al. (2006), and Koch
\& McWilliam (2008).

A difficulty in the comparison with previous studies is that there is
variance between the earlier works in the 0.1 to 0.2 dex range.  This
not surprising, given the variety of techniques and assumptions
employed by previous studies to measure chemical abundances for
47~Tuc.  If we compare our mean abundance ratios with the mean of the
previously published studies listed in
Table~\ref{table4}, we find that, of 14 species measured
by two or more previous studies, our values are higher by 0.08 dex
with a standard deviation of 0.17 dex.  Only two [X/Fe] ratios
measured here differ by more than two sigma from the mean literature
values: Zr and Eu.  These elements are represented by one weak line
each.  In the case of Zr the line occurs at the ends of two adjacent
orders, where the noise is large.  However, since our [Zr/Fe] ratio,
near zero, is similar to our other heavy elements it seems possible
that the value found here is correct.  The single Eu~II line in our
spectrum, at 6645\AA\ , is quite weak (EW$\sim$16 m\AA ), so it is
possible that our EW is artificially low due to noise.

It is notable that our [Na/Fe] ratio, at $+$0.45 dex, is larger than
the mean of previous studies by 0.2--0.3 dex, depending on whether the
Wylie et al.  (2006, henceforth W06) result is included.  This
difference may not be significant, given the 0.17 dex rms scatter in
our Na abundances, however part of the difference likely results from
the low solar photospheric abundance for Na given by Asplund et
al. (2005), which is 0.10 dex lower than their meteoritic value
(itself lower by 0.06 dex than earlier estimates of the solar
meteoritic Na abundance).  Yet another possibility is that the
integrated-light Na abundance may really be higher than in individual
red giant and turnoff stars if a significant fraction of the AGB star
population in 47~Tuc shows proton burning products in their
atmospheres. In this regard it is interesting that the sample of W06
is dominated by AGB stars and is unusually enhanced in Na, with a
mean [Na/Fe]=$+$0.65 dex.  If our high Na values reflect proton
burning products in the envelopes of luminous evolved stars in the
cluster, then we might also expect to see depletions of O, and
possibly Mg, combined with enhancements of Al in the integrated-light
analysis.  We were unable to measure O in this work, however we do
find that Mg is lower than the average of the published studies by 0.18 dex,
and Al is higher by 0.15 dex.  More detailed work would be required
to investigate this possibility.

Besides these comparisons with other studies of 47~Tuc we may
investigate how well our integrated-light abundances compare with the
composition of the Galaxy in general.  This is particularly useful for
elements studied here that were not included in previous work on
47~Tuc (e.g. Cu, Mn).  We are interested to know whether our
integrated-light abundances are consistent with Galactic stars and
clusters with metallicity similar to 47~Tuc.
For example, the mean [X/Fe] for alpha elements studied here (Mg, Si,
Ca, Ti), at $+$0.34 dex, compares well with the halo average of
$+$0.35 (e.g. see McWilliam 1997).

As discussed above the [Na/Fe] and [Al/Fe] ratios are higher (by
$\sim$0.2 and 0.1 dex respectively) than typically seen for the 47~Tuc
metallicity, but this is likely due to proton burning products in the
atmospheres of the most luminous stars in the cluster.
Within the measurement uncertainties our 47~Tuc [Mn/Fe] ratio, at
$-$0.44 dex, is lower than, but consistent with, the deficient values
seen in Galactic stars and clusters at the same [Fe/H], near 
$\sim$$-$0.3 dex (e.g. Sobeck et al. 2006; Johnson 2002; McWilliam, Rich \&
Smecker Hane 2003; Carretta et al. (2004).

In the disk and halo of the Galaxy the [Cu/Fe] ratio declines roughly
linearly with decreasing [Fe/H], reaching a value of $\sim$$-$0.6 to
$-$0.7 dex by [Fe/H]$\sim$$-$1.5 (Mishenina et al. 2002; Simmerer et
al. 2003).  At the metallicity of 47~Tuc previous studies indicate
[Cu/Fe] near $-$0.1 to $-$0.2 dex.  This is entirely consistent with
the [Cu/Fe] ratio found here, at $-$0.13 dex, from our integrated-light
analysis.

W06 abundances for the light s-process elements Y and Zr are enhanced
by $\sim$0.6 to 0.7 dex, in contrast to the solar-like ratios found
here; unfortunately, there is insufficient data from other studies of 47~Tuc
to draw a conclusion regarding these two elements.  W06 selected a sample
of AGB and RGB stars from 47~Tuc, so it may be that their s-process
abundance enhancements simply reflect AGB evolution of the stars
themselves; however, they found similar s-process enhancements in both
their RGB and AGB populations.  Previous studies of Y and
Zr in nearby  Galactic disk and halo stars (e.g. Edvardsson et al. 1993; 
Gratton \& Sneden 1994) 
show that [Y/Fe] remains at the solar ratio to the metallicity of 47~Tuc, and
that [Zr/Fe] is very slightly enhanced, near 0.1 to 0.2 dex, consistent
with the results for these two elements found here.  We suggest that the enhancements
found for Y and Zr by W06 are most likely due to either systematic errors, or 
result from s-process enhancements in their AGB stars.

The [Eu/Fe] ratio in Galactic metal-poor stars follows a trend similar
to the alpha elements, increasing with decreasing [Fe/H] in the solar 
neighborhood, to a value near 0.3 to 0.4 dex and roughly flat below
[Fe/H]$\sim$$-$1 (McWilliam \& Rich 1994; Woolf et al. 1995).  At the
47~Tuc metallicity the measured [Eu/Fe] ratio of Galactic stars ranges
from 0.2 to 0.3 dex.  Thus, our rather low [Eu/Fe] ratio, at 0.04 dex,
is not only lower than other reported measurements in individual
47~Tuc stars, but is also at odds with the general trend seen in
Galactic stars in general.  With a central line depth of only $\sim$3
percent it is hardly surprising that the Eu~II line at 6645\AA\ gives
discordant results.

Table~\ref{table4} indicates a 0.25 dex enhancement in
[Sc~II/Fe], based on one line, but a normal solar-scaled Sc abundance
from a single Sc~I line.  However, the Sc~II enhancement in our
integrated-light analysis is in qualitative agreement with claims of
mild Sc enhancements with decreasing metallicity (e.g. Bai et
al. 2004; Nissen et al. 2000); but see Prochaska \& McWilliam (2000).

The remaining elements roughly scale with [Fe/H], similar to the trends seen
in the Galaxy (e.g. Edvardsson et al 1993; Sneden \& Gratton 1994; McWilliam 1997).

The above abundance trends indicate that our integrated-light 47~Tuc
abundances are consistent with the composition of Galactic stars of
similar metallicity.  This supports the idea that we have successfully
managed to perform detailed chemical abundance analysis on the
integrated light spectrum of the core of 47~Tuc, using the observed
color-magnitude diagram as an essential input ingredient.

\subsection{Integrated-Light Abundance Analysis With Theoretical Isochrones}

For GCs beyond the Local Group it is not currently possible to resolve
individual stars to obtain empirical {\em cmds}.  Thus, the next
step in developing an IL abundance analysis strategy for
extra-galactic clusters is to use theoretical isochrones with
parameters constrained by the IL spectrum.  In this paper, we have
employed theoretical isochrones from two groups: Padova and Teramo.  The 
Padova isochrones of Girardi et al. (2000) and Salasnich et al. (2000) 
were obtained from the group's website.\footnote{Padova isochrones were
obtained from the web site {\tt
    http://pleiadi.pd.astro.it/isoc\_photsys.00/isoc\_ubvrijhk}.} The
isochrones we use are the set computed with convective overshoot and a
constant mixing length.  For the Padova isochrones, our IL abundance
calculations were made with scaled solar composition only, due to the
very limited selection of alpha-enriched models.  The competing
isochrones from the Teramo\footnote{Teramo (BaSTI) isochrones were
  obtained from the web site {\tt http://www.te.astro.it/BASTI/}.}
group (e.g. Cassisi, Salaris \& Irwin 2003; Pietrinferni et al. 2006)
were available for a variety of assumptions at the start of our work,
including alpha-enhanced or scaled solar composition, with or without
convective overshooting, two mass-loss rates (Riemer's mass-loss
parameter $\eta$=0.2 and 0.4), and normal or extended AGB.  Maraston
(2005) investigated the merits of the these various Teramo isochrones,
as well as those from the Padova group, by comparing to observed
clusters and favored the Teramo (BaSTI) isochrones.  Guided by the
Maraston (2005) and the BaSTI web site recommendations, we selected
the Teramo classical evolutionary tracks with no overshooting for the
treatment of the core, a metallicity-dependent mixing length
parameter, an extended AGB, and a mass-loss parameter of $\eta$=0.40.
Most recently the Teramo isochrones have been updated for corrections
to the alpha-enhanced opacities initiated by Ferguson et al. (2005).
While the Teramo isochrones offer a greater parameter flexibility than
the Padova isochrones, we believe that it is useful to compare the IL
abundance results obtained using both the Teramo and Padova
isochrones.

Both the Teramo and Padova isochrones are provided without luminosity
or mass functions. We therefore computed the frequency of each point
on the theoretical isochrones using the IMF recommended by Kroupa
(2002), and the initial masses indicated for each isochrone
point.  A simple program was used to bin points along the theoretical
isochrones into {\em cmd} boxes, each containing at least 3.5\% of the
total V-band flux.  For each box the V-band flux-weighted model
atmosphere parameters were computed according to the equations 4, 5
and 6, above.  In this way, the theoretical isochrones provided an
input file of atmosphere parameters for our IL abundance analysis
program  with 21--27 {\em cmd} boxes, similar to the input file
used for the observed {\em cmd} analysis (see Table~\ref{table3}).

For the most luminous stars in the cluster, near the tip of the giant
branch, small number statistics can lead to incomplete sampling if the
cluster has a low total luminosity or if only a small fraction of the
cluster is scanned with the spectrograph. The latter is true for our
spectrum of 47~Tuc.  In order to address the issue of this statistical
incompleteness in the observed IL spectrum of the 47~Tuc core we used
the theoretical isochrones and Kroupa IMF to estimate the number of
stars in each {\em cmd} box.  To make this calculation it was
necessary to input an estimated M$_{\rm v}$ for the 32$\times$32 arc
second region of the core scanned with the spectrograph slit.  We did
this by summing the observed V-band fluxes of the stars in the HST
core photometry and correcting for our adopted distance modulus and
foreground reddening. This procedure resulted in a value of
M$_{\rm v}$(core) =$-$6.25.  As the giant branch tip is approached,
the number of stars in the high luminosity {\em cmd } boxes decreases.
We truncated the top of the {\em cmd} luminosity function at the point
where the integrated probability of finding a single star fell below 
0.5, since it was more probable that no stars were present in the cluster 
beyond this point.

\begin{figure}
\plotone{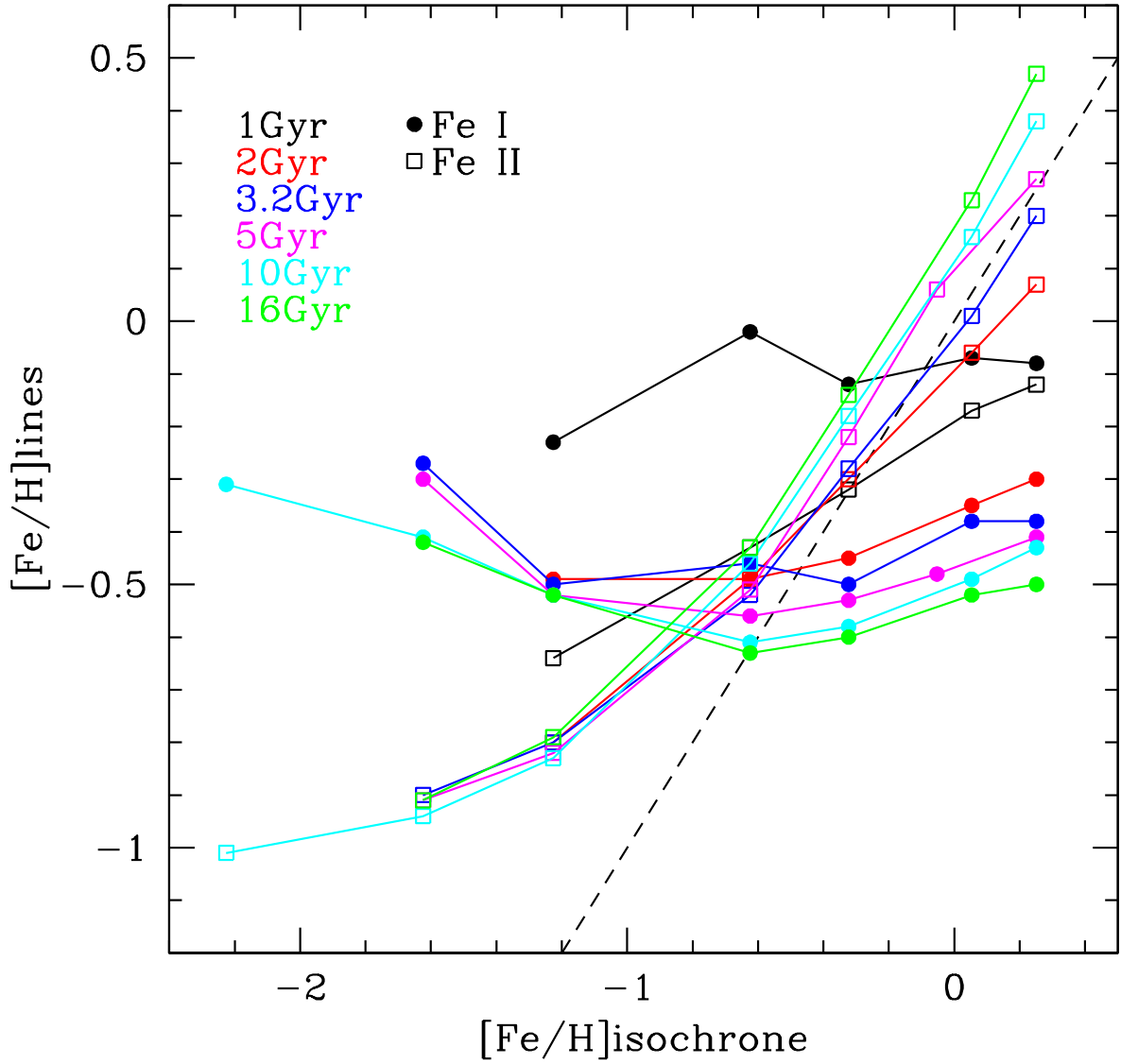}
\caption{Computed abundances from Fe~I and Fe~II lines in 47~Tuc
  (filled circles and open squares respectively) for Padova isochrones
  of various ages and metallicities.  Note the different sensitivity
  to isochrone metallicity of the Fe~I and Fe~II lines, and the
  dependence on age.  The dashed line is the locus of [Fe/H] from the Fe lines
  equal to the value in the input isochrone.}
\label{fig-ilabunds.padova.sun}
\end{figure}

\begin{figure}
\plotone{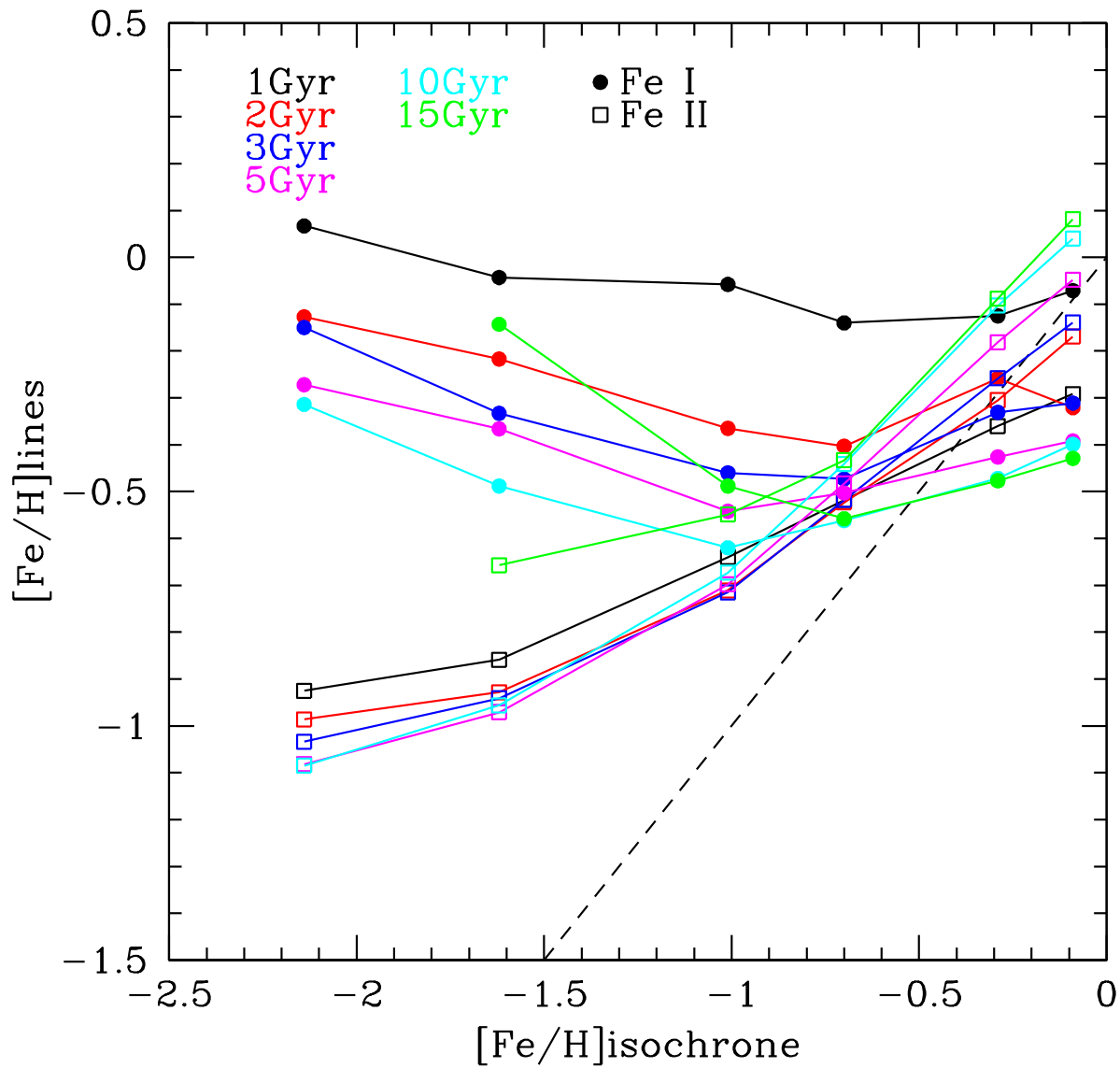}
\caption{Computed abundances, from Fe~I and Fe~II lines (filled
  circles and open squares respectively) in 47~Tuc, for various ages
  and metallicities, using Teramo (BaSTI) alpha-enhanced,
  AGB-enhanced, classical isochrones with mass loss parameter
  $\eta$=0.40.  Note the different sensitivity to isochrone
  metallicity of the Fe~I and Fe~II lines, and the dependence on age.}
\label{fig-ilabunds.basti.alpha}
\end{figure}

Following the same analysis procedures as before we computed Fe~I and 
Fe~II abundances from the 47~Tuc IL EWs using theoretical isochrones, from 
the Padova and Teramo groups, in place of the observed {\em cmds}.  However, 
because the age and metallicity were not known in advance we computed the 
line abundances for a range of assumed age and metallicity.  For the Padova
scaled solar isochrones we used scaled solar composition Kurucz model
atmospheres with the ODFNEW opacity distribution function; the results are
presented in Figure~\ref{fig-ilabunds.padova.sun}.  From the Teramo (BaSTI)
group, we used the alpha-enhanced isochrones throughout, corrected for
the Ferguson et al. (2005) alpha opacities, together with the
alpha-enhanced Kurucz stellar model atmosphere grid with the AODFNEW
opacity distribution functions (Castelli \& Kurucz 2004).  For the
Teramo (BaSTI) isochrones, spanning a large range of age and
metallicity, the iron abundance results are shown in
Figure~\ref{fig-ilabunds.basti.alpha}.

In Figure~\ref{fig-ilabunds.padova.sun} and
\ref{fig-ilabunds.basti.alpha}, the Fe~I abundances
that we identify using the 
isochrones at any given age
change by less than 0.2 dex over the whole range of input isochrone
metallicities.  The Fe~I abundances agree with the input isochrone metallicity 
near [Fe/H]$\sim$$-$0.5 to $-$0.6 dex for the 15 Gyr models.
The figures show that significantly higher
[Fe/H] values are obtained from the lines if very young ages are
adopted; at 1Gyr, agreement between isochrone metallicity and line
abundances occurs near $-$0.1 dex, while for old ages the result
hardly changes at all between 10 and 15 Gyr.  For iron abundance
uncertainties of $\sim$0.1 dex the convergence of the stellar
isochrones result in an insensitivity to age older than about 5 Gyr.

The Fe~II line abundances in Figures~\ref{fig-ilabunds.padova.sun} and
\ref{fig-ilabunds.basti.alpha} increase much more strongly with
isochrone metallicity than the Fe~I lines.  We interpret the strong
dependence of the Fe~II abundance with adopted input isochrone
metallicity as follows: when the adopted isochrone metallicity is
overestimated, the computed continuum opacity is artificially enhanced
due to increased H$^-$ from increased electrons from the ionization of
the extra metals.  In this case, to match the integrated-light Fe~II
line strength requires an over-estimate of the true metal content, as
this depends on the line to continuum opacity.  In this way, the
computed Fe~II abundances increase with the adopted isochrone Z.  For
Fe~I lines, on the other hand, an overestimated isochrone metallicity
results in enhanced recombination of Fe~II to Fe~I in the model
atmospheres. This increase in the Fe~I number is matched by a similar
increase in the H$^-$ opacity and, as a result, the computed Fe~I line
strengths remain relatively unchanged for a given input isochrone
metallicity.  At very low isochrone Z, around 1/100 solar, the Fe~II
abundance trends in Figures~\ref{fig-ilabunds.padova.sun} and
\ref{fig-ilabunds.basti.alpha} become less sensitive to metallicity.
We believe that this results from the fact that the electron density
in very metal-poor stellar atmospheres no longer depends upon
ionization of metals, but is dominated by the ionization of hydrogen.
We note that the Fe~II abundances in
Figures~\ref{fig-ilabunds.padova.sun} and
\ref{fig-ilabunds.basti.alpha} also show a sensitivity to isochrone
age, with the largest change for the youngest ages; but the direction
of the derived abundance changes is such that younger isochrones give
lower Fe~II abundances, whereas Fe~I abundances are increased.

It must be noted that the initial impression from
Figures~\ref{fig-ilabunds.padova.sun} and
\ref{fig-ilabunds.basti.alpha} is that the self consistency appears to
be better for the Padova iron abundance than for the Teramo results,
because the Fe~II and Fe~I abundances and the isochrone [Fe/H] value
agree at similar values, near $-$0.6 dex as mentioned above.  On the
other hand, in these figures, the Fe~I and Fe~II agreement for the
Teramo-based analysis agree at much lower [Fe/H] than Fe~I and the
isochrone [Fe/H].  This could easily be due to higher adopted
[$\alpha$/Fe] ratios for the Teramo isochrones and Kurucz
alpha-enhanced models than actually present in 47~Tuc.  Furthermore,
the whole issue is somewhat confused by the recent changes in the
solar composition by Asplund et al. (2005) and the delay between such
solar results and the compositions used to compute stellar model
atmospheres and isochrones.  While the Padova agreement is better we
must remember that 47~Tuc really is enhanced in alpha elements
(e.g. see Table~\ref{table4}), as expected from its overall
metallicity, so alpha-enhanced models are appropriate.  Thus, it may
be that the Fe~I, Fe~II and isochrone [Fe/H] agreement found by use of
the solar-composition Padova isochrones was simply fortuitous.

\subsubsection{Isochrone Problems?}

In order to identify the cause of the differences between 
the iron abundance results from observed {\em cmd} and
theoretical isochrones, we have compared the
observed and predicted luminosity functions for 47~Tuc in detail.
To do so, we adopt the 47~Tuc distance modulus of 13.50$\pm$0.08
and age, of 11.2$\pm$1.1 Gyr, from Gratton et al. (2003).  We note
that more recent results reviewed by Koch \& McWilliam (2008) 
indicate a mean distance modulus of 13.22, but this short
scale leads to a change in $\log{g}$ of $\sim$0.1 dex, and does not
significantly affect the resultant abundances.  For the reddening we
averaged the two E(B$-$V) values given by Gratton et al. (2003), at
0.021 and 0.035, with the value of 0.032 from Schlegel et al. (1998),
for a mean of 0.03 magnitudes.  For this comparison, we adopted the
alpha-enhanced, AGB-enhanced, Teramo isochrones with a metallicity of
Z=0.0080 (corresponding to [Fe/H]=$-$0.70 according to the Teramo web
site), and a mass-loss parameter of $\eta$=0.40.  We note that the
assumed conversion between [Fe/H] and Z for alpha-enhanced composition
varies significantly in the literature; this is likely due to the
recent changes to the best estimates for the solar oxygen and iron
abundances (e.g. Asplund et al. 2005).

\begin{figure}
\plotone{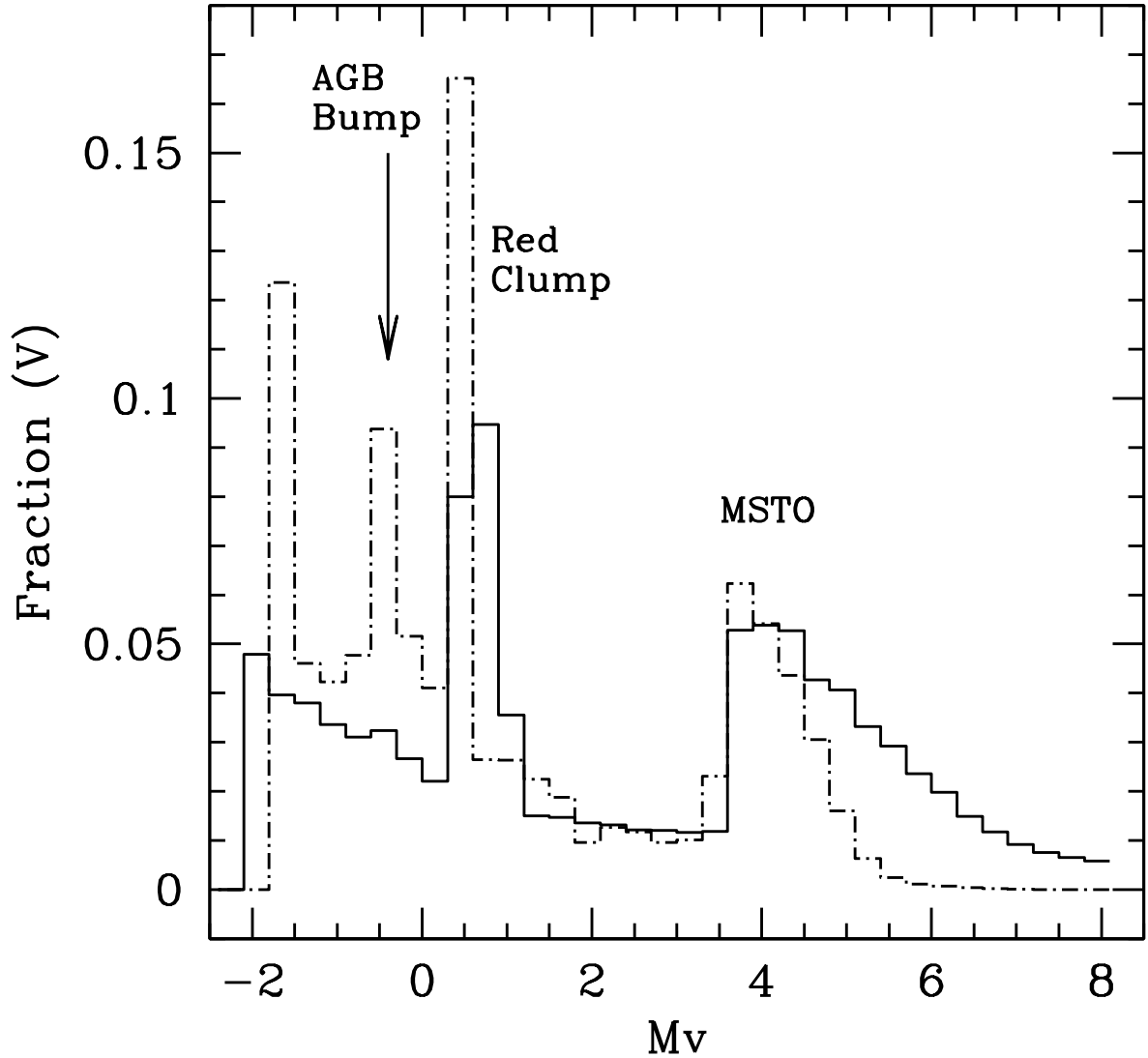}
\caption{A comparison of the observed V-band luminosity function for
  the 47~Tuc core (dashed line) with the 11 Gyr alpha-enhanced, AGB
  enhanced, Teramo (BaSTI) isochrone, with no overshooting, Z=0.0080
  and $\eta$=0.40 (solid line).  The observations indicate a paucity
  of stars below the main sequence turnoff, but more red giants, in
  particular the AGB bump region near M$_{\rm v}$=$-$0.5.  }
\label{fig-basti.11gyr.compare}
\end{figure}

In Figure~\ref{fig-basti.11gyr.compare}, we compare the observed and
theoretical V-band luminosity functions for 47~Tuc.  We note that
while the adopted age of 11 Gyr provides a good match to the {\em cmd}
at the main sequence turnoff and the Red Clump, it is also possible to
obtain good matches with the observed luminosity function using older
ages and a shorter distance modulus for the cluster. For this reason,
we cannot confidently measure the cluster age.  For the isochrones
which produce good matches to the observed luminosity function, three
major differences are still apparent: a deficit of stars below the
turnoff in the observations; a general excess of observed giants above
the Red Clump, highlighted by the spike in stars at the AGB bump; and an
excess in the observed {\em cmd} at the very tip of the giant branch.

While it might be reasonable to assume that the paucity of stars
observed below the main sequence turnoff is due to incomplete
photometry, Howell et al. (2000) has noted this in the data and
attributed it to a real deficit of low-mass stars, presumably caused
by dynamical mass-segregation processes.  Mass segregation was first
observed in the 47~Tuc core by Paresce et al. (1995) and is known in
numerous other cluster cores (e.g. King, Sosin \& Cool 1995; Ferraro
et al. 1997; Gouliermis et al. 2004).  The striking excess of AGB bump
stars in the observed core V-band luminosity function corresponds to
the group of stars near V$\sim$13.2 and (B$-$V)$\sim$1.0 in the {\em
  cmd} shown in Figure~\ref{fig-47tuccmd}.  We note that this same
group of stars is present in the Kaluzny (1997) V,V$-$I 47~Tuc
color-magnitude diagram of an outer field.  We investigated a number
of parameters for the Teramo isochrones, but we could find no
combination of ages, from 10 to 14 Gyr, or mass-loss rate that gave a
spike corresponding to the AGB bump.  We conclude that the models are
deficient in predicting the AGB bump, despite the use of ``AGB
enhanced'' models.  This under-prediction is reminiscent of the
Schiavon et al. (2002) 0.4 dex deficiency of the theoretical AGB
numbers compared with observations; however, here we require an
enhancement for the AGB bump only.  Finally, while there is an excess
of flux at the tip of the observed giant branch luminosity function,
compared to the predictions, the total flux at the tip comes from only
5 stars; thus, at $\sim$30\% this flux excess is within the 1$\sigma$
Poisson noise of $\sim$45\%.  Therefore, while the excess flux from
the giant branch tip is real, it does not indicate a problem with the
theoretical isochrone.

\begin{figure}
\plotone{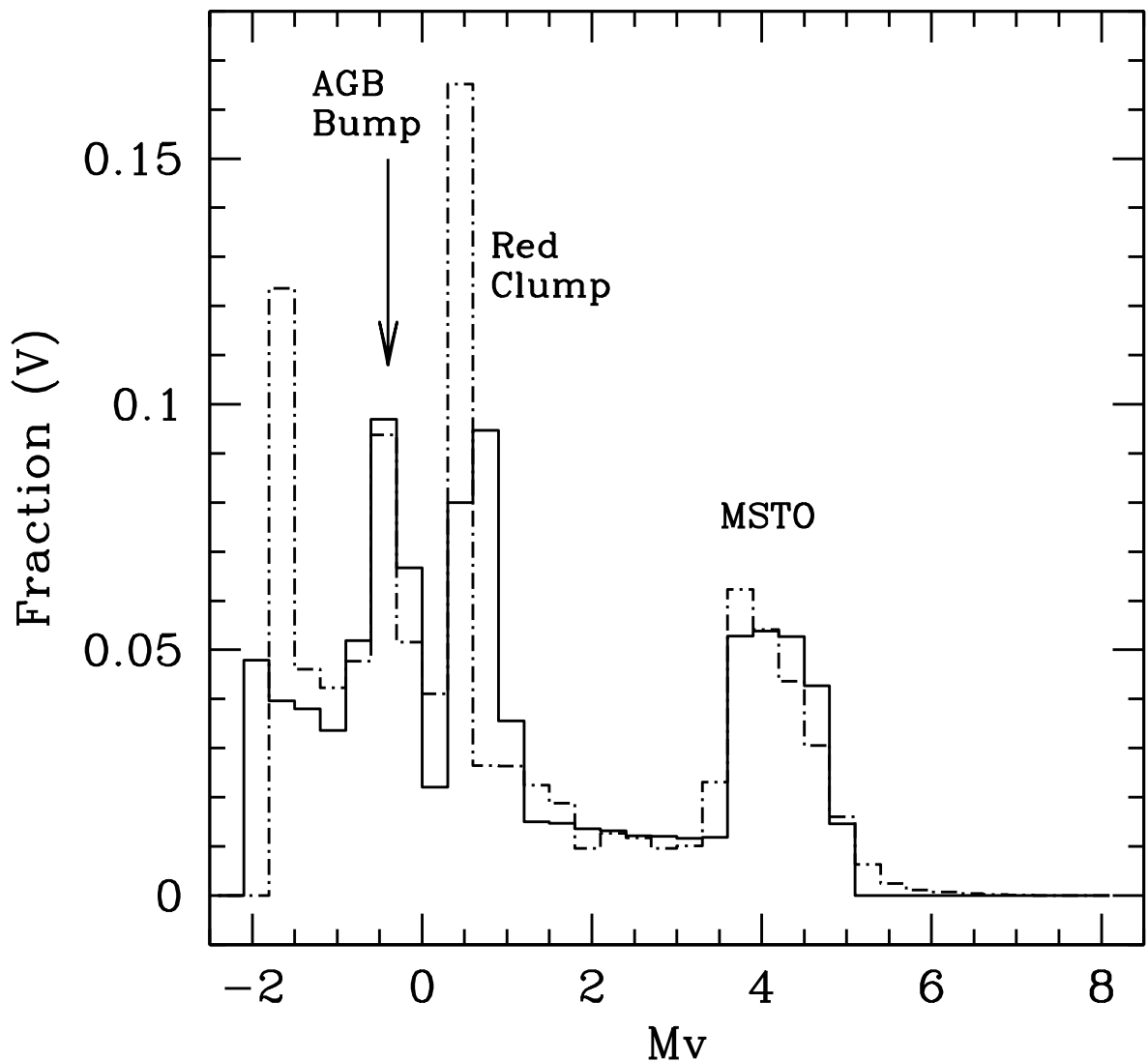}
\caption{A comparison of the observed V-band luminosity function for
  the 47~Tuc core (dashed line) with a the modified 11 Gyr
  alpha-enhanced, AGB enhanced, Teramo (BaSTI) isochrone, with no
  overshooting, Z=0.0080 and $\eta$=0.40 (solid line).  Modifications
  include a cutoff for low luminosity stars below M$_{\rm v}$=$+$4.90,
  and an enhancement, by a factor of 3.0, for stars in the between
  M$_{\rm v}$ of $-$0.10 and $-$0.70.  The modified isochrone is a
  much better fit to the observed {\em cmd}, but there are still more
  giants in the real cluster than the predictions.  }
\label{fig-altered.11gyr.compare}
\end{figure}

We have applied two corrections to the theoretical Teramo isochrone to
make the corresponding V-band luminosity function appear more like the
observed function.  First, to approximately match the main sequence
turnoff region, we ignore all stars with M$_{\rm v}$ more than
$+$4.90.  Second, to match the observed AGB bump region, we apply an
enhancement factor of 3.0 to stars in the M$_{\rm v}$ range from
$-$0.10 to $-$0.70.  Figure~\ref{fig-altered.11gyr.compare} compares
the observed luminosity function with the corrected Teramo isochrone;
the match is significantly improved, although it is clear that the
real cluster still has slightly more flux coming from the giants than
the model.  Although not plotted here, we note that the theoretical
luminosity function for Padova isochrones with age and abundance
parameters like 47~Tuc show differences with the 47~Tuc luminosity
function that are very similar to those discussed above; they also
show a deficit in the predicted AGB bump by a factor of 3, and RGB tip
deficit slightly more than for the Teramo isochrones, and a
near-identical overestimate of dwarf stars below the turnoff.

\begin{figure}
\plotone{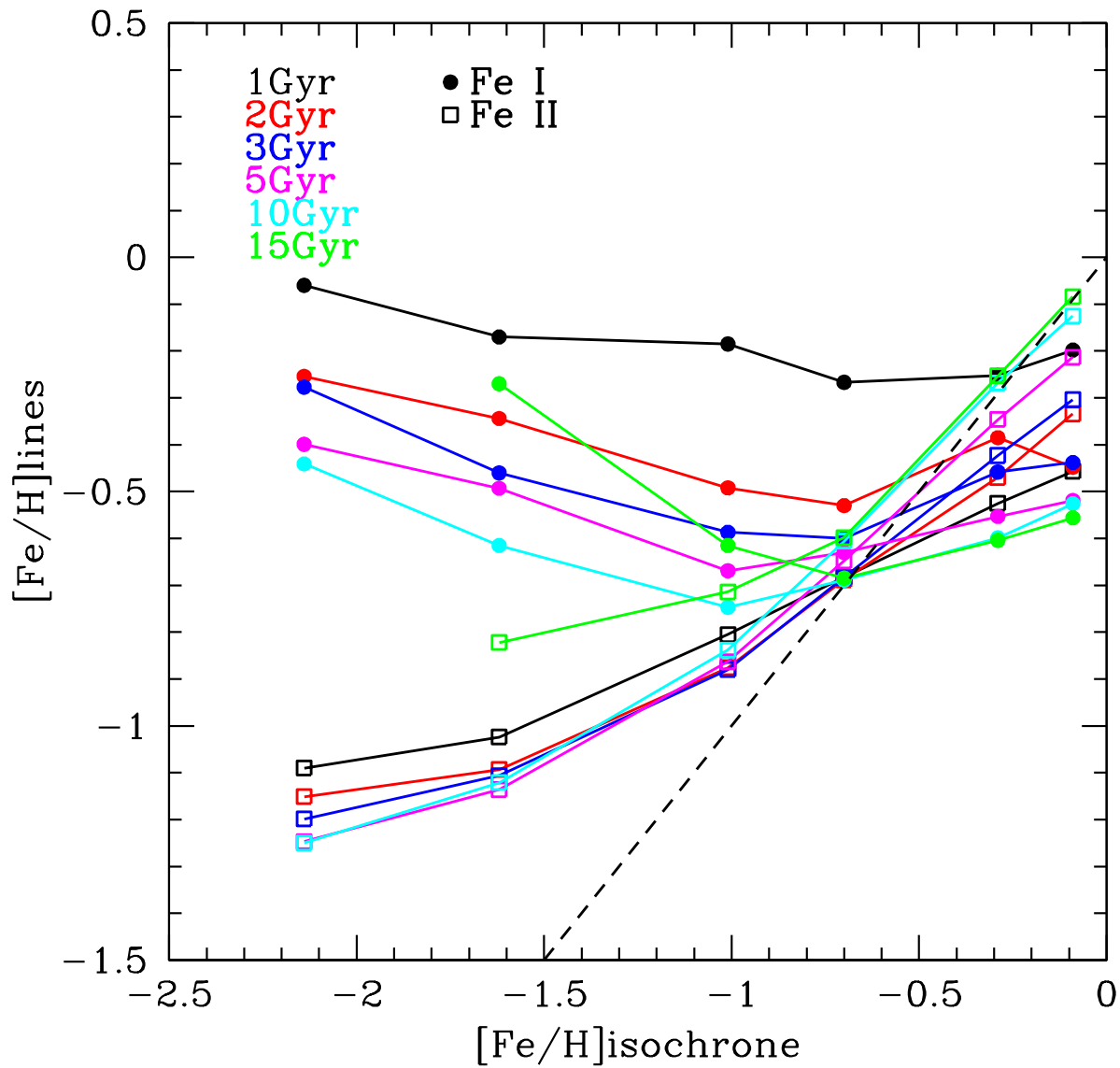}
\caption{Computed abundances for Fe~I and Fe~II (filled circles and
  open squares respectively) in 47~Tuc integrated light, using Teramo
  (BaSTI) isochrones with corrections of $-$0.125 and $-$0.165 dex
  respectively; similar to Figure~\ref{fig-ilabunds.basti.alpha}.
  Note that for 10 and 15 Gyr Fe~I and Fe~II abundances are equal
  $\sim$0.3 dex lower than the isochrone metallicity; we do not
  understand this inconsistency. }
\label{fig-tracks1.corrected.basti.alpha}
\end{figure}

A computation of the IL abundances from the measured Fe~I and Fe~II
lines gives different results for the original 47~Tuc Teramo isochrone
and our modified version of the isochrone with the MS and AGB bump
corrections.  For our list of Fe~I lines the mean difference is
$-$0.125 dex, while for the Fe~II lines the difference is $-$0.165
dex, where these differences are corrected-isochrone minus
uncorrected-isochrone abundance.  To first order the differences
provide zero-point offsets to be applied to uncorrected-isochrone
abundances.  We note that these offsets are valid only for one specific 
age and metallicity.  Given the variability of turnoff, RGB, HB and AGB 
stars in GCs with different ages and metallicities it is unlikely that 
these offsets will be universal.
When these corrections are applied to the abundances in
Figure~\ref{fig-ilabunds.basti.alpha} we obtain a corrected plot,
Figure~\ref{fig-tracks1.corrected.basti.alpha}, showing the dependence
of [Fe~I/H] and [Fe~II/H] on assumed age and isochrone metallicity.
We note that there is a systematic trend of decreasing derived Fe~I
abundance with increasing age, similar to the Padova isochrone results
in Figure~\ref{fig-ilabunds.padova.sun}, but at 15 Gyr the Fe~I
abundances show an increase over the 10 Gyr model; a similar
turn-around occurs in the direction of the derived Fe~II abundances.
This effect appears to result from a change in the core helium-burning
stars in the isochrone from red clump to blue horizontal branch,
and is related to age, metallicity and the mass-loss parameter,
$\eta$.  

As shown in
Figure~\ref{fig-tracks1.corrected.basti.alpha}, the Fe~I lines give
$-$0.70 dex for the 10--15 Gyrs isochrones with [Fe/H] $-$0.70 dex.
The satisfying agreement between isochrone [Fe/H] and the value
derived from Fe~I lines indicates an older age range (10--15 Gyr) for
47~Tuc, as expected from color-magnitude diagrams (e.g. Gratton et
al. 2003); ages of 3 Gyr or less are completely inconsistent with the
Fe~I line abundances.  The Fe~II lines give higher abundances than by
0.08 and 0.09 dex, respectively, for the 10-15 Gyr isochrones with
[Fe/H]=$-$0.70 dex.  While the difference of $\sim$0.08 dex between Fe~I
and Fe~II abundances are within measurement uncertainties, there is a
systematic difference which bears some thought.  We suggest three
possibilities that might explain these systematically higher Fe~II
abundances derived from the theoretical isochrones.  First, there may
be systematic errors in the measured equivalent widths of Fe~II lines,
which are weak and often blended in the 47~Tuc IL spectrum.  Second,
the alpha enhancement adopted for the theoretical isochrone may be
higher than the actual alpha enhancement of the cluster.  This may be
related to the reduction in the solar oxygen abundance by Asplund et
al. (2005). Third, there may be a statistical variance, or other
mechanism, that increased the representation of stars at the tip of
the giant branch in excess of those predicted by the Teramo
isochrones.

Given that our Padova isochrones employed solar composition rather
than including the alpha enhancements necessary for to 47~Tuc, more
detailed comparisons of the merits of the two sets of theoretical
isochrones is not possible at this time.

\subsubsection{Isochrone Diagnostics and Abundances for Unresolved GCs}

A significant difficulty in the use of the theoretical isochrones for
IL abundance analysis is the choice of isochrone parameters of age,
metallicity, and alpha enhancement.  Fortunately, for 47~Tuc we can
assume a roughly old age, and be assured that our Fe~I abundances will
be close to the truth.  For unresolved clusters, the integrated colors
could be used to constrain the combined effects of age and metallicity;
however, the well-known age-metallicity
degeneracy seen in cluster {\em cmds} makes determination of the
isochrone parameters impossible solely on the basis of photometry.
This is similar to the covariance between temperature and metallicity
in the abundance analysis of individual stars.  For full spectroscopic
analysis of individual stars, various abundance diagnostics are used
to constrain the atmosphere parameters.  Likewise, we propose to use
diagnostic spectral features to constrain the GC isochrone parameters,
resulting in an isochrone choice that is completely consistent with
the spectra.  The most obvious example of such a diagnostic is that
the isochrone metallicity must be consistent with the abundances
computed from the absorption lines; this may include the alpha
enhancement, which should agree with the abundances returned from the
alpha element lines.  We discuss a variety of other diagnostics below.

In general the abundance computed for an individual line depends on
the isochrone age and metallicity, but also on the excitation potential
and ionization stage of the line.  For example, if an isochrone with an
incorrect age is adopted in the analysis, abundances derived from low
excitation Fe~I lines will be different than for high excitation Fe~I
lines due to the inappropriate stellar temperatures, resulting in a non-zero
slope in the plot of iron abundance versus excitation potential.  Our work (see
below) bears out this expectation, however we also note that this use
of excitation potential versus iron Fe~I line abundances as an age
diagnostic may be complicated by the presence of a hot horizontal
branch that might be confused with hot main sequence stars in old, low
metallicity clusters.  The Fe~II lines should also provide a
diagnostic, because they are sensitive to the electron density (unlike
the Fe~I lines), which in individual stars is a strong function of
gravity.  Thus, a constraint on the isochrone parameters comes
from the consistency of the abundance results for lines from both
ionization stages of iron.

Another potential spectral diagnostic is derived from the computed
abundance of individual Fe lines as a function of wavelength. A
sub-population of relatively hot stars within a cluster will have its
greatest contribution to the total flux at blue wavelengths.  Thus, we
may only expect agreement between iron abundances from blue and red
wavelengths with the correct mix of hot and cool stars.  
This constraint of consistency with wavelength is similar
to the the method used by Maraston et al. (2006), who found that
thermally pulsing AGB stars contribute significantly to the total infrared
fluxes of stellar populations in the $\sim$1 Gyr age range.

\begin{figure}
\plottwo{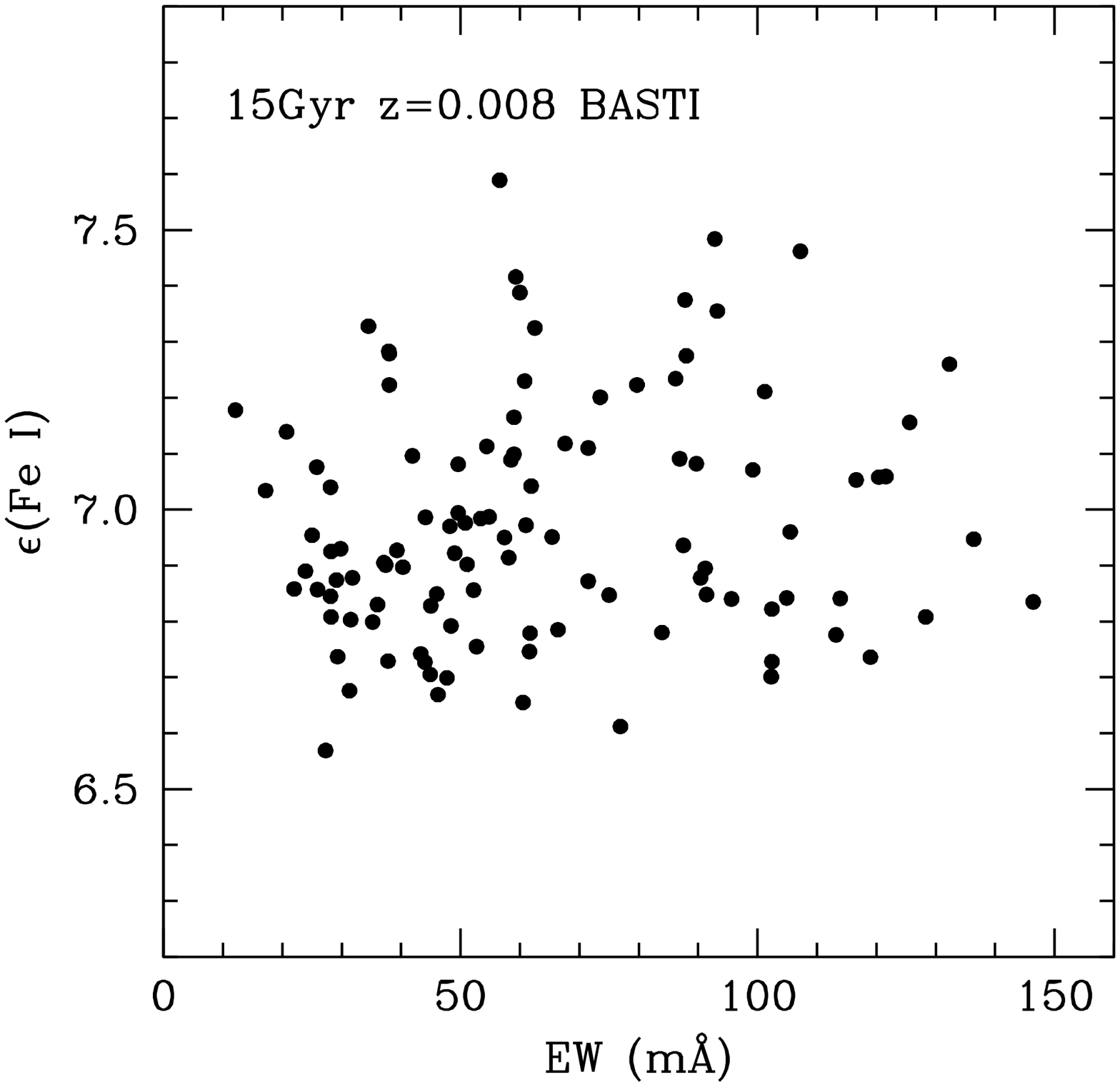}{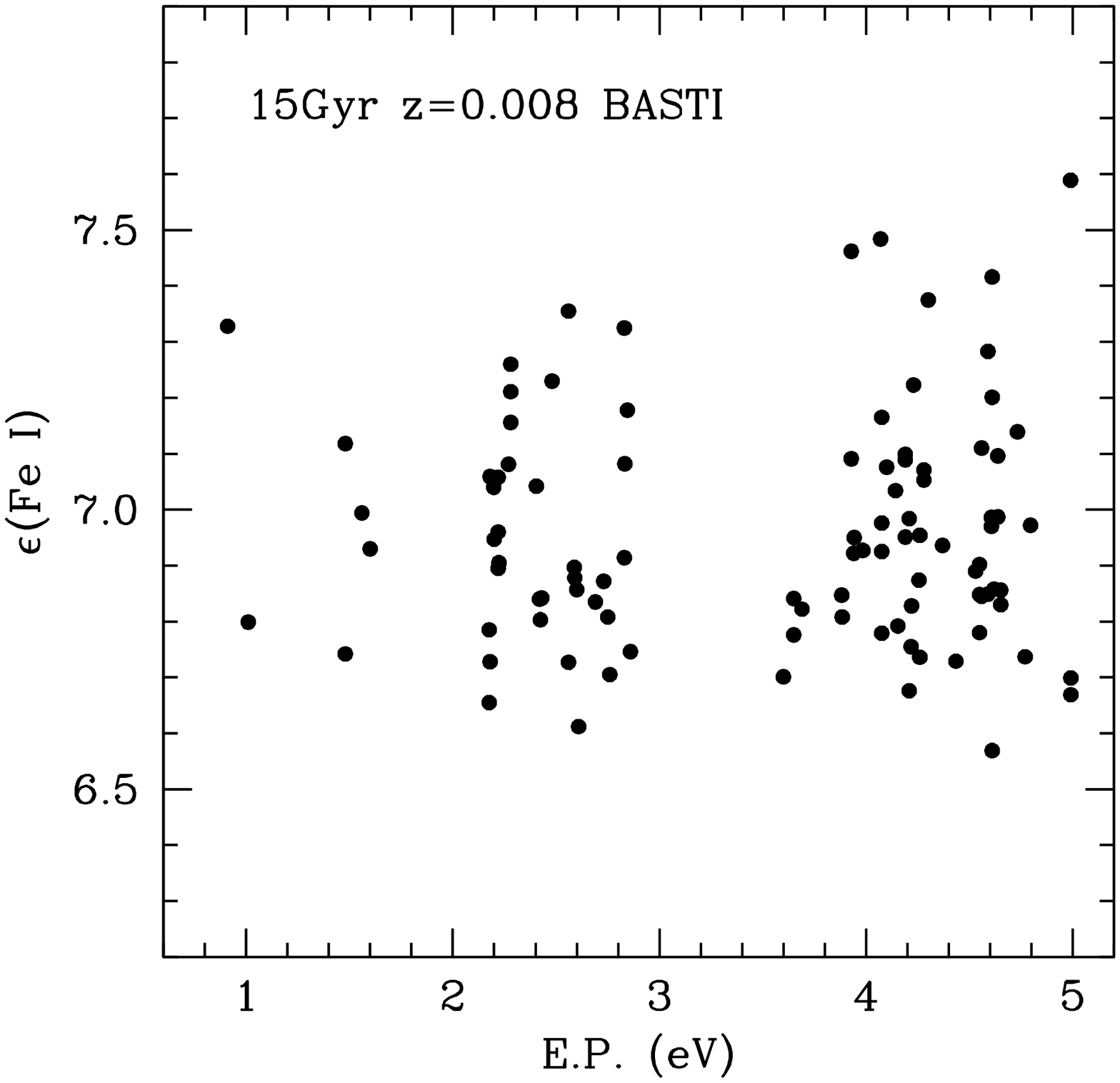}
\caption{ Plots of iron abundance versus EW and EP, computed using the
  Teramo 15Gyr, z=0.008, alpha-enhanced isochrone.  {\bf Left
    Panel:\quad} The iron abundance is independent of EW, indicating
  that the adopted mix of microturbulent velocities is consistent with
  the observed cluster.  \quad{\bf Right Panel:\quad} The roughly
  independent run of iron abundance with EP indicates that the mixture
  of stellar temperatures is consistent with the spectrum.}
\label{fig-ewepiso15z008}
\end{figure}

To investigate spectral diagnostics of the isochrone parameters we begin with
Figure~\ref{fig-ewepiso15z008}, which shows iron abundances 
 versus EW and excitation potential (EP) for the lines 
in Table~2. Abundances for this plot were
derived using a theoretical, alpha-enhanced, Teramo {\em cmd}
for an age of 15 Gyr.   Similar
plots are used as diagnostics for studies of individual stars.
Figure~\ref{fig-ewepiso15z008} shows that the iron abundance is
approximately independent of EW and EP, which indicates that the
microturbulent velocities and temperatures employed in the abundance
calculations were consistent with the measured line EWs.

\begin{figure}
\plottwo{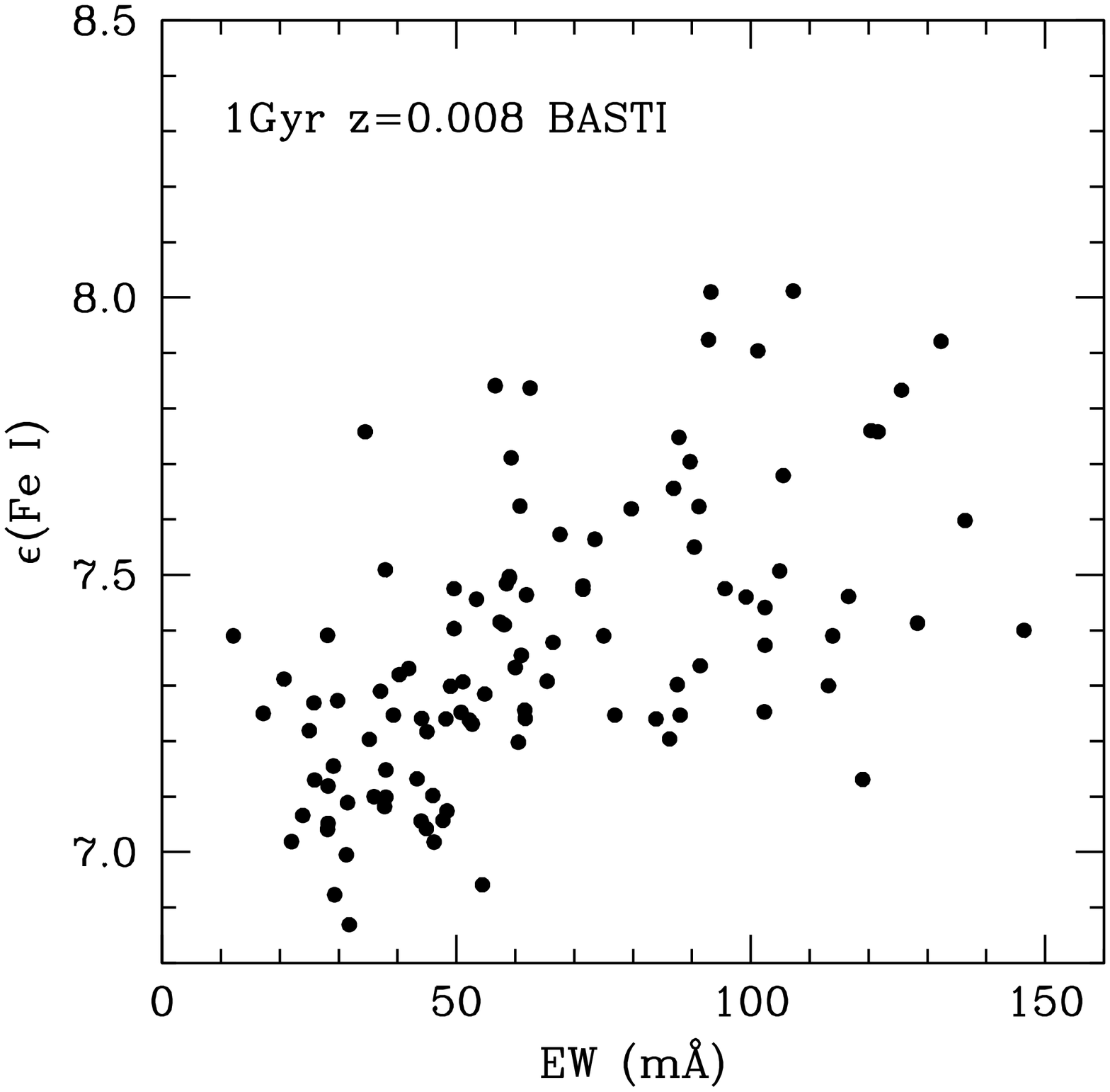}{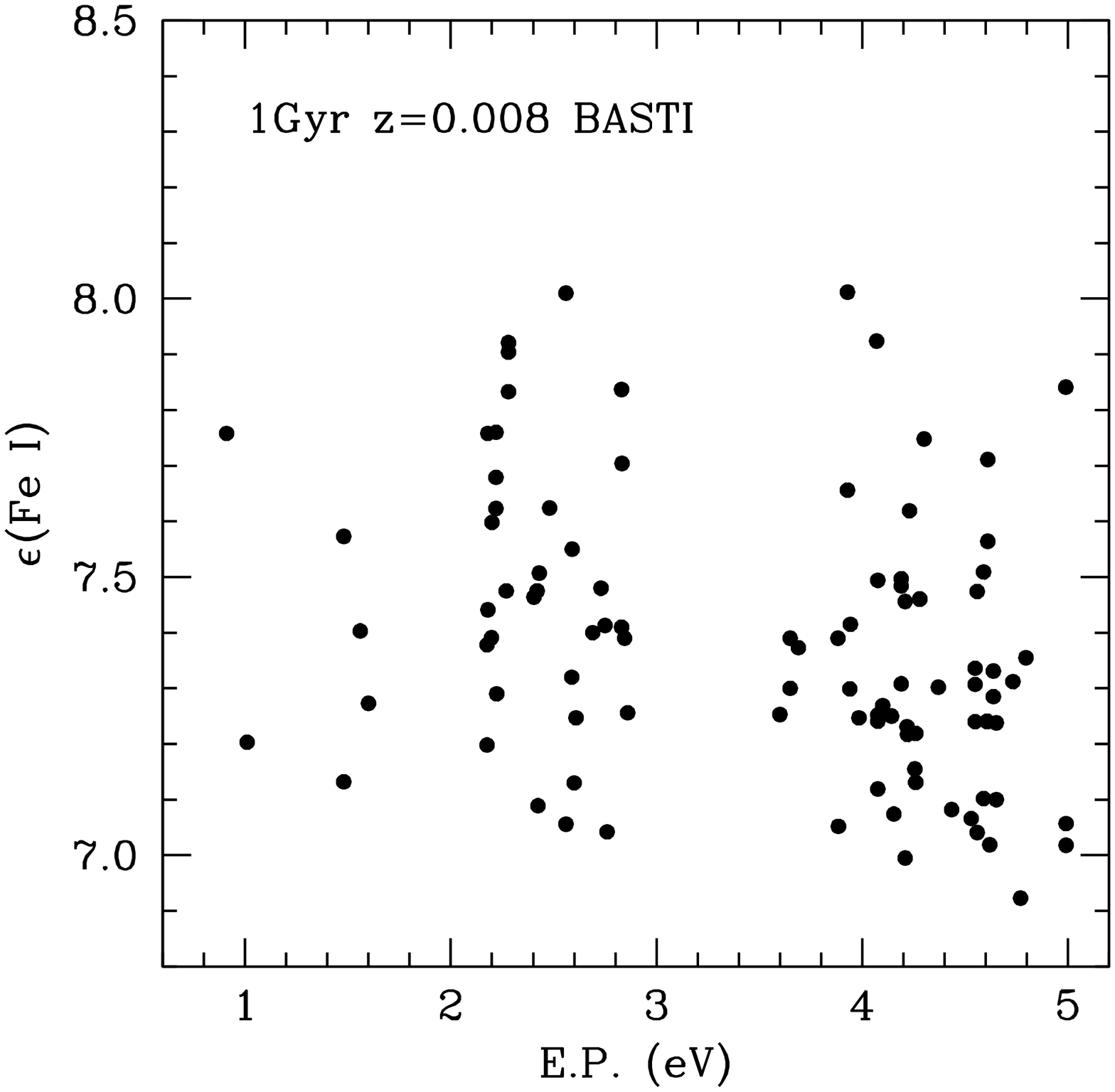}
\caption{ Plots of iron abundance versus EW and EP, computed using the
  Teramo 1Gyr, z=0.008, alpha-enhanced isochrone and measured EWs.
  {\bf Left Panel:\quad} The positive slope of iron abundance with EW
  indicates that the actual {\em cmd} contains stars with smaller microturbulent
  velocities than the observed cluster; presumably this results from
  the large fraction of main sequence stars in the 1Gyr isochrone, which
  have high gravities and low microturbulent velocities.  
\quad{\bf Right Panel:\quad} The decreasing iron abundance with line
  excitation potential indicates that the actual Globular Cluster is
  cooler than the input isochrone; this is expected if the adopted isochrone
  contains more young, hot main sequence stars than the cluster.}
\label{fig-ewepiso1z008}
\end{figure}

Figure~\ref{fig-ewepiso1z008} shows the results of our abundance
calculations using a Teramo isochrone for an age of only 1Gyr.  In this
case the plot of iron abundance versus EW indicates that abundances derived
from strong and weak lines do not agree, with strong lines giving higher
abundances.  In an analysis of a single star's spectrum, this observation
would indicate that the adopted microturbulent velocity parameter is
too small.  Since Figures~\ref{fig-47tuc.ewab.bv} and our analysis
using the empirical {\em cmd} has already demonstrated that our
adopted microturbulent velocity law is roughly correct, the
disagreement between strong and weak lines in
Figure~\ref{fig-ewepiso1z008} suggests that the input isochrone has
too many stars with low microturbulent velocity.  This suggests that
the input isochrone contains more dwarf stars than the cluster,
because dwarf stars have lower microturbulent
velocities than giants.  The plot of iron abundance with excitation
potential for the 1 Gyr isochrone shows a decrease in abundance with
increasing EP.  For single stars this would indicate that the adopted
temperature is too low.  In the integrated light analysis this
suggests that there is an excess of hot stars in the input isochrone
compared to the real cluster, which is consistent with the idea that
there is a larger fraction of dwarf stars in the 1Gyr isochrone than
in the observed cluster.  Thus, both the EW and EP plots suggest that
the older, 10--15 Gyr isochrones provide better consistency among
computed iron line abundances than do very young isochrones ($\sim$1
Gyr).

Figures~\ref{fig-ewepiso15z008} and \ref{fig-ewepiso1z008} are used
here only to demonstrate diagnostic possibilities; clearly careful
measurement of the deviations from zero slope in the EW and EP plots
offers the possibility to determine the best fit isochrone age.  The
measured 1$\sigma$ uncertainty in the slopes of these two figures, at
0.00063 dex/m\AA\ and 0.019 dex/eV, respectively, is sufficient to
provide 3$\sigma$ constraints on ages younger than 2 Gyr only.
However, a factor of 2 improvement in the abundance measurement uncertainties
would permit 3$\sigma$ measurement of ages up to 5 Gyr.  This reduced
abundance scatter could reasonably be expected by mitigation of uncertainties
due to $gf$ values, blends, and use of higher S/N spectra.  In this regard a
differential abundance technique, relative to a standard giant or a known
cluster, and a careful study of each line would be fruitful.
Indeed, results with a GC training set (Bernstein et al. 2008) suggests that
it will be possible to provide age constraints for the current plots
to ages up to $\sim$3--5 Gyr.  Older than 5 Gyr the abundance slope
differences are too small to constrain the age.

In principle, iron ionization equilibrium (consistency of the Fe~I and 
Fe~II abundances and isochrone metallicity) can also be used to constrain
the isochrone age.  This is particularly sensitive to the Fe~II abundance,
which requires the additional constraint of the isochrone alpha enhancement
and proper accounting for the most luminous stars in the isochrone, since the
Fe~II lines are strongest at the lowest gravities.  In
Figure~\ref{fig-tracks1.corrected.basti.alpha} we see that the 5 Gyr isochrone
for our 47~Tuc model is closest to consistency of Fe~I, Fe~II abundances and 
the isochrone metallicity; however, the 47~Tuc {\em cmd} suggests a significantly
older age, close to 12 Gyr (e.g. Koch \& McWilliam 2008).

Another age diagnostic may be obtained from a plot of iron abundance
with wavelength.  In this case, excessively young ages would give
different abundances for lines at blue wavelengths.  In
practice, both our 1Gyr and 10--15 Gyr give noisy, approximately flat,
trends of abundance with wavelength for our spectra; although 
much bluer spectral lines may have shown sensitivity to age.

\begin{figure}
\plotone{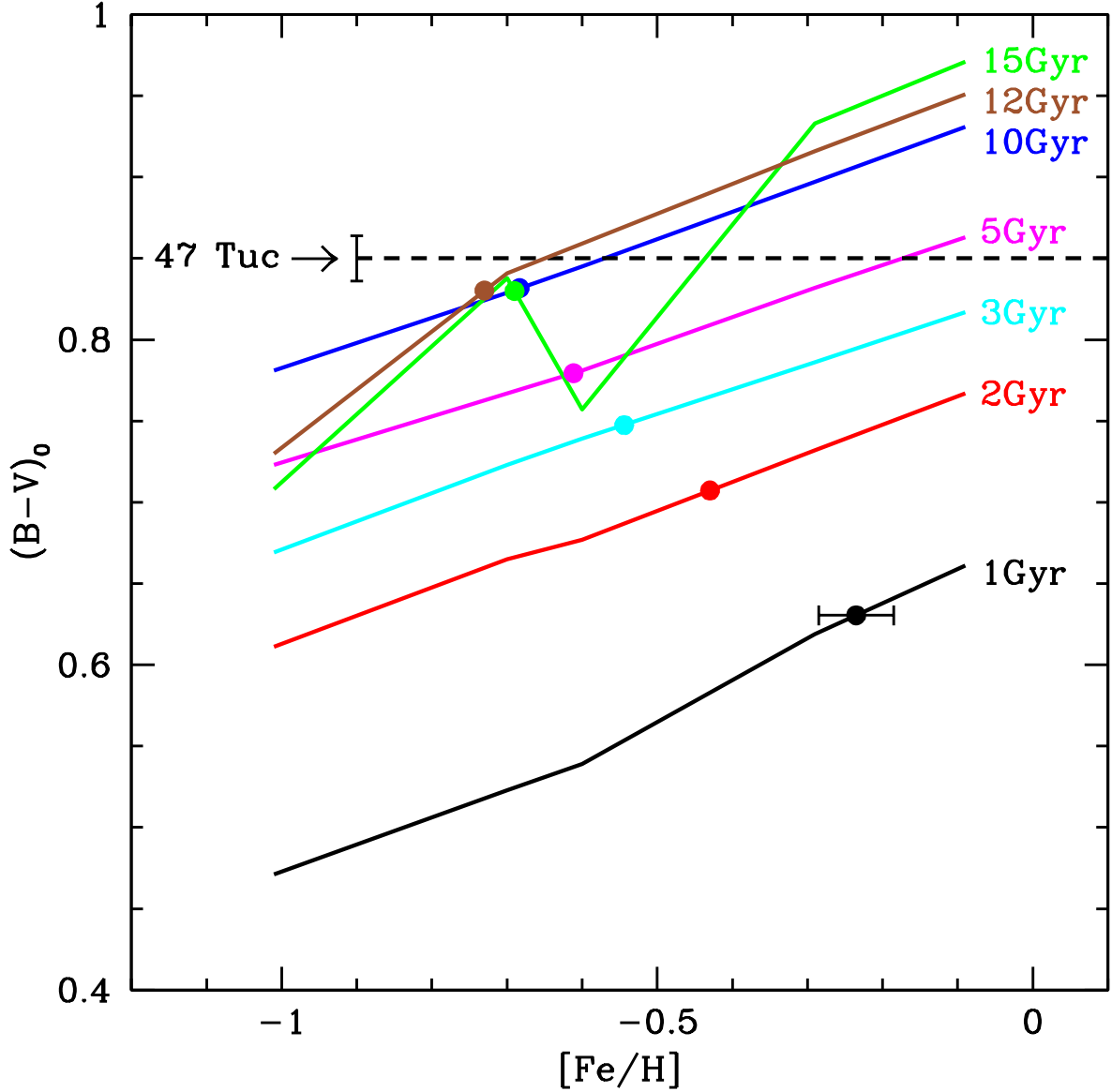}
\caption{Solid lines indicate our computed integrated-light (B$-$V)$_0$ GC colors,
based on Teramo isochrones (canonical alpha-enhanced models with $\eta$=0.4),
for a range of age and [Fe/H].  Filled circles indicate the [Fe/H] values
derived from 100 Fe~I lines in the IL spectrum, depending on assumed age.
The dashed horizontal line shows the observed (B$-$V)$_0$ color for the inner
47~Tuc region included in our IL spectrum, as measured by Peterson (1986).  
Error bars indicate $\pm$1$\sigma$ uncertainties in the observed color and
on the mean [Fe/H].  The intersection of the observed (B$-$V)$_0$ color with derived
[Fe/H] point indicates a self-consistent solution at [Fe/H]=$-$0.70$\pm$0.056 
and age $\geq$10 Gyr.  Note that the 15 Gyr isochrone color trend differs from younger
isochrones, due to the presence of a blue horizontal branch at 15 Gyr, instead
of a red clump.}
\label{fig-bvfeh}
\end{figure}

Given the limitation associated with spectral age diagnostics we now
include broad-band, B$-$V colors, combined with the abundances
derived from the spectra, to improve constraints on both age and metallicity.  
Essentially we use the different sensitivity to age of the integrated-light
colors and abundances derived from the Fe absorption lines to simultaneously
constrain age and [Fe/H].
In Figure~\ref{fig-bvfeh} we show theoretical integrated-light B$-$V colors, for
alpha-enhanced, $\eta$=0.4, Teramo isochrones with ages ranging from 1 to 15 Gyr.
To calculate the cluster B$-$V color we integrated the predicted B and V fluxes for
each point on the Teramo isochrone, appropriately weighted by the Kroupa IMF.
We note that our IL B$-$V colors depend upon the reliability of the Teramo
isochrones, and as we have already seen the AGB bump in the Teramo predictions
are under-estimated compared to the 47~Tuc {\em cmd}.  Fortunately, when we
enhance the AGB bump of the Teramo isochrones the effect on our IL B$-$V color
is negligible, with a change of only $-$0.004 magnitudes; therefore, we should
lower the loci in Figure~\ref{fig-bvfeh} by 0.004 magnitudes.  We note that
the effect of $\alpha$-enhancement on the integrated B$-$V color seems to be rather
small for 47~Tuc parameters: for 12 Gyr and [Fe/H]=$-$0.70 our solar composition 
calculations give B$-$V 0.012 magnitudes bluer than the alpha-enhanced isochrone;
thus, an uncertainty of $\sim$0.1 dex in the adopted alpha-enhancement would change
the predicted B$-$V color by only 0.002 magnitudes.  Another isochrone
morphology effect occurs for the oldest tracks with the choice of mass-loss parameter,
$\eta$, such that the helium burning stars at high mass-loss and low masses reside
on the Horizontal Branch instead of the Red Clump.  Indeed, the unusual behavior of
the 15 Gyr track in Figure~\ref{fig-bvfeh} is due to the emergence of the
Horizontal Branch compared to younger isochrones; notably, the 15 Gyr isochrone
with a lower mass-loss rate ($\eta$=0.2) behaves more regularly.
Another isochrone systematic that can affect the our colors is the
relationship between color, temperature and metallicity adopted by the Teramo
group: the B-band is particularly challenging for spectrum synthesis, due to the 
large number of strong absorption lines in red giants, and the M star spectra 
are especially complex.

Figure~\ref{fig-bvfeh} indicates the computed [Fe/H] for each age track, where
the isochrone metallicity and [Fe/H] from the Fe~I lines are consistent, as
derived from Figure~\ref{fig-tracks1.corrected.basti.alpha}.  These points
form a locus that intersects the observed B$-$V color for the isochrone whose
age and metallicity are consistent with both the photometric and spectroscopic
data.  The horizontal dashed line in the figure shows the observed IL B$-$V
color of 47~Tuc from Peterson (1986), de-reddened for E(B$-$V)=0.03; the
$\pm$1$\sigma$ error bar of the observed (B$-$V)$_0$, at 0.014 mag., is the
quadrature sum of 0.01 magnitude uncertainties in both the observed color and 
the reddening.

The random error on the mean integrated-light [Fe/H] value, determined from the
Fe~I lines is 0.021 dex.  The systematic uncertainties, as discussed in 
section \S\ref{sec:Iron_Abundances}, give a total random + systematic uncertainty
on the [Fe~I/H] values in Figure~\ref{fig-bvfeh}, of 0.050 dex; this is indicated by 
the horizontal $\pm$1$\sigma$ error bars on the 1Gyr point.  When we also include
the 0.025 dex uncertainty, due to the choice of isochrone age (estimated from
Figure~\ref{fig-bvfeh}), the total 1$\sigma$ systematic + random uncertainty on
unresolved IL [Fe/H] from Fe~I lines becomes $\pm$0.056 dex.

Other potential sources of uncertainty include non-LTE corrections, and
errors in the input isochrones, including the adopted helium mass fraction, Y,
the assumptions about mixing length, mass-loss and innumerable other parameters
that went into the Teramo isochrone calculations.

Figure~\ref{fig-bvfeh} constrains the 47~Tuc isochrone age in the range 10 to 15 Gyr,
with [Fe/H]=$-$0.70 $\pm$0.056 dex. 
The B$-$V color and [Fe/H] uncertainties indicates that the 8 Gyr isochrone is ruled out
at the 2.7$\sigma$ level.  This integrated-light result is completely consistent with
the {\em cmd} best fit of 12 Gyr with the Teramo isochrones by Koch \& McWilliam (2008) and 
[Fe/H]=$-$0.76 $\pm$0.01~$\pm$0.04 from echelle spectra of individual K giants, as well
as our value of [Fe/H]=$-$0.75$\pm$0.026$\pm$0.045 dex, obtained for the
IL analysis which employed the known 47~Tuc {\em cmd}.  

The excellent agreement between IL [Fe/H] and age with other studies is tempered by the
recognition that the isochrone parameters (e.g. mass-loss parameter, helium mass fraction,
no convective overshooting) were chosen that best match Galactic globular clusters; thus, 
it is not surprising that the 47~Tuc (an important Galactic GC) {\em cmd} is well matched
by the Teramo isochrones employed here.  However, since intermediate age and young GCs
are not found in the Galaxy the constraints on that part of parameter space is limited.
Fortunately for this analysis of 47~Tuc, the helium core burning stars occupy a well defined
red clump. Analysis of GCs with prominent blue horizontal branches may be much
more difficult, although some horizontal branch tests with GCs have been performed (e.g. 
Cassisi \& Salaris 1997).

Finally, we note that the EW and EP plots in
Figures~\ref{fig-ewepiso15z008} and \ref{fig-ewepiso1z008} show that
there is a possibility to detect the difference between hot main
sequence stars and hot horizontal branches.  If a cluster contains
more hot horizontal giant branch stars than are present in the
isochrone being used for the analysis, the EP--abundance plot would
show a slope resulting from hotter than predicted stellar temperatures.
While this EP plot may not distinguish between the high temperatures of young dwarf
stars and horizontal branch stars, the EW plot would differ for these
two cases, because the Horizonatal Branch stars have higher microturbulent
velocities than dwarf stars.  This diagnostic possibility will be explored in 
a later paper.

\section{Summary and Conclusions}

We are developing a method for analyzing the detailed chemical
abundances of extragalactic, spatial unresolved globular clusters
using high resolution spectroscopy.  To do so, we have taken spectra
of a ``training set'' of resolved GCs in the Milky Way and LMC. In
this paper, we take the first step in developing the analysis method
by exploring the analysis of an integrated light spectrum of
the core of the Galactic globular cluster 47~Tuc.  The spectra
discussed here are high signal to noise ratio ($\sim$100 per 
pixel) echelle spectra that we obtained by scanning over the central
32$\times$32 arcsecond region of the cluster.  Further work on the training 
set will be presented in later papers.

We have explored two methods for detailed abundance analysis of 
IL GC spectra: first, with observed {\em cmds}
and subsequently using theoretical isochrones to calculate
integrated-light EWs from model atmosphere spectrum synthesis.  We
have modified the stellar spectrum synthesis program, MOOG, in order
to compute abundances from IL EWs of single lines, and for the general
case of blended features, including features arising from hyperfine
structure.  The main purpose of calculating abundances using the
observed {\em cmd} was to verify the basic strategy associated with
modeling an light--weighted equivalent width, but it would also be of
value for measuring abundances from ground-based spectra of clusters
in nearby galaxies (e.g. LMC, M31, and M33) that are resolved by
space-based telescopes.

We have demonstrated that GCs similar to 47~Tuc (M$_v$$\sim$$-$9) can
be used to derive abundances for many elemental species, with
accuracies similar to extant studies of individual stars.  Given the
current faint limit of echelle spectrographs on the world's largest
telescopes, this suggests that we may derive detailed chemical
abundances for clusters to a distance of $\sim$5 Mpc.

Our {\em cmd}-based IL spectral abundance analysis of the 47~Tuc core
provides [Fe~I/H] and [Fe~II/H] abundances of $-$0.75$\pm$0.026$\pm$0.045
and $-$0.72$\pm$0.056$\pm$0.064 dex respectively (random and systematic errors).
These [Fe/H] values are well within the
[Fe/H] range, of $-$0.60 to $-$0.81 dex, reported by recent high
resolution studies of individual stars in 47~Tuc.

Diagnostic plots of iron abundance versus EW, line excitation
potential, and wavelength reveal no systematic trends for the Fe~I
lines in our {\em cmd} analysis.  In the study of single stars, these
diagnostic plots are used to constrain stellar atmosphere parameters.
For our integrated-light analysis the EW plot indicates that our
adopted run of microturbulent velocity with gravity is approximately
correct.  The wavelength and excitation potential plots suggests that
the mix of stellar temperatures of the adopted isochrone model is
correct.

Comparison of our {\em cmd}-based 47~Tuc IL abundance results for
various elements are consistent with previous studies within the range
of reported [X/Fe] values.  It is notable that elements whose
abundances, relative to Fe, are measured higher or lower than the
solar ratio in previous studies (e.g. Mg, Si, Ca, Ti, Mn) are
confirmed in this work, while elements whose abundances scale with
[Fe/H] (e.g. s-process elements, V, Cr, Ni, Co) follow the scaling
here.  Our [Cu/Fe] ratio, at $-$0.13 dex, while not measured by any
study of individual 47~Tuc stars (perhaps due to the complications of
hfs analysis) is completely consistent with the small deficiency seen
in Galactic stars having the same [Fe/H] as 47~Tuc; indeed, our [X/Fe]
ratios for 47~Tuc are generally consistent with expectations based on
its [Fe/H].  The notable exceptions to the abundance consistencies are
our higher values for Na and Al, our low value for Eu, and a slightly
low value for Mg (but Mg is still enhanced).  Given the very small
central depth of the single Eu~II line available we believe that the
low Eu result is simply an artifact of the spectrum noise.  However,
we believe that the pattern of Na, Mg and Al abundances indicates the
presence of proton burning products, presumably in the envelopes of
the luminous AGB star population within the cluster.  If this is
correct then it may be a general feature to be found in other old
populations, such as extra-galactic bulges and elliptical galaxies.  A
difference between abundance results here and in the literature that
we are not able to understand are the enhanced abundances of light
s-process elements found by Wylie et al. (2006).  It is possible that
Wylie et al. (2006) detected genuine s-process abundance enhancements
in the stars, formed during AGB evolution, that is not present in the
majority of the cluster.

Table~\ref{table4} shows that the median rms scatter of our
integrated-light abundances is 0.18 dex, in good agreement with the
0.17 dex scatter with published abundances.  For our 102 Fe~I lines
the rms scatter is slightly higher, at 0.26 dex, probably due to the
inclusion of blended lines, or lines with somewhat uncertain $gf$
values.  The formal error on our mean Fe~I abundance is 0.03 dex,
while most other species have errors on the mean in the range 0.07 to
0.10 dex.

A break-down of the fractional EW contributed by each {\em cmd} region
indicates that the integrated-light Fe~I lines are formed mostly in
the upper part of the RGB, increasing towards the most luminous stars,
with a non-negligible contribution from the AGB and Red Clump regions.
The low excitation potential Fe~I line formation is skewed even more
to the most luminous red giants.  On the other hand the Fe~II lines
have the greatest contribution from the AGB and Red Clump regions with
significant formation in the RGB, but not especially skewed to the
most luminous RGB stars.

We have explicitly investigated the potential complications in IL
analysis that might arise due to M giants by performing spectrum
synthesis calculations, including 5.4 million TiO lines, for two M
giants contained in the scanned core region of 47~Tuc.  From the
synthetic spectra we measured pseudo-EWs and included these as
corrections to the flux-weighted EWs for the cluster analysis.  This
led to negligible abundance corrections for 47~Tuc.  The mild impact
of M stars on the 47~Tuc IL analysis was partly due to the low M giant
fraction, resulting from the low metallicity of the cluster, and also
because the heavy TiO blanketing greatly reduced the M giant flux
contribution for the lines.  However, we note that a TiO window, in
the region 7300--7600\AA , will allow flux from M stars to have a
greater influence on the integrated light than at other wavelengths.
Thus, we suggest that Fe~I line abundances from the TiO window,
compared to other wavelengths, would provide a diagnostic of the
M giant fraction; this will be especially important for clusters
more metal-rich than 47~Tuc.

Although integrated light abundance analysis of resolved clusters
present interesting possibilities for clusters in some Local Group
galaxies, most applications of IL abundance analysis will be for
unresolved clusters in more distant galaxies.  Therefore, we have
extended our integrated light analysis technique to the case of
unresolved clusters, using theoretical isochrones.  We include scaled
solar abundance theoretical isochrones of the Padova group (Girardi et
al. 2000; Salasnich et al.2000), but we focus on the alpha-enriched
isochrones of the Teramo group with mass loss parameter $\eta$=0.4 and
extended AGB treatment (e.g. Cassisi, Salaris \& Irwin 2003;
Pietrinferni et al. 2006) that employ the recently corrected
alpha-element opacities of Ferguson et al. (2005).  For both sets of
theoretical isochrones we used the Kroupa (2002) IMF to set the number
of stars at each mass.

Abundances were computed using the measured integrated-light EWs and a
grid of theoretical isochrones ranging from 1 to 15 Gyr.  However,
because the Teramo isochrones include choices of mass-loss parameter,
alpha enhancement, and AGB treatment, where the Padova models do not
have as much flexibility, and following the recommendations of
Maraston (2005), we favor the Teramo isochrones.

Plots of isochrone [Fe/H] versus [Fe/H] derived from the Fe~I and
Fe~II lines in the integrated-light spectrum show different behaviors
for Fe~I and Fe~II lines.  For a given age, the Fe~I line abundances
are insensitive to the input isochrone metallicity, with a range of
only $\sim$0.2 dex over 2 dex in isochrone metallicity.  Above
[Fe/H]$\sim$$-$1.5, the derived Fe~II abundances scale with the adopted
input isochrone metallicity (for a given age), with unit slope; below
this metallicity limit the Fe~II abundances are roughly independent of
isochrone metallicity, similar to the Fe~I lines.  
The Fe~I line abundances do show a strong dependence on the adopted
isochrone age. This is apparently modulated by the mean temperature of
the cluster stars, which increases with decreasing age.  
For old ages, from 10 to 15Gyr, the
derived abundances vary by only $\sim$0.1 dex, presumably because the
red giant branches are very similar.

Our diagnostic plots of Fe abundance versus line EP, and line
EW, show distinct trends when an isochrone age is selected that is
significantly younger than the observed cluster: there is a slope in
the EW versus abundance plot due to the low mean microturbulent
velocity of the extra dwarf stars in the young isochrone, and there is
a slope in the excitation versus abundance plot due to the high
average temperature of the extra hot dwarfs in the young isochrone.
By use of these diagnostics we may certainly eliminate the ages of 1
to 2 Gyr for the 47~Tuc spectrum. Careful differential abundance
analyses may push this age discrimination up to 5 Gyr, but the
abundance trends are too subtle for anything older.  Our mild age
discriminant may be of limited use; however, for old GCs the exact
age has very little effect on the derived abundances.

These diagnostics suggest to us a possible discriminant between blue
horizontal giant branch (BHB) stars in an old cluster and hot dwarf
stars in a young cluster.  The BHB stars will affect the abundance
versus EW plot differently than dwarf stars in a young cluster,
because the BHB stars have higher microturbulent velocities than dwarf
stars.  Thus, while the excitation versus abundance plot may show that
there is a population of hot stars in the cluster, the EW versus
abundance plot will indicate whether these hot stars are low gravity
BHB stars, or high gravity dwarfs.

Comparing the luminosity function of the theoretical Teramo
isochrones with the best fit parameters of Gratton et al.(2003), we
find that the theoretical isochrone over-predicted stars below the
main sequence turnoff and under-predicted the luminosity of the AGB
bump; there is also a slight under-prediction for the tip of the
giant branch, but this could reasonably be due to Poisson noise
effects.  The lack of stars below the main sequence turnoff is a well
known result of mass-segregation in cluster cores.  We are unaware of
the source of the under-predicted AGB bump luminosity, but we note
that Schiavon et al. (2002) found a similar under-prediction 47~Tuc,
but his correction covered all luminous giants; it appears that both
Teramo and Padova theoretical isochrones fall short of an accurate
description of the AGB bump.  In order to reproduce HST observations
of the 47~Tuc core we corrected the theoretical luminosity function by
increasing the contribution of the AGB bump region by a factor of 3.0,
and applied a cut to eliminate main sequence stars less luminous than
M$_{\rm v}$=4.90.

We find that the corrected theoretical isochrone for 47~Tuc core
reduces the derived IL abundances of Fe~I and Fe~II lines by 0.125 and
0.165 dex, respectively.  We have applied these as zero-point
corrections to IL abundances derived for Teramo isochrones; but,
strictly speaking, they are only valid for only one age and metallicity.

When the corrected theoretical isochrones are used the [Fe/H] based on 
Fe~I lines is consistent with the input isochrone [Fe/H] at $-$0.70 dex and
an age of 10 to 15 Gyr; however, in this analysis the trend of line abundance
with excitation potential only rules-out ages younger than $\sim$2 Gyr.

In order to improve our constraints on our age estimate from integrated-light,
we computed theoretical integrated-light B$-$V colors, as a function of metallicity
and age, based on the Teramo isochrones and the Kroupa IMF.  We found that the
observed B$-$V color combined with the self consistent [Fe/H], from the spectral 
line abundance analysis, indeed results in superior constraints on both age and 
metallicity than from the spectra alone.  The B$-$V color plus spectral lines give an
age in the range 10 to 15 Gyr, with 8 Gyr ruled out at at the 2.7$\sigma$ level,
and [Fe/H]=$-$0.70$\pm$0.021$\pm$0.045 dex, based on the Fe~I lines.  This
integrated-light result is completely consistent with the {\em cmd} best fit of 
12 Gyr with the Teramo isochrones by Koch \& McWilliam (2008) and
[Fe/H]=$-$0.76 $\pm$0.01~$\pm$0.04 from echelle spectra of individual K giants.
Indeed, based on Fe~I lines, the mean of 5 high resolution abundance
studies of individual stars in 47~Tuc gives a mean of [Fe/H]=$-$0.70
$\pm$0.04 dex, with a scatter about the mean of 0.086 dex, in perfect
agreement with our value derived using theoretical isochrones alone.

One slight inconsistency is that [Fe/H] derived from the Fe~II lines
is higher than the intersection of Fe~I and isochrone values by
$\sim$0.08 dex.  While this difference may simply be due to
measurement error, other explanations include the possibility that the
input alpha enhancement for the isochrones and model atmospheres is
higher than for the cluster, or that Poisson noise at the tip of the
giant branch increased the number of tip giants over the predicted
value.  In any event, the difference is small and similar to the
dispersion between studies.

We have shown that detailed chemical abundance analysis of the
integrated-light of Galactic globular cluster 47~Tuc provides results
similar to the analysis of its individual stars.  This relies on the
low velocity dispersion of this rather luminous cluster.  Our work
suggests that we may employ techniques outlined in this paper to study
the composition of globular clusters in distant galaxies, out to about
5~Mpc, using existing spectrograph-telescope combinations.  
Thus, it now appears possible to use GCs to study the detailed chemical 
composition of galaxies far beyond the Local Group.

\acknowledgements

We are grateful to Steve Shectman for not only building the du~Pont
echelle spectrograph and writing the du~Pont telescope control
software, with which the data for this paper was obtained, but also
for volunteering to modify the du~Pont guider control software to
permit scanning of the echelle slit over a large enough area of sky to
obtain the echelle format integrated light spectra of 47 Tuc and other
Galactic globular clusters.  We also thank Raja Guhathakurta and
Ricardo Schiavon for making available to us their HST photometry of
the 47 Tuc core.  RAB acknowledges support from the NSF through grant
number AST-0507350.

\clearpage

\input{tab1.tex}                         

\input{tab2.tex}

\input{tab3.tex}


\input{tab4.tex}

\clearpage

\clearpage

\clearpage

\clearpage

\clearpage

\clearpage

\clearpage

\clearpage

\clearpage

\clearpage

\clearpage

\clearpage

\clearpage

\clearpage

\clearpage

\clearpage

\clearpage

\clearpage

\clearpage

\clearpage

\end{document}

%% file: tab1.tex

\begin{deluxetable}{lcccccc} 
\tablecolumns{7}
\tablewidth{0pc}
\tablecaption{Log of observations and estimated S/N. \label{table1}}
\tablehead{
 \colhead{Object} & \colhead{Exp. (sec)} &
 \colhead{UT Date} & \quad &
 \multicolumn{3}{c}{Approximate S/N (per pix)}  \\
     & & & & \colhead{4500\AA} & \colhead{5500\AA} & \colhead{6450\AA} }
\startdata
47 Tuc  &    11,030 &  18\&19 July 2000 & &  60  &  100  &  140   \\
\enddata
\end{deluxetable}


%% file: tab2.tex

\begin{deluxetable}{lcccr} 
\tablecolumns{5} 
\tablewidth{0pc} 
\tablecaption{Line Parameters and Integrated-Light Equivalent Widths for 47 Tuc 
\label{table2} }
\tablehead{
 \colhead{Species} & \colhead{Wavelength} & \colhead{E.P.}  &
 \colhead{ log~gf}  & \colhead{EW} \\
      &  (m\AA ) &  (eV)  &        &  (m\AA )  \\
  }
\startdata
Na I &  6154.230 &  2.100 & $-$1.570 &  43.5 \\
Na I &  6154.230 &  2.100 & $-$1.570 &  63.2 \\
Na I &  6160.753 &  2.100 & $-$1.270 &  72.0 \\
\\
Mg I &  7387.700 &  5.750 & $-$0.870 &  62.1 \\
\\
Al I &  6696.032 &  3.140 & $-$1.481 &  62.7 \\
Al I &  6698.669 &  3.140 & $-$1.782 &  43.7 \\
\\
Si I &  5690.433 &  4.930 & $-$1.750 &  40.0 \\
Si I &  6237.328 &  5.610 & $-$1.010 &  44.9 \\
Si I &  6721.844 &  5.860 & $-$1.169 &  41.0 \\
Si I &  6976.504 &  5.950 & $-$1.044 &  34.5 \\
Si I &  7226.208 &  5.610 & $-$1.440 &  27.2 \\
Si I &  7405.790 &  5.610 & $-$0.660 &  95.7 \\
\\
Ca I &  5260.390 &  2.520 & $-$1.719 &  40.5 \\
Ca I &  5581.979 &  2.520 & $-$0.555 & 106.7 \\
Ca I &  5588.764 &  2.520 & $+$0.358 & 156.7 \\
Ca I &  5590.126 &  2.520 & $-$0.571 & 106.9 \\
Ca I &  5867.572 &  2.930 & $-$1.640 &  43.1 \\
Ca I &  6161.295 &  2.520 & $-$1.266 &  92.6 \\
Ca I &  6166.440 &  2.520 & $-$1.142 &  92.8 \\
Ca I &  6455.605 &  2.520 & $-$1.290 &  79.6 \\
Ca I &  6471.668 &  2.520 & $-$0.686 & 101.2 \\
Ca I &  6572.795 &  0.000 & $-$4.310 & 120.8 \\
Ca I &  7326.160 &  2.930 & $-$0.230 & 144.3 \\
\\
Sc I &  6210.671 &  0.000 & $-$1.570 &  39.7 \\
\\
Sc II & 6604.600 &  1.360 & $-$1.480 &  56.0 \\
\\
Ti I  & 5282.402 &  1.053 & $-$1.300 &  57.6 \\
Ti I  & 5282.402 &  1.053 & $-$1.300 &  62.9 \\
Ti I  & 5295.780 &  1.067 & $-$1.633 &  49.0 \\
Ti I  & 5295.780 &  1.067 & $-$1.633 &  68.8 \\
Ti I  & 5453.650 &  1.443 & $-$1.610 &  51.2 \\
Ti I  & 5648.570 &  2.495 & $-$0.260 &  32.5 \\
Ti I  & 6303.767 &  1.440 & $-$1.566 &  59.9 \\
Ti I  & 6554.238 &  1.440 & $-$1.218 &  77.5 \\
Ti I  & 6556.077 &  1.460 & $-$1.074 &  93.5 \\
Ti I  & 6743.127 &  0.900 & $-$1.630 &  76.5 \\
Ti I  & 7209.468 &  1.460 & $-$0.500 & 100.4 \\
Ti I  & 7216.190 &  1.440 & $-$1.150 &  85.9 \\
Ti I  & 7251.717 &  1.430 & $-$0.770 & 100.2 \\
\\
Ti II & 5336.780 &  1.582 & $-$1.700 & 103.4 \\
Ti II & 5381.010 &  1.570 & $-$2.080 & 100.6 \\
Ti II & 5381.010 &  1.570 & $-$2.080 &  88.8 \\
\\
V I   & 6274.658 &  0.270 & $-$1.670 &  50.3 \\
V I   & 6285.165 &  0.280 & $-$1.510 &  70.7 \\
V I   & 6531.429 &  1.220 & $-$0.840 &  39.3 \\
\\
Cr I  & 6330.096 &  0.940 & $-$2.920 &  67.0 \\
Cr I  & 6881.716 &  3.440 & $-$0.450 &  28.3 \\
Cr I  & 7400.188 &  2.900 & $-$0.111 & 100.5 \\
\\
Mn I  & 5399.479 &  3.853 & $-$0.287 &  12.9 \\
Mn I  & 5537.750 &  2.187 & $-$2.017 &  52.6 \\
Mn I  & 6013.497 &  3.073 & $-$0.251 &  73.4 \\
Mn I  & 6016.647 &  3.073 & $-$0.216 &  83.6 \\
\\
Fe I  & 5522.446 &  4.209 & $-$1.450 &  31.3 \\
Fe I  & 5543.936 &  4.218 & $-$1.040 &  52.7 \\
Fe I  & 5560.212 &  4.435 & $-$1.090 &  37.8 \\
Fe I  & 5586.771 &  3.370 & $-$0.120 & 152.1 \\  
Fe I  & 5618.632 &  4.209 & $-$1.275 &  53.4 \\
Fe I  & 5633.946 &  4.991 & $-$0.230 &  46.2 \\
Fe I  & 5633.946 &  4.991 & $-$0.230 &  47.7 \\
Fe I  & 5652.318 &  4.260 & $-$1.850 &  25.0 \\
Fe I  & 5741.848 &  4.256 & $-$1.672 &  29.1 \\
Fe I  & 5752.032 &  4.549 & $-$0.864 &  51.1 \\
Fe I  & 5775.081 &  4.220 & $-$1.297 &  45.0 \\
Fe I  & 5778.453 &  2.588 & $-$3.440 &  40.3 \\
Fe I  & 5811.915 &  4.143 & $-$2.330 &  17.2 \\
Fe I  & 5862.357 &  4.549 & $-$0.051 &  91.4 \\
Fe I  & 5905.671 &  4.652 & $-$0.690 &  52.2 \\
Fe I  & 5934.655 &  3.929 & $-$1.070 & 107.2 \\
Fe I  & 5934.655 &  3.929 & $-$1.070 &  86.9 \\
Fe I  & 5943.578 &  2.198 & $-$4.456 &  28.1 \\
Fe I  & 5976.777 &  3.943 & $-$1.503 &  57.4 \\
Fe I  & 6003.011 &  3.882 & $-$1.120 &  75.0 \\
Fe I  & 6012.210 &  2.223 & $-$4.038 &  37.1 \\
Fe I  & 6024.058 &  4.549 & $-$0.120 &  83.9 \\
Fe I  & 6027.051 &  4.076 & $-$1.089 &  61.7 \\
Fe I  & 6078.491 &  4.796 & $-$0.481 &  61.0 \\
Fe I  & 6079.008 &  4.652 & $-$1.020 &  36.0 \\
Fe I  & 6096.664 &  3.984 & $-$1.830 &  39.3 \\
Fe I  & 6151.617 &  2.176 & $-$3.300 &  60.5 \\
Fe I  & 6151.617 &  2.176 & $-$3.300 &  66.4 \\
Fe I  & 6173.341 &  2.220 & $-$2.863 &  91.2 \\
Fe I  & 6180.209 &  2.730 & $-$2.615 &  71.5 \\
Fe I  & 6187.995 &  3.940 & $-$1.673 &  49.0 \\
Fe I  & 6200.313 &  2.609 & $-$2.386 &  76.9 \\
Fe I  & 6213.437 &  2.220 & $-$2.490 & 120.4 \\
Fe I  & 6226.734 &  3.884 & $-$2.120 &  28.2 \\
Fe I  & 6229.232 &  2.830 & $-$2.821 &  58.1 \\
Fe I  & 6246.327 &  3.600 & $-$0.796 & 102.3 \\
Fe I  & 6254.253 &  2.280 & $-$2.435 & 125.6 \\
Fe I  & 6254.253 &  2.280 & $-$2.435 & 132.3 \\
Fe I  & 6265.141 &  2.180 & $-$2.532 & 102.4 \\
Fe I  & 6265.141 &  2.180 & $-$2.532 & 121.6 \\
Fe I  & 6270.231 &  2.860 & $-$2.543 &  61.6 \\
Fe I  & 6297.799 &  2.220 & $-$2.681 & 105.5 \\
Fe I  & 6301.508 &  3.650 & $-$0.701 & 113.9 \\
Fe I  & 6311.504 &  2.830 & $-$3.150 &  62.5 \\
Fe I  & 6322.694 &  2.590 & $-$2.438 &  90.4 \\
Fe I  & 6330.848 &  4.733 & $-$1.640 &  20.7 \\
Fe I  & 6335.337 &  2.200 & $-$2.175 & 136.4 \\
Fe I  & 6336.830 &  3.690 & $-$0.840 & 102.4 \\
Fe I  & 6353.849 &  0.910 & $-$6.350 &  34.5 \\
Fe I  & 6380.750 &  4.190 & $-$1.366 &  58.5 \\
Fe I  & 6380.750 &  4.190 & $-$1.366 &  59.0 \\
Fe I  & 6393.612 &  2.430 & $-$1.505 & 160.8 \\ 
Fe I  & 6411.658 &  3.650 & $-$0.646 & 113.2 \\
Fe I  & 6421.360 &  2.280 & $-$1.979 & 152.6 \\ 
Fe I  & 6430.856 &  2.180 & $-$1.954 & 151.8 \\ 
Fe I  & 6475.632 &  2.560 & $-$2.929 &  93.2 \\
Fe I  & 6481.878 &  2.280 & $-$2.980 & 101.2 \\
Fe I  & 6518.366 &  2.832 & $-$2.397 &  89.7 \\
Fe I  & 6533.928 &  4.559 & $-$1.360 &  28.1 \\
Fe I  & 6546.252 &  2.750 & $-$1.536 & 128.3 \\
Fe I  & 6593.871 &  2.430 & $-$2.377 & 104.9 \\
Fe I  & 6597.571 &  4.770 & $-$0.970 &  29.3 \\
Fe I  & 6608.044 &  2.270 & $-$3.939 &  49.6 \\
Fe I  & 6627.560 &  4.530 & $-$1.559 &  23.9 \\
Fe I  & 6646.966 &  2.600 & $-$3.917 &  25.9 \\
Fe I  & 6648.121 &  1.010 & $-$5.720 &  35.2 \\
Fe I  & 6677.997 &  2.690 & $-$1.395 & 146.4 \\
Fe I  & 6703.576 &  2.760 & $-$3.059 &  44.9 \\
Fe I  & 6705.105 &  4.610 & $-$1.050 &  27.3 \\
Fe I  & 6710.323 &  1.480 & $-$4.807 &  43.3 \\
Fe I  & 6715.386 &  4.590 & $-$1.540 &  37.9 \\
Fe I  & 6725.364 &  4.100 & $-$2.227 &  25.8 \\
Fe I  & 6726.666 &  4.607 & $-$1.087 &  44.1 \\
Fe I  & 6733.153 &  4.620 & $-$1.479 &  22.0 \\
Fe I  & 6739.524 &  1.560 & $-$4.801 &  49.6 \\
Fe I  & 6750.164 &  2.420 & $-$2.592 &  95.6 \\
Fe I  & 6810.262 &  4.607 & $-$0.992 &  48.2 \\
Fe I  & 6820.371 &  4.638 & $-$1.214 &  41.9 \\
Fe I  & 6828.592 &  4.638 & $-$0.843 &  54.8 \\
Fe I  & 6839.835 &  2.560 & $-$3.378 &  44.0 \\
Fe I  & 6841.341 &  4.610 & $-$0.733 &  73.5 \\
Fe I  & 6851.652 &  1.600 & $-$5.247 &  29.8 \\
Fe I  & 6855.161 &  4.559 & $-$0.741 &  71.5 \\
Fe I  & 6857.249 &  4.076 & $-$2.050 &  28.2 \\
Fe I  & 6858.155 &  4.590 & $-$0.939 &  46.0 \\
Fe I  & 6859.480 &  2.845 & $-$4.447 &  12.1 \\
Fe I  & 6911.512 &  2.424 & $-$3.967 &  31.5 \\
Fe I  & 6916.680 &  4.154 & $-$1.359 &  48.4 \\
Fe I  & 6988.524 &  2.404 & $-$3.519 &  61.9 \\
Fe I  & 7022.957 &  4.190 & $-$1.148 &  65.4 \\
Fe I  & 7068.423 &  4.070 & $-$1.319 &  92.8 \\
Fe I  & 7090.390 &  4.230 & $-$1.109 &  79.7 \\
Fe I  & 7114.549 &  2.692 & $-$3.937 &  16.0 \\
Fe I  & 7132.987 &  4.076 & $-$1.635 &  59.0 \\
Fe I  & 7145.312 &  4.610 & $-$1.240 &  59.3 \\
Fe I  & 7151.464 &  2.480 & $-$3.657 &  60.8 \\
Fe I  & 7180.004 &  1.480 & $-$4.707 &  67.6 \\
Fe I  & 7219.682 &  4.076 & $-$1.617 &  50.8 \\
Fe I  & 7411.162 &  4.280 & $-$0.287 & 116.6 \\
Fe I  & 7445.758 &  4.260 & $+$0.053 & 119.0 \\
Fe I  & 7491.646 &  4.301 & $-$1.067 &  87.8 \\
Fe I  & 7531.153 &  4.370 & $-$0.557 &  87.5 \\
Fe I  & 7568.900 &  4.280 & $-$0.600 &  99.2 \\
\\
Fe II & 5197.576 &  3.230 & $-$2.240 &  81.0 \\
Fe II & 6238.390 &  3.889 & $-$2.690 &  35.0 \\
Fe II & 6247.557 &  3.892 & $-$2.380 &  37.0 \\
Fe II & 6416.928 &  3.890 & $-$2.740 &  36.0 \\
Fe II & 6432.683 &  2.890 & $-$3.630 &  27.0 \\
Fe II & 6456.383 &  3.903 & $-$2.130 &  50.0 \\
Fe II & 6516.080 &  2.891 & $-$3.440 &  44.0 \\
\\
Co I  & 6188.998 &  1.710 & $-$2.450 &  68.0 \\
Co I  & 6814.961 &  1.960 & $-$1.900 &  53.2 \\
Co I  & 7052.870 &  1.960 & $-$1.620 & 112.4 \\
\\
Ni I  & 6327.604 &  1.680 & $-$3.150 &  60.7 \\
Ni I  & 6482.809 &  1.930 & $-$2.630 &  77.0 \\
Ni I  & 6586.319 &  1.950 & $-$2.810 &  62.6 \\
Ni I  & 6643.638 &  1.680 & $-$2.300 & 120.8 \\
Ni I  & 6767.784 &  1.830 & $-$2.170 &  96.9 \\
Ni I  & 6772.321 &  3.660 & $-$0.980 &  43.3 \\
Ni I  & 7110.905 &  1.930 & $-$2.980 &  67.0 \\
Ni I  & 7393.609 &  3.610 & $-$0.270 &  90.0 \\
Ni I  & 7422.286 &  3.630 & $-$0.140 &  91.1 \\
Ni I  & 7522.778 &  3.660 & $-$0.400 & 111.1 \\
Ni I  & 7525.118 &  3.630 & $-$0.520 &  78.4 \\
Ni I  & 7555.607 &  3.850 & $+$0.110 & 105.1 \\
\\
Cu I  & 5105.541 &  1.390 & $-$1.520 & 116.9 \\
\\
Y I   & 6435.049 &  0.070 & $-$0.820 &  30.5 \\
Y II  & 5402.780 &  1.840 & $-$0.520 &  20.6 \\
\\
Zr I  & 6143.183 &  0.070 & $-$1.100 &  30.4 \\
Zr I  & 6143.183 &  0.070 & $-$1.100 &  30.0\rlap{:} \\
\\
Ba II & 5853.688 &  0.604 & $-$1.010 &  93.2 \\
Ba II & 6141.727 &  0.700 & $-$0.077 & 137.9 \\
Ba II & 6141.727 &  0.700 & $-$0.077 & 151.9 \\
Ba II & 6496.908 &  0.600 & $-$0.377 & 161.8 \\ 
Ba II & 6496.908 &  0.600 & $-$0.377 & 168.4 \\ 
\\
La II & 6390.480 &  0.320 & $-$1.520 &  30.7 \\
La II & 6774.260 &  0.130 & $-$1.810 &  13.6 \\
\\
Nd II & 5319.820 &  0.550 & $-$0.194 &  44.8 \\
\\
Eu II & 6645.127 &  1.380 & $+$0.204 &  15.8 \\
\\
\enddata 
\end{deluxetable} 


%% file: tab3.tex

\begin{deluxetable}{rccccrcr} 
\tablecolumns{8} 
\tablewidth{0pc} 
\tablecaption{47 Tuc Core BV CMD Parameters\label{table3}}
\tablehead{
 \colhead{$\overline{V_0}$}  & \colhead{$\overline{(B-V)_0}$} & 
 \colhead{T$_{\rm eff}$}  &
 \colhead{ log~g}  & \colhead{$\xi$} & \colhead{R} &  \colhead{f(V)$_a$} & 
 \colhead{N$_{\rm star}$}
\\  }
\startdata
  11.867 &   ...  & 3350 & 0.300 &  1.88 & 237.30 &  0.0149 &      1\rlap{$_b$} \\
  12.127 &   ...  & 3600 & 0.320 &  1.88 & 122.10 &  0.0117 &      1\rlap{$_b$} \\
\\
  11.723 &  1.537 & 3899 & 0.558 &  1.83 &  77.85 &  0.0507 &      3 \\
  11.796 &  1.488 & 3957 & 0.648 &  1.81 &  70.24 &  0.0325 &      2 \\
  11.963 &  1.452 & 4001 & 0.760 &  1.78 &  61.74 &  0.0405 &      3 \\
  12.135 &  1.317 & 4175 & 0.979 &  1.74 &  47.98 &  0.0229 &      2 \\
  12.416 &  1.296 & 4204 & 1.116 &  1.71 &  40.97 &  0.0356 &      4 \\
  12.779 &  1.199 & 4342 & 1.365 &  1.65 &  30.75 &  0.0445 &      7 \\
  13.100 &  1.130 & 4446 & 1.566 &  1.61 &  24.39 &  0.0379 &      8 \\
  13.498 &  1.053 & 4568 & 1.803 &  1.56 &  18.57 &  0.0489 &     15 \\
  14.088 &  0.966 & 4714 & 2.126 &  1.49 &  12.81 &  0.0266 &     14 \\
  14.500 &  0.921 & 4792 & 2.334 &  1.45 &  10.08 &  0.0429 &     33 \\
  15.051 &  0.855 & 4915 & 2.616 &  1.39 &   7.28 &  0.0365 &     47 \\
  15.949 &  0.804 & 5014 & 3.025 &  1.30 &   4.55 &  0.0312 &     94 \\
  16.730 &  0.757 & 5127 & 3.389 &  1.22 &   2.99 &  0.0097 &     58 \\
  17.051 &  0.708 & 5272 & 3.582 &  1.18 &   2.40 &  0.0163 &    131 \\
  17.174 &  0.578 & 5695 & 3.796 &  1.14 &   1.87 &  0.0477 &    429 \\
  17.417 &  0.541 & 5833 & 3.940 &  1.11 &   1.59 &  0.0399 &    448 \\
  17.632 &  0.539 & 5842 & 4.029 &  1.09 &   1.43 &  0.0390 &    534 \\
  17.891 &  0.549 & 5803 & 4.120 &  1.07 &   1.29 &  0.0389 &    678 \\
  18.244 &  0.580 & 5691 & 4.221 &  1.05 &   1.15 &  0.0388 &    939 \\
  18.915 &  0.670 & 5389 & 4.374 &  1.01 &   0.96 &  0.0128 &    614 \\
\\
  12.967 &  1.074 & 4533 & 1.570 &  1.61 &  24.28 &  0.0374 &      7 \\
  13.040 &  0.974 & 4700 & 1.699 &  1.58 &  20.93 &  0.0450 &      9 \\
  13.868 &  0.818 & 4986 & 2.182 &  1.48 &  12.01 &  0.0808 &     35 \\
  13.915 &  0.775 & 5081 & 2.245 &  1.47 &  11.17 &  0.0600 &     27 \\
  13.973 &  0.720 & 5237 & 2.339 &  1.45 &  10.02 &  0.0466 &     22 \\
  16.209 &  0.276 & 7306 & 3.886 &  1.12 &   1.69 &  0.0049 &     19 \\
  15.267 &  0.185 & 7693 & 3.609 &  1.18 &   2.32 &  0.0050 &      8 \\
\enddata 
\tablecomments{\\
a: \quad fractional V-band flux in each box\\
b: \quad M star }
\end{deluxetable} 


%% file: tab4.tex

\begin{deluxetable}{lccrcc|ccccc} 
\tablecolumns{6} 
\tablewidth{0pc} 
\tablecaption{Average Abundances for 47 Tuc Integrated Light (with HST cmd) \label{table4}}

\tablehead{
 \colhead{Species} & \colhead{log$_{\rm 10}$$\epsilon$(M)} & 
 \colhead{{\large $\sigma$}(M)} &
 \colhead{N$_{\rm lines}$}  &
 \colhead{[M/Fe]$^1$}  & \colhead{Notes} &
 \colhead{BW92}  & \colhead{C04}   & \colhead{A05} & \colhead{W06} & \colhead{KM08} \\
}
\startdata

Na~I   &  5.97 & 0.17 &   3 &  $+$0.45 &        & $+$0.11         & $+$0.23 & $+$0.03 & $+$0.65  &  $+$0.22 \\
Mg~I   &  7.05 & ...  &   1 &  $+$0.22 &        & $+$0.49         & $+$0.40 & $+$0.24 &   ...    &  $+$0.46 \\
Al~I   &  6.20 & 0.02 &   2 &  $+$0.53 &        & $+$0.67         & $+$0.23 & $+$0.13 &   ...    &  $+$0.47 \\
Si~I   &  7.18 & 0.23 &   6 &  $+$0.37 &        & $+$0.39         & $+$0.30 & $+$0.24 &   ...    &  $+$0.40 \\
Ca~I   &  5.92 & 0.24 &  10 &  $+$0.31 &        & $+$0.06         & $+$0.20 & $+$0.00 &   ...    &  $+$0.35 \\
Sc~I   &  2.37 & ...  &   1 &  $+$0.02 &        &   ...           &   ...   &   ...   &   ...    &    ...   \\
Sc~II  &  2.60 & ...  &   1 &  $+$0.25 &        & $-$0.01         & $+$0.13 &   ...   &   ...    &    ...   \\
Ti~I   &  4.61 & 0.24 &  13 &  $+$0.41 &        & $+$0.16         & $+$0.26 & $+$0.21 &   ...    &  $+$0.36 \\
Ti~II  &  4.74 & 0.16 &   3 &  $+$0.54 &        &   ...           & $+$0.38 & $+$0.28 &   ...    &  $+$0.38 \\
V~I    &  3.22 & 0.16 &   3 &  $-$0.08 &     2  &   ...           & $+$0.05 &   ...   &   ...    &    ...   \\
Cr~I   &  4.92 & 0.16 &   3 &  $-$0.02 &        &   ...           & $+$0.11 &   ...   &   ...    &    ...   \\
Mn~I   &  4.25 & 0.18 &   4 &  $-$0.44 &     2  &   ...           & $-$0.29 &   ...   &   ...    &    ...   \\
Fe~I   &  6.77 & 0.26 & 102 &  $-$0.75\rlap{$^1$} &  3 & $-$0.81         & $-$0.67 & $-$0.66 & $-$0.60  &  $-$0.76 \\ 
Fe~II  &  6.73 & 0.15 &   7 &  $-$0.72\rlap{$^1$} &  3 &                 & $-$0.56 & $-$0.69 & $-$0.64  &  $-$0.84 \\
Co~I   &  4.56 & 0.36 &   3 &  $+$0.34 &     2  &   ...           &   ...   &   ...   &   ...    &    ...   \\
Co~I   &  4.19 & ...  &   1 &  $-$0.04 &   2,4  &   ...           &   ...   &   ...   &   ...    &    ...   \\
Ni~I   &  5.53 & 0.20 &  12 &  $+$0.00 &        &   ...           & $+$0.06 &   ...   &   ...    &    ...   \\
Cu~I   &  3.38 & ...  &   1 &  $-$0.13 &     2  &   ...           &   ...   &   ...   &   ...    &    ...   \\
Y~I    &  1.73 & ...  &   1 &  $+$0.22 &        &   ...           &   ...   &   ...   & $+$0.65  &    ...   \\
Y~II   &  1.32 & ...  &   1 &  $-$0.13 &        & $+$0.48         &   ...   &   ...   & $+$0.65  &    ...   \\
Zr~I   &  1.94 & ...  &   1 &  $+$0.05 &     2  & $-$0.22         &   ...   & $-$0.17 & $+$0.69  &    ...   \\
Ba~II  &  1.49 & 0.03 &   3 &  $+$0.02 &     2  & $-$0.22\rlap{:} &   ...   & $+$0.31 &   ...    &    ...   \\
La~II  &  0.58 & 0.27 &   2 &  $+$0.15 &     2  & $+$0.24         &   ...   & $+$0.05 & $+$0.31  &    ...   \\
Nd~II  &  0.79 & ...  &   1 &  $+$0.04 &     2  &   ...           &   ...   &   ...   & $+$0.42  &    ...   \\
Eu~II  & \llap{$-$}0.14 & ...  &   1 &  $+$0.04 &    2  & $+$0.36 &   ...   & $+$0.33 & $+$0.14  &    ...   \\
 \\
\enddata 
\tablecomments{\\ {\bf 1:} The photospheric Solar Abundance Distribution of
Asplund, Grevesse \& Sauval (2005) was used to determine [X/Fe], except Fe for which we assumed an abundance of 7.50.
Our [Fe/H] values were computed differential to the same lines in the sun.\\
{\bf 2:} Lines of these elements are strongly affected by hfs; hfs calculations
         were employed for these abundances.\\
{\bf 3:} For Fe the number in the [X/Fe] column is [Fe/H].\\
{\bf 4:} This abundance was derived from the Co~I 6814\.9\AA\ line, which is cleaner
         than the other two Co~I lines.\\
{\bf References:} BW92 (Brown \& Wallerstein 1992), C04 (Carretta et al. 2004), 
A05 (Alves-Brito et al. 2005), W06 (Wylie et al. 2006), KM08 (Koch \& McWilliam 2008)\\
 }
\end{deluxetable} 


%% file: ms.bbl
\begin{thebibliography}{}

\bibitem[Alonso et al.~(1999)]{a99}
Alonso, A., Arribas, S. \& Martinez-Roger, C. 1999, A\&AS, 140, 261

\bibitem[Alonso et al.~(2001)]{a01}
Alonso, A., Arribas, S. \& Martinez-Roger, C. 2001, A\&AS, 376, 1039


\bibitem[Alves-Brito et al.~(2005)]{a05_01}
Alves-Brito, A., Barbuy, B., Ortolani, S., Momany, Y., Hill, V., Zoccali, M., Renzini, A. et al. 
2005, A\&A, 435, 657

\bibitem[Asplund et al.~(2005)]{a05_02}
Asplund, M., Grevesse, N. \& Sauval, A.J. 2005,
``Cosmic Abundances as Records of Stellar Evolution and Nucleosynthesis'', ASP Conference Series vol. 336, 2005, p.~25;
Austin, Texas; eds. F.N. Bash \& T.G. Barnes (also astro-ph/0410214).


\bibitem[Bai et al. (2004)]{1704}
Bai, G.S., Zhao, G., Chen, Y.Q., Shi, J.R., Klochkova, V.G., Panchuk, V.E., Qui, H.M., \&
Zhang, H.W. 2004, A\&A 425, 671

\bibitem[Bernstein \& McWilliam (2005)]{bm05}
Bernstein, R.A., \& McWilliam, A. 2005, in ``Resolved Stellar Populations'', ASP Conf Series,
                 eds. D. Valls-Gabaud and M. Chavez  (astro-ph/0507042)

\bibitem[Brodie \& Huchra (1990)]{bh90}
Brodie, J.P., \& Huchra, J.P. 1990, ApJ, 362, 503

\bibitem[Brown et al. (1994)]{bsl94}
Briley, M.M., Smith, V.V., \& Lambert, D.L. 1994, ApJ, 424, L119

\bibitem[Brown \& Wallerstein (1992)]{bw92}
Brown, J.A., \& Wallerstein, G. 1992, AJ, 104, 1818

\bibitem[Carretta et al.~2004]{cg47t}
Carretta, E., Gratton, R.G., Bragaglia, A., Bonifacio, P., \& Pasquini, L. 2004, A\&A, 416, 925

\bibitem[Cohen 1978]{cohen78}
Cohen, J.G., 1978, ApJ, 223, 487

\bibitem[Cassisi \& Salaris 1997]{cs97}
Cassisi, S., \& Salaris, M. 1997, \mnras, 285, 593

\bibitem[Cassisi, Salaris & Irwin 2003]{1727}
Cassisi, S., Salaris, M. \& Irwin, A.W. 2003, ApJ, 588, 862

\bibitem[Castelli \& Kurucz~2004]{1730}
Castelli, F., \& Kurucz, R.L. 2004, IAU Symposium 210, ``Modelling of Stellar Atmospheres'',
eds. N. Piskanov et al., poster A20 (astro-ph/0405087)

\bibitem[Castilho et al.~2000]{castilho}
Castilho, B.V., Pasquini, L., Allen, D.M., Barbuy, B., \& Molaro, P. 2000, A\&A, 361, 92

\bibitem[Edvardsson (1993)]{1737}
Edvardsson, B., Andersen, J., Gustafsson, B., Lambert, D.L., Nissen, P.E., \& Tomkin, J.
    1993, \aap, 275, 101

\bibitem[Elson \& Santiago (1996)]{1741}
Elson R.A.W., \& Santiago, B.X. 1996, \mnras, 278, 617

\bibitem[Faber (1973)]{1744}
Faber, S.M. 1973, ApJ, 179, 731

\bibitem[Faber~\&~Jackson~(1976)]{1747}
Faber, S.M., \& Jackson R.E. 1976, ApJ, 204, 668

\bibitem[Faber et al.~(1985)]{faber85}
Faber, S.M., Friel, E.D., Burstein, D.,, \& Gaskell, C.M. 1985, ApJS, 57, 711

\bibitem[Forte, Strom \& Strom 1981]{fss81}
Forte, J.C., Strom, S.W., \& Strom, K.M. 1981, ApJ, 245, L9

\bibitem[Ferguson et al. 2005]{1756}
Ferguson, J.W., Alexander, D.R., Allard, F., Barman, T., Bodnarik, J.G., Hauschildt, P.H., Heffner-Wong, A., Tamanai, A.
2005, ApJ, 623, 585.


\bibitem[Ferraro et al. 1997]{1761}
Ferraro, F.R., Carretta, E., Bragaglia, A., Renzini, A., Ortolani, S.

\bibitem[Fulbright, McWilliam \& Rich 2006]{fmr06}
Fulbright, J., McWilliam, A., \& Rich, R.M. 2006, ApJ, 636, 821

\bibitem[Geisler et al. 1996]{g96}
Geisler, D., Lee, M.G., \& Kim, E. 1996, AJ, 111, 1529

\bibitem[Girardi et al. 2000]{1770}
Girardi et al. 2000, AAS 141, 371

\bibitem[Gouliermis et al. 2004]{1773}
Gouliermis, D., Keller, S.C., Kontizas, M., Kontizas, E., Bellas-Velidis, I. 2004, A\&A, 416, 137

\bibitem[Salasnich et al. 2000]{1776}
Salasnich et al. 2000, AA 361, 1023

\bibitem[Gratton et al. 1994]{1779}
Gratton, R.G. \& Sneden, C. 1994, \aap, 287, 927

\bibitem[Gratton et al. 2003]{1782}
Gratton, R.G., Bragaglia, A., Carretta, E., Clementini, G., Desider, S., Grundahl, F., \&
       Lucatello, S. 2003, AA, 408, 529

\bibitem[Grevesse \& Sauval (1999)]{1786}
Grevesse, N., \& Sauval, A.J. 1999, A\&A, 347, 348

\bibitem[Guhathakurta et al. 1992]{1789}
Guhathakurta, P., Yanny, B., Schneider, D.P., \& Bahcall, J.N. 1992, AJ, 104, 1790

\bibitem[Harris 1991]{1792}
Harris, W.E. 1991, ARA\&A, 29, 543

\bibitem[Houdashelt et al. 2000]{1795}
Houdashelt, M.L., Bell, R.A., Sweigart, A.V., \& Wing, R.F. 2000, ApJ, 119, 1424

\bibitem[Howell et al. 2000]{1798}
Howell, J.H., Guhathakurta, P., \& Gilliland, R.L. 2000, PASP, 112, 1200

\bibitem[Johnson 2002]{1801}
Johnson, J.A. 2002, ApJS, 139, 219

\bibitem[Kaluzny 1997]{1804}                 
Kaluzny, J. 1997, A\&AS, 122, 1

\bibitem[Kaluzny et al. 1998]{1807}                 
Kaluzny, J., Kubiak, M., Szymanski, M., Udalski, A., Krzeminski, W., Mateo, M. \& Stanek, K.Z.
        1998, A\&AS, 128, 19

\bibitem[Kaufer et al.~2004]{1811}                   
Kaufer, A., Venn, K.A., Tolstoy, E., Pinte, C., Kudritzki, R.P. 2004, AJ, 127, 2723

\bibitem[King et al.~1995]{1814}                   
King, I.R., Sosin, C., \& Cool, A.M. 1995, ApJ, 452, L36

\bibitem[Koch \& McWilliam (2008)]{1817}                   
Koch, A., \& McWilliam, A. 2007, in preparation

\bibitem[Kroupa~2002]{1820}                   
Kroupa, P., 2002, Science, 295, 82

\bibitem[Ku\check{c}inskas et al.~2006]{1823}                   
Ku$\check{c}$inskas, A., Hauschildt, P.H., Brott, I., Vansevi$\check{c}$ius, V., Lindegren, L., Tanab$\acute{e}$, T., 
\& Allard, F. 2006, A\&A, 452, 1021

\bibitem[Letarte et al. 2006]{let06}                   
Letarte, B., Hill, V., Jablonka, P., Tolstoy, E., Francois, P., \& Meylan, G. 2006, A\&A, 453, 547

\bibitem[Lebzelter \& Wood 2005]{1830}                   
Lebzelter, T. \& Wood, P.R. 2005, A\&A, 441, 1117

\bibitem[Luck \& Bond (1985)]{lb85}
Luck, R.E., \& Bond, H.E. 1985, ApJ, 292, 559

\bibitem[Maraston 2005]{1836}
Maraston, C. 2005, MNRAS, 362, 799

\bibitem[Maraston et al. 2006]{1839}
Maraston, C., Daddi, E., Renzini, A., Cimatti, A., Dickinson, M., Papovich, C., 
           Pasquali, A., \&  Pirzcal, N. 2006, ApJ, 652, 85

\bibitem[McWilliam 1997]{1843}
McWilliam, A. 1997, ARAA, 35, 503

\bibitem[McWilliam \& Bernstein 2002]{1846}
McWilliam, A., \& Bernstein, R.A. 2002, in ``Extragalactic Star Clusters'', IAU Symposium 207,
       Held in Pucon, Chile March 12-16, 2001. 
       Edited by D. Geisler, E.K. Grebel, and D. Minniti.
       San Francisco: Astronomical Society of the Pacific, 2002., p.739

\bibitem[McWilliam, Rich \& Smecker Hane 2003]{1852}
McWilliam, A., Rich, R.M., Smecker Hane, T.A. 2003, ApJ, 592, L21

\bibitem[Mishenina et al. 2002]{1855}
Mishenina, T.V., Kovtyukh, V.V., Soubiran, C., Travaglio, C., \& Busso, M. 2002, 
A\&A, 396, 189

\bibitem[Nissen et al. (2000)]{1859}
Nissen, P.E., Chen, Y.Q., Schuster, W.J., \& Zhao, G. 2000, A\&A, 353, 722

\bibitem[Paresce et al. 1995]{1862}
Paresce, F., de Marchi, G., Jedrzejewski, R. 1995, ApJ, 442, L57

\bibitem[Peng, Ford \& Freeman (2004)]{1865}
Peng, E.W., Ford, H.C., \& Freeman, K.C. 2004, \apj, 602, 705

\bibitem[Peterson (1986)]{peterson86}
Peterson, C.J. 1986, \pasp, 98, 192

\bibitem[Pietrinferni et al. (2006)]{1868}
Pietrinferni, A., Cassisi, S., Salaris, M., \& Castelli, F. 2006,  ApJ, 642, 797

\bibitem[Pilachowski et al. 1980]{p80}
Pilachowski, C.A., Canterna, R., \& Wallerstein, G. 1980, ApJ, 235, L21

\bibitem[Plez 1998]{1874}
Plez, B. 1998, A\&A, 337, 495

\bibitem[Prochaska \& McWilliam (2000)]{1877}
Prochaska, J.X., \& McWilliam, A. 2000, ApJ, 537, 57


\bibitem[Pryor \& Meylan 1993]{1881}
Pryor, C. \& Meylan, G. 1993, ASP Conference Series, vol. 50, 357.
             S.G. Djorgovski and G. Meylan (eds.)


\bibitem[Racine, Oke \& Searle 1978]{1886}
Raceine, R., Oke, J.B., \& Searle, L. 1978, ApJ, 223, 82

\bibitem[Schiavon et al. 2002]{1889}
Schiavon, R.P., Faber, S.M., Rose, J.A., \& Castilho, B.V. 2002, ApJ, 580, 873

\bibitem[Schlegel et al. 1998]{1892}
Schlegel, D.J., Finkbeiner, D.P., \& Davis, M. 1998, ApJ, 500, 525

\bibitem[Simmerer et al.(2003)]{1895}
Simmerer, J., Sneden, C., Ivans, I.I., Kraft, R.P., Shetrone, M.D., \& 
Smith, V.V. 2003, AJ, 125, 2018

\bibitem[Smith \& Lambert (1985)]{1899}
Smith, V.V. \& Lambert, D.L., 1985, ApJ, 294, 326

\bibitem[Sneden, C. 1973]{1902}
Sneden, C. 1973, ApJ, 184, 839

\bibitem[Sneden et al. 1973]{s91}
Sneden, C., Kraft, R.P., Prosser, C.F., \& Langer, G.E. 1991, AJ, 102, 2001

\bibitem[Sobeck et al. 2006]{1908}
Sobeck, J.S., Ivans, I.I., Simmerer, J.A., Sneden, C., Hoeflich, P., Fulbright, 
J.P., \& Kraft, R.P. 2006, ApJ, 131, 2949

\bibitem[Trager 2004]{1912}                   
Trager, S.C. 2004, in {\em Origin and Evolution of the Elements}, Carnegie Observatories 
                 Centennial Symposia, vol. 4. Cambridge University Press. 
                 Edited by A. McWilliam and M. Rauch, 2004, p. 391.


\bibitem[Venn et al.~2001]{1918}                   
Venn, K.A., Lennon, D.J., Kaufer, A., McCarthy, J.K., Przybilla, N., Kudritzki, R.P.,
Lemke, M., Skillman, E.D., \& Smartt, S.J. 2001, ApJ, 547, 765


\bibitem[Wallerstein 1962]{w62}                   
Wallerstein, G. 1962, ApJS, 6, 407

\bibitem[Whitmore et al. 1995]{1926}                   
Whitmore, B.C., Sparks, W.B., Lucas, R.A., Macchetto, F.D., \& Biretta, J.A. 1995, \apj, 454, L73

\bibitem[Winkler 1997]{1929}                   
Winkler, 1997, MNRAS, 287, 481

\bibitem[Worthey et al.~(1994)]{worthey94}
Worthey, G., Faber, S.M., Gonzalez, J.J., \& Burstein D. 1994, ApJS, 94, 687

\bibitem[Wylie et al. (2006)]{1935}                   
Wylie, E.C., Cottrell, P.L., Sneden, C.A., \& Lattanzio, J.C. 2006, ApJ, 649, 248

\bibitem[Zinn (1985)]{1938}                   
Zinn, R. 1985, ApJ, 293, 424

\end{thebibliography}
